\documentclass[10pt,journal,compsoc]{IEEEtran}
\ifCLASSOPTIONcompsoc
  \usepackage[nocompress]{cite}
\else
  \usepackage{cite}
\fi
\ifCLASSINFOpdf
\else
\fi
\usepackage[cmex10]{amsmath}
\usepackage{amssymb}
\usepackage{algorithm}
\usepackage{algcompatible}

\usepackage{graphicx}
\usepackage{caption}
\usepackage{subcaption}

\usepackage{stfloats}

\usepackage{xcolor}

\begin{document}
%
% paper title
% Titles are generally capitalized except for words such as a, an, and, as,
% at, but, by, for, in, nor, of, on, or, the, to and up, which are usually
% not capitalized unless they are the first or last word of the title.
% Linebreaks \\ can be used within to get better formatting as desired.
% Do not put math or special symbols in the title.
%\title{Towards Real-Time Detection and Tracking of Blob-Filaments in Fusion Plasma Big Data}
\title{Towards Real-Time Detection and Tracking of Spatio-Temporal Features: Blob-Filaments in Fusion Plasma
}
%
%
% author names and IEEE memberships
% note positions of commas and nonbreaking spaces ( ~ ) LaTeX will not break
% a structure at a ~ so this keeps an author's name from being broken across
% two lines.
% use \thanks{} to gain access to the first footnote area
% a separate \thanks must be used for each paragraph as LaTeX2e's \thanks
% was not built to handle multiple paragraphs
%
%
%\IEEEcompsocitemizethanks is a special \thanks that produces the bulleted
% lists the Computer Society journals use for "first footnote" author
% affiliations. Use \IEEEcompsocthanksitem which works much like \item
% for each affiliation group. When not in compsoc mode,
% \IEEEcompsocitemizethanks becomes like \thanks and
% \IEEEcompsocthanksitem becomes a line break with idention. This
% facilitates dual compilation, although admittedly the differences in the
% desired content of \author between the different types of papers makes a
% one-size-fits-all approach a daunting prospect. For instance, compsoc 
% journal papers have the author affiliations above the "Manuscript
% received ..."  text while in non-compsoc journals this is reversed. Sigh.

\author{Lingfei~Wu,
        Kesheng~Wu, 
        Alex~Sim, 
        Michael~Churchill, \\
        Jong~Y.~Choi, 
        Andreas~Stathopoulos,
        CS~Chang, 
        and~Scott~Klasky %~\IEEEmembership{Member,~IEEE}% <-this % stops a space
\IEEEcompsocitemizethanks{
\IEEEcompsocthanksitem L. Wu and A. Stathopoulos are with the Department of Computer Science, College of William and Mary, Williamsburg, VA, 23185.\protect\\
% note need leading \protect in front of \\ to get a newline within \thanks as
% \\ is fragile and will error, could use \hfil\break instead.
E-mail: \{lfwu,andreas\}@cs.wm.edu
\IEEEcompsocthanksitem K. Wu and A. Sim are with Lawrence Berkeley National laboratory, Berkeley, CA, 94720.\protect\\
E-mail: \{KWu,ASim\}@lbl.gov
\IEEEcompsocthanksitem M. Churchill and C. Chang are with Princeton Plasma Physics Laboratory, Princeton, NJ, 08536.\protect\\
E-mail: rmchurch@mit.edu, cschang@pppl.gov
\IEEEcompsocthanksitem J. Choi and S. Klasky are with Oak Ridge National Laboratory, Oak Ridge, TN, 37831.\protect\\
E-mail: \{choij,klasky\}@ornl.gov}% <-this % stops an unwanted space
\thanks{Manuscript received in May 4th, 2015. Revised manuscript received in April 18th, 2016.}
}

% note the % following the last \IEEEmembership and also \thanks - 
% these prevent an unwanted space from occurring between the last author name
% and the end of the author line. i.e., if you had this:
% 
% \author{....lastname \thanks{...} \thanks{...} }
%                     ^------------^------------^----Do not want these spaces!
%
% a space would be appended to the last name and could cause every name on that
% line to be shifted left slightly. This is one of those "LaTeX things". For
% instance, "\textbf{A} \textbf{B}" will typeset as "A B" not "AB". To get
% "AB" then you have to do: "\textbf{A}\textbf{B}"
% \thanks is no different in this regard, so shield the last } of each \thanks
% that ends a line with a % and do not let a space in before the next \thanks.
% Spaces after \IEEEmembership other than the last one are OK (and needed) as
% you are supposed to have spaces between the names. For what it is worth,
% this is a minor point as most people would not even notice if the said evil
% space somehow managed to creep in.

% The paper headers
\markboth{IEEE Transactions on Big Data,~Vol.~XX, No.~X, June~2016}
{Wu \MakeLowercase{\textit{et al.}}: Spatio-temporal Features}
% The only time the second header will appear is for the odd numbered pages
% after the title page when using the twoside option.
% 
% *** Note that you probably will NOT want to include the author's ***
% *** name in the headers of peer review papers.                   ***
% You can use \ifCLASSOPTIONpeerreview for conditional compilation here if
% you desire.

% The publisher's ID mark at the bottom of the page is less important with
% Computer Society journal papers as those publications place the marks
% outside of the main text columns and, therefore, unlike regular IEEE
% journals, the available text space is not reduced by their presence.
% If you want to put a publisher's ID mark on the page you can do it like
% this:
%\IEEEpubid{0000--0000/00\$00.00~\copyright~2014 IEEE}
% or like this to get the Computer Society new two part style.
%\IEEEpubid{\makebox[\columnwidth]{\hfill 0000--0000/00/\$00.00~\copyright~2014 IEEE}%
%\hspace{\columnsep}\makebox[\columnwidth]{Published by the IEEE Computer Society\hfill}}
% Remember, if you use this you must call \IEEEpubidadjcol in the second
% column for its text to clear the IEEEpubid mark (Computer Society jorunal
% papers don't need this extra clearance.)

% use for special paper notices
%\IEEEspecialpapernotice{(Invited Paper)}

% for Computer Society papers, we must declare the abstract and index terms
% PRIOR to the title within the \IEEEtitleabstractindextext IEEEtran
% command as these need to go into the title area created by \maketitle.
% As a general rule, do not put math, special symbols or citations
% in the abstract or keywords.
\IEEEtitleabstractindextext{%
\begin{abstract}
%  This work was originally motivated the need to identify and track blob
%  filaments, a feature often associated with instability in magnetically
%  confined fusion plasma.  Understanding and mitigating such instability
%  would improve fusion reactors and make fusion a truly inexhaustible
%  source of clean energy.  
A novel algorithm and implementation of real-time identification and tracking of blob-filaments in fusion reactor data is presented.
  Similar spatio-temporal features are important
  in many other applications, for example, ignition kernels in
  combustion and tumor cells in a medical image.  This work presents an
  approach for extracting these features by dividing the overall task
  into three steps: local identification of feature cells, grouping
  feature cells into extended feature, and tracking movement of feature
  through overlapping in space.  Through our extensive work in
  parallelization, we demonstrate that this approach can effectively
  make use of a large number of compute nodes to detect
  and track blob-filaments in real time in fusion plasma.  On a set of 30GB fusion simulation
   data, we observed linear speedup
  on 1024 processes and completed blob detection in less than three
  milliseconds using Edison, a Cray XC30 system at NERSC.
\end{abstract}

% Note that keywords are not normally used for peerreview papers.
}

% make the title area
\maketitle

% To allow for easy dual compilation without having to reenter the
% abstract/keywords data, the \IEEEtitleabstractindextext text will
% not be used in maketitle, but will appear (i.e., to be "transported")
% here as \IEEEdisplaynontitleabstractindextext when the compsoc 
% or transmag modes are not selected <OR> if conference mode is selected 
% - because all conference papers position the abstract like regular
% papers do.
\IEEEdisplaynontitleabstractindextext
% \IEEEdisplaynontitleabstractindextext has no effect when using
% compsoc or transmag under a non-conference mode.

% For peer review papers, you can put extra information on the cover
% page as needed:
% \ifCLASSOPTIONpeerreview
% \begin{center} \bfseries EDICS Category: 3-BBND \end{center}
% \fi
%
% For peerreview papers, this IEEEtran command inserts a page break and
% creates the second title. It will be ignored for other modes.
\IEEEpeerreviewmaketitle

\IEEEraisesectionheading{\section{Introduction}\label{sec:introduction}}
% Computer Society journal (but not conference!) papers do something unusual
% with the very first section heading (almost always called "Introduction").
% They place it ABOVE the main text! IEEEtran.cls does not automatically do
% this for you, but you can achieve this effect with the provided
% \IEEEraisesectionheading{} command. Note the need to keep any \label that
% is to refer to the section immediately after \section in the above as
% \IEEEraisesectionheading puts \section within a raised box.

% The very first letter is a 2 line initial drop letter followed
% by the rest of the first word in caps (small caps for compsoc).
% 
% form to use if the first word consists of a single letter:
% \IEEEPARstart{A}{demo} file is ....
% 
% form to use if you need the single drop letter followed by
% normal text (unknown if ever used by IEEE):
% \IEEEPARstart{A}{}demo file is ....
% 
% Some journals put the first two words in caps:
% \IEEEPARstart{T}{his demo} file is ....
% 
% Here we have the typical use of a "T" for an initial drop letter
% and "HIS" in caps to complete the first word.
\IEEEPARstart{A} wide variety of ``big data'' such as simulations of
diesel combustion and images of tissue from biopsy, are spatiotemporal in
nature~\cite{Cressie:2011:SST, wu2011finding}.  When analyzing these
data sets, a common task is to find coherent structure in space and
time, for example, ignition kernels in combustion, and cancerous cells
in medical images.  There are many possible approaches to identify such
a feature based on the application requirements~\cite{chandola2009anomaly, Han2011Data, Gupta2014Outlier, yang2005generalized}.  However, when
faced with tight time constraints many of these techniques are too slow
to produce an satisfactory answer.

Our work was originally motivated by the need to detect spatio-temporal
feature associated with instabilities in fusion plasma.  Magnetic
confinement fusion has the potential to be an inexhaustible source of clean
energy; and billions of dollars have been invested in developing fusion
reactors, like the ITER project \cite{aymar2002iter}. 
However, steady-state plasma confinement
is often interrupted by blob filaments driven by the edge turbulence.  A
blob filament (or blob) is a magnetic-field-aligned plasma structure
that appears near the edge of the confined plasma, and has significantly
higher density and temperature than the surrounding plasma
\cite{d2011convective}.  Blobs can also be considered as outliers
because they are rare events that convect filaments of plasma outwards
towards the containment wall, causing substantial heat loss, degradation
of the plasma confinement, and erosion of the containment wall.  By
identifying and tracking these blob filaments from fusion plasma data
streams, physicists can improve their understanding of the dynamics and
interactions of such coherent structures (blobs) with edge turbulence.

Fusion experiments and simulations could easily produce many terabytes
per second; and features such as blobs have to be detected in
milliseconds in order for the control system to have a chance to take
mitigating actions.  Though there are many well known feature extraction
methods for detecting outliers, they often have some shortcomings.
Classical multi-dimensional outlier detection techniques are designed to
detect global outliers.  However, these techniques do not distinguish
between non-spatial attributes and spatial attributes and do not
consider apriori information about the statistical distribution of the
data \cite{shekhar2003unified}. Since spatio-temporal data types have
unique characteristics and their relations are more complicated than
ordinary data, dedicated outlier detection techniques are typically
required to examine anomalies in data across space and
time\cite{Gupta2014Outlier}.  In this work, we propose an approach for
detecting and tracking spatio-temporal features such as blobs by breaking
down the process into three steps: (1) find cells that satisfying
application specific requirements, (2) group cells into spatial features,
and (3) track features by the amount of overlap in space.  By
varying the first step, this procedure could be applying to different
applications.  Earlier, this
approach was applied to data from regular meshes~\cite{wu2009fastbit,
  wu2011finding}.  In this work, we will demonstrated that it can also
be effectively applied to irregular mesh data.

\begin{figure}[!h]
\centering
\includegraphics[width=3.4in]{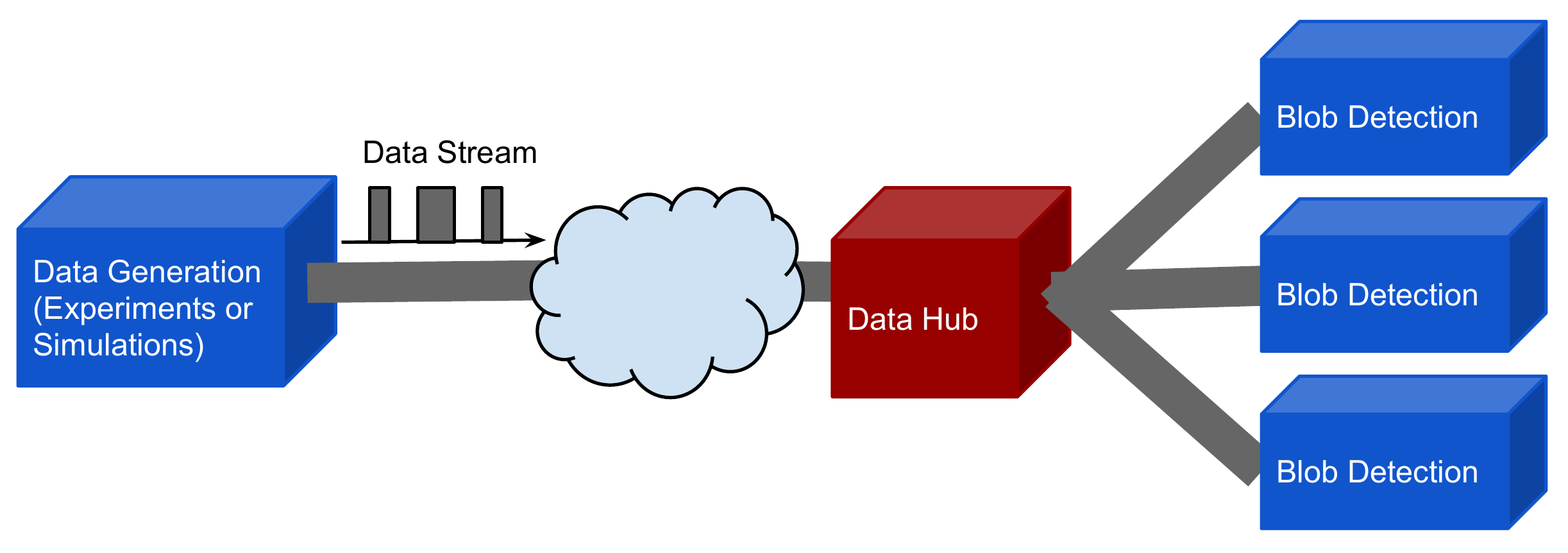}
 \caption{A real-time data analysis frame for finding blob-filaments in fusion plasma data streams}
\label{fig:fusion data stream}
\end{figure}

This work addresses several challenges exemplified by
the detection of blobs in fusion plasma. First off, fusion experiments
and numerical simulations can easily generate massive amounts of data
per run. During a magnetic fusion device experiment (or "shot"),
terabytes of data are generated over short time periods (on the order of
hundreds of seconds). In the XGC1 fusion simulation
\cite{chang2009compressed,ku2009full}, a few tens of terabytes can be
generated per second. Timely access to this amount of data can already
be a challenge \cite{Dong2013Expediting,Dong2013SDS}, but analyzing all
this data in real time is impractical. Currently, there are three types
of analyses in most of fusion experiments: in-shot-analysis,
between-shot-analysis, and post-run-analysis. All existing blob
detection methods address post-run-analysis, but in this work, we focus
on the more challenging first two cases to provide a real-time analysis
so that scientists can monitor the progress of fusion
experiments. Figure \ref{fig:fusion data stream} presents a real-time
analysis frame for finding blob-filaments in fusion plasma data
streams. To perform this data analysis in real time, we utilize
effectively modern supercomputers to address the high volume and
velocity challenges arising from fusion plasma big data.

This work has been integrated into the International Collaboration Framework for Extreme Scale Experiments (ICEE), a wide-area in-transit data analysis framework for near real-time scientific applications \cite{choi2013icee}. ICEE takes advantage of an efficient IO solution ADIOS \cite{lofstead2008flexible}, and a cutting-edge indexing solution FastBit \cite{wu2009fastbit}, to design and construct a real-time remote data processing framework over wide-area networks for international collaborations such as the ITER project. In this system, a blob detection algorithm is used to monitor the health of fusion experiments at the Korea Superconducting Tokamak Advanced Research (KSTAR). However, existing data analysis approaches are often single-threaded, only for post-run analysis, and take a long time to produce results. Also, compared to the simulation data, the resolution of the raw camera data may be coarse, but interesting features can still be identified after normalization. In order to meet real-time feedback requirement, we develop a real-time blob detection method, which can leverage in-situ raw data in the ICEE server and find blob-filaments efficiently during fusion experiments. Our blob detection algorithm is not limited to KSTAR only, and can be applied to other fusion experiments and simulations.

In this research, we apply the three-step approach to detect and track blob structures in 
fusion data, with the goal of achieving millisecond response time on terabytes of data.  
With this response space, it is possible for the control system of the magnetic confinement
fusion reactor to implement mitigating strategies in real time.  To the
best of our knowledge, this is the first time a blob detection method
could satisfy the \emph{millisecond time requirement}.  Additional
contributions of this work include:
\begin{itemize}
\item We illustrate how to adopt the three-step approach to detect and track blob-filaments as an example of spatial-temporal feature on an irregular mesh.
\item We propose a two-phase region outlier detection method for finding blob-filaments. In the first phase, we apply a distribution-based outlier detection scheme to identify blob candidate points. In the second phase, we adopt a fast two-pass \emph{connected component labeling} (CCL) algorithm from \cite{wu2009optimizing} to find different region outliers on an irregular mesh. 
\item We develop a high-performance blob detection approach to meet real-time feedback requirements by exploiting many-core architectures in a large cluster.  %This is the first work to achieve real-time blob detection in \emph{only a few milliseconds}. 
\item We propose a scheme to efficiently track the movement of region outliers by linking the centers of the region outlier over consecutive frames.
\item We have implemented our blob detection algorithm with hybrid MPI/OpenMP, and demonstrated the effectiveness and efficiency of our implementation with a set of data from the XGC1 fusion simulations. Our tests show that we can achieve \emph{linear time speedup} and complete blob detection in \emph{two or three milliseconds} using a cluster at NERSC. In addition, we demonstrate that our method is more robust than recently developed state-of-the-art blob detection methods in \cite{davis2014fast,myra2013edge}.
\end{itemize}

The rest of paper is organized as follows. In Section II, we give the problem formulation of the blob detection and discuss related work. In Section III we describe in detail our three-step approach consisting of a two-phase region outlier detection algorithm and a tracking scheme for identifying and tracking blobs. We then present a real-time blob detection approach by leveraging MPI/OpenMP parallelization in a large cluster in Section IV. The blob detection and tracking results and its real time evaluation are shown in Section V. We conclude the paper, and give our future plans in Section VI.

\section{Problem Definition and Related Work}
\label{sec:Problem Definition and Related Work}
In this section, we introduce our problem definition and discuss previous work related to our study. For related work, we first discuss existing research work on outlier detection, and then review previous work on blob detection in fusion plasma domain. 

\subsection{Problem Definition}
\label{subsec:Problem Definition}
Extracting spatial-temporal features play an important role in analyzing scientific and engineering applications, including behavior recognition \cite{dollar2005behavior}, bioinformatics \cite{winkler2010quantitative}, video analysis c\cite{le2011learning}, and health informatics \cite{aminian2002spatio}. Depending on the applications, mining spatial features in one time frame and relationships among spatial objects in and across time frames are extremely challenging tasks due to three reasons. First, the extent and shape of a feature could be an important indicator in determining its influence. However, due to various data type (regular and irregular), it is not easy to apply a generic approach for all applications. Second, effectively incorporating the temporal information in the overall analysis is a necessity to uncover interesting upcoming events. Finally, how to process very large data sets in real time demands appropriately responding to extreme scale computing and big data challenges. In this work, we attack this problem by presenting a three-step approach for detecting and tracking spatio-temporal features in the context of blob-filament detection in fusion plasma.  

\begin{figure}[!t]
\centering
\includegraphics[width=3.0in]{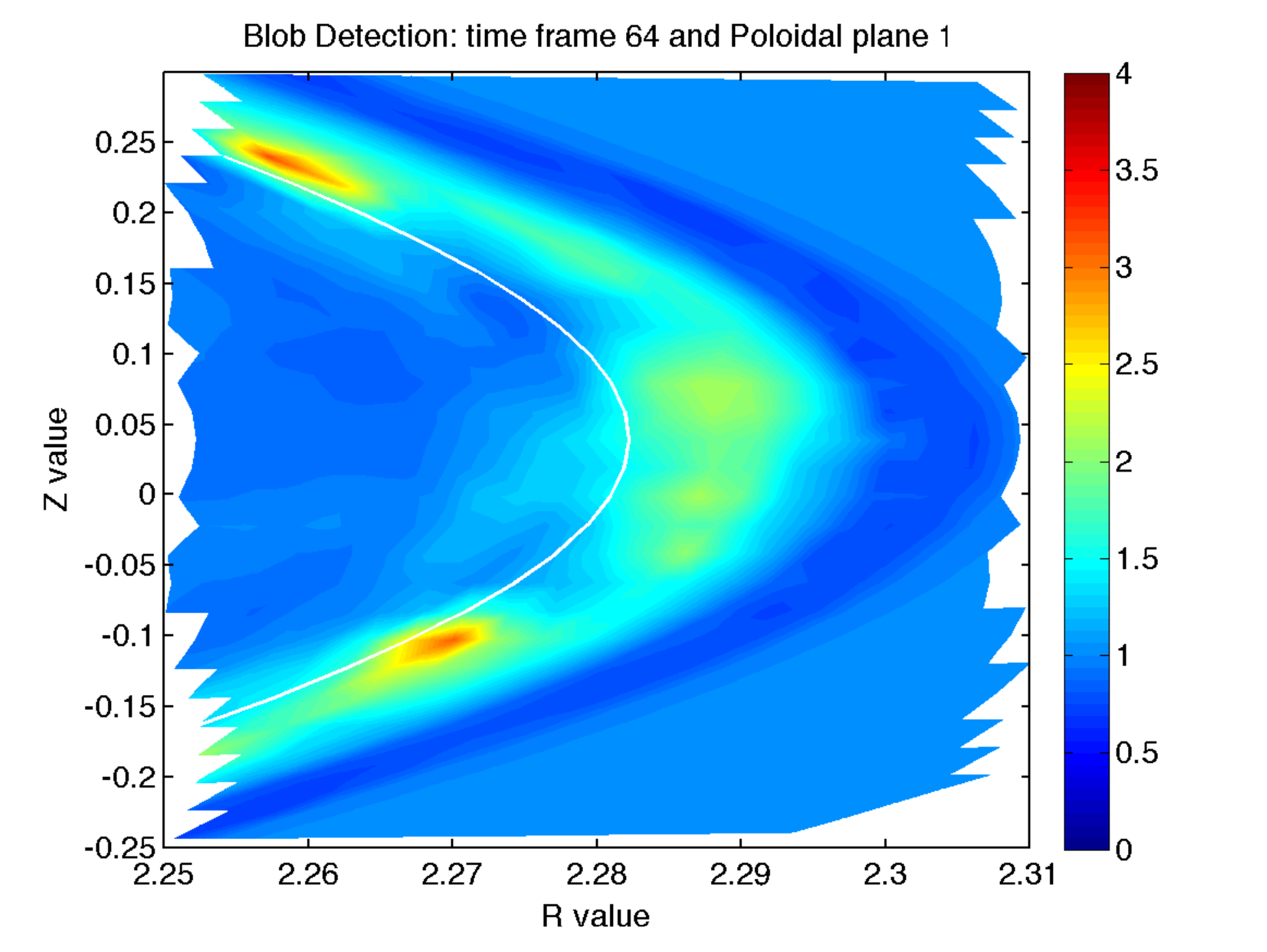}
 \caption{A contour plot of the local normalized density in the region of interests in one time frame in fusion experiments or numerical simulations. A cross-section of the torus is called a poloidal plane. $R$ and $Z$ are cylindrical coordinates and the major radius of the torus is denoted by $R$. }
\label{fig:density_regions}
\end{figure}

The definition of a blob is varied in the literature depending on fusion experiments or simulations as well as available diagnostic information for measurements \cite{d2011convective}. This makes blob detection a challenging task. Figure \ref{fig:density_regions} plots local normalized density distribution in the regions of interest in one time frame. We can observe that there are two reddish spots located at the left portion of the figure, which are associated with blob-filaments and are significantly different from their surrounding neighbors. It is clear that a reddish spot is not a single point but a group of connected points or a region. Therefore, we formulate the blob detection problem as a region outlier detection problem. Similar to the spatial outlier \cite{shekhar2003unified}, a region outlier is a group of spatial connected objects whose non-spatial attribute values are significantly different from those of other spatial surrounding objects in its spatial neighborhood. Figure \ref{fig:density_regions} shows blobs are region outliers. The number of region outliers detected is determined by pre-defined criteria provided by domain experts.

The problem is to design an efficient and effective approach to detect and track different shapes of region outliers simultaneously in fusion plasma data streams. By identifying and monitoring these blob-filaments (region outliers), scientists can gain a better understanding about this phenomena. In addition, a data stream is an ordered sequence of data that arrives continuously and has to be processed online. Due to the high arrival rate of data, the blob detection must finish processing before the next data chunk arrives \cite{sadik2014research}. Therefore, another critical problem is to develop a high-performance blob detection approach in order to meet the real-time requirements. 

\subsection{Outlier Detection}
\label{subsec:Outlier Detection}
The problem of outlier detection has been extensively studied and can be generally classified into four categories: distance-based, density-based, clustering-based, and distribution-based approaches \cite{hodge2004survey,chandola2009anomaly}. 

Distance-based methods \cite{knox1998algorithms} use a distance metric to measure the distances among data points. If the number of data points within a certain distance from the given point is less than pre-defined threshold, then this point is determined as an outlier. This approach could be very useful with accurate pre-defined threshold. However, it may not be proper to use a simple threshold if different densities in various regions of the data exhibit across space or time. 

Density-based methods \cite{breunig2000lof} assign a local outlier factor (LOF) to each sample based on their local density. The LOF determines the degree of outlierness, where samples with high LOF value are identified as outliers. This approach does not require any prior knowledge of underlying distribution of the data. However, it has a high computational complexity since pair-wise distances have to be computed to obtain each local density value.

Clustering-based methods \cite{he2003discovering} conduct clustering-based techniques on the sample points of the data to characterize the local data behavior. Since this method does not focus on outlier detection, the outliers are produced as by-products and it is not optimized for outlier detection. 

Distribution-based methods \cite{shekhar2003unified} applies machine learning techniques to estimate a probability distribution over the data and develop a statistical test to detect outliers. These methods use all dimensions to define a neighborhood for comparison and typically do not distinguish non-spatial attributes from spatial attributes. 

In the context of data streams, a line of research has been devoted to develop efficient outlier detection techniques \cite{subramaniam2006online, pokrajac2007incremental, elahi2008efficient, angiulli2007detecting, aggarwal2011outlier, sadik2014research}. But their main focus is to solve the problem of event detection in sensor network \cite{subramaniam2006online}, query processing \cite{angiulli2007detecting, pokrajac2007incremental}, clustering \cite{elahi2008efficient}, and graph outliers \cite{aggarwal2011outlier}. Therefore, these methods cannot be easily generalized to region outlier detection problems. In addition, the problem of blob detection presents a special challenge, because the spatiotemporal attributes of the blob-filaments has to be considered together to study their various characteristics including speed, direction, movement, and size.  More importantly, these methods are mostly single-threaded which cannot cope with real-time requirements in fusion plasma. 

A number of distributed outlier detection methods have also been studied in \cite{subramaniam2006online, dutta2007distributed, lozano2005parallel, otey2006fast, hung2002parallel}. Most of these methods are seeking an efficient way to parallelize classical outlier detection methods such as distance-based outliers \cite{hung2002parallel,lozano2005parallel}, distribution-based outliers \cite{subramaniam2006online}, density-based outliers \cite{lozano2005parallel}, density-based outliers \cite{otey2006fast}, and PCA-based techniques \cite{dutta2007distributed}. However, there methods are not generally applicable to region outlier detection and tracking. In particular, in order to tackle high volume and velocity challenges arising from fusion plasma big data, specialized outlier detection scheme and suitable high performance computing technique are demanded to complete blob detection in the order of milliseconds.  

In the first two steps of our proposed approach, we apply distribution-based outlier detection to detect outlier points by considering only non-spatial attributes and then leverage fast CCL to construct the region outliers by taking into account spatial-attributes. We choose distribution-based outlier detection since it can solve the problem of finding outliers efficiently if an accurate approximation of a data distribution can be properly found \cite{shekhar2003unified,subramaniam2006online}. Normally the distribution of the stream data may change over time \cite{Gupta2014Outlier}. However, this assumption may not hold in fusion experiments since a fusion experiment lasts very short time period from a few seconds to hundreds of seconds. Therefore, we consider the simpler problem of fixed distribution parameters, noting that several fusion devices have shown similar distribution functions of blob events. Then we can perform exploratory data analysis to compute best fitted distribution parameters offline and then build an accurate online distribution model. We leave the more complicated problem of real-time distribution estimation for future work.

\subsection{Blob Detection in Fusion Plasma}
\label{subsec:Blob Detection in Fusion Plasma}
Independently, fusion blob detection problems have been researched by the physics community in the context of coherent structures in fusion plasma \cite{d2011convective}. Various post-run blob detection methods have been proposed to identify and track these structures, to study the impact of the size, movement and dynamics of blobs. A plasma blob is most commonly determined by some threshold, computed statistically in the local plasma density signal
\cite{xu2012turbulent,fuchert2013influence,zweben1985search}. %\cite{xu2012turbulent,fuchert2013influence,zweben1985search,muller2006probabilistic}.
 However, the exact criteria have varied from one experiment to another, which reflects the intrinsic variability and complexity of the blob structures. In \cite{xu2012turbulent}, a conditional averaging approach is applied to analyze spatio-temporal fluctuation data.
% obtained from a two-dimensional probe array inside the last closed flux surface (LCFS) of the HL-2A tokamak. 
When the vorticity is larger than one standard deviation at some time frame, a blob is considered to be detected by the probe.
 In \cite{fuchert2013influence}, the conditional averaging technique is also used to study the evolution of the blob-filaments using Langmuir probes and a fast camera. %If a reference signal, with a certain sampling interval, has large fluctuation amplitude greater than a specified trigger condition, a blob structure is declared at that time frame. 

Without using a conditional averaging technique, \cite{zweben1985search} searches for blob structures can be done using local measurements of the 2D density data obtained from a 2D probe array. Identification of a blob is based on the choices of several constraints such as the threshold intensity level, the minimum distance of blob movement, and the maximum allowed blob movement between successive frames. The trajectories of the different blobs can be computed with the blob centers based on identification results in each time frame. The seminal work by Zweben, et. al.\cite{zweben1985search} was the first attempt to take only individual time frame data into account to detect blobs and track their movements, although the process of identification of a blob was somewhat arbitrary and oversimplified. 
%In \cite{muller2006probabilistic}, an analysis method was presented in terms of object-related observables to allow a sound probabilistic analysis. After preprocessing the signals from 2D imaging data to form signal matrix, a threshold-segmentation approach is used to identify blob structures when the local density is greater than an appropriately chosen threshold. Bounding polygons are also employed to track blob movements and compute their trajectories. 

Due to the emergence of fast cameras and beam emission spectroscopy in the last decade, the situations of insufficient diagnostic access and limited spatial and temporal resolution have been greatly improved. In the context of computer version, a number of methods have been developed to tackle blob detection problem, which is aimed to detect points or regions in the image that either brighter or darker than the surrounding \cite{kong2013generalized}. Among them, scale-space methods based on the Laplacian of Gaussian \cite{laptev2003space, lindeberg1998feature, collins2003mean} and Watershed detection methods based on local extrema in the intensity landscape \cite{vincent1991watersheds} are two main classes of blob detectors. In \cite{love2007image}, Love and Kumath made the first attempt to apply an image analysis using Watershed techniques for identifying blobs in fusion plasma. The images are first processed to remove the noise spikes, followed by further smoothing using a Gaussian filter, and then identified by various image segmentation techniques.  
However, due to noise and lack of a ground truth image, this approach can be sensitive to the setting of parameters, and it is hard to use generic method for all images. In addition, the output from visualization is not convenient to feed into other analysis \cite{wu2011finding}. The regions of interest computed from this work can be more conveniently fed into other analyses. For instance, one can compute blobs in the regions of interest very quickly and transmit these compact meta information over internet to remote domain scientists for real-time analysis.

Recently, several researchers \cite{davis2014fast,kube2013blob,myra2013edge} have developed a blob-tracking algorithm that uses raw fast camera data directly with GPI technique. In \cite{davis2014fast,myra2013edge}, they leverage a contouring method, database techniques and image analysis software to track the blob motion and changes in the structure of blobs. After normalizing each frame by an average frame created from roughly one thousand frames around the target time frame, the resulting images are contoured and the closed contours satisfying certain size constraints are determined as blobs. Then, an ellipse is fitted to the contour midway between the smallest level contours and the peak. All information about blobs are added into a SQL database for more data analysis. This method is close to our approach but it can not be used for real-time blob detection since they compute time-averaged intensity to normalize the local intensity. Additionally, only closed contours are treated as blobs, which may miss blobs at the edges of the regions of interest. Finally, these methods are still post-run-analysis, which cannot provide real-time feedback in fusion experiments.

\section{Our proposed approach}
\label{sec:Our proposed approach}
%In this section, we provide a detailed description of our proposed approach for finding and tracking blobs over time. 
Given a fusion data stream, which consists of a time ordered sequence of sample frames that arrive continuously from fusion experiments or numerical simulations through remote direct memory access protocols. Our data sets are simulated electron density from the fusion simulation code XGC1 \cite{chang2009compressed,ku2009full}. In the present data sets, simulation data is captured every 2.5 microseconds for a total time window of 2.5 milliseconds. Each point $s_i \in S$ in a time frame $t$ has a spatial attribute $(r,z,t)$ which defines its location in a triangulated measurement grid, and some non-spatial attributes including all important plasma quantities such as electron density $n_e(r,z,t)$  as well as connectivity information in a poloidal plane. The spatial neighborhoods are defined for each point from the connectivity database in a triangulated grid. Formally, an region outlier responding to a blob is defined as a spatial area in the regions of interest where a subset $B_i \subseteq S$ is a group of connected points $s_i$.

\begin{figure}[!t]
\centering
\includegraphics[width=1.5in]{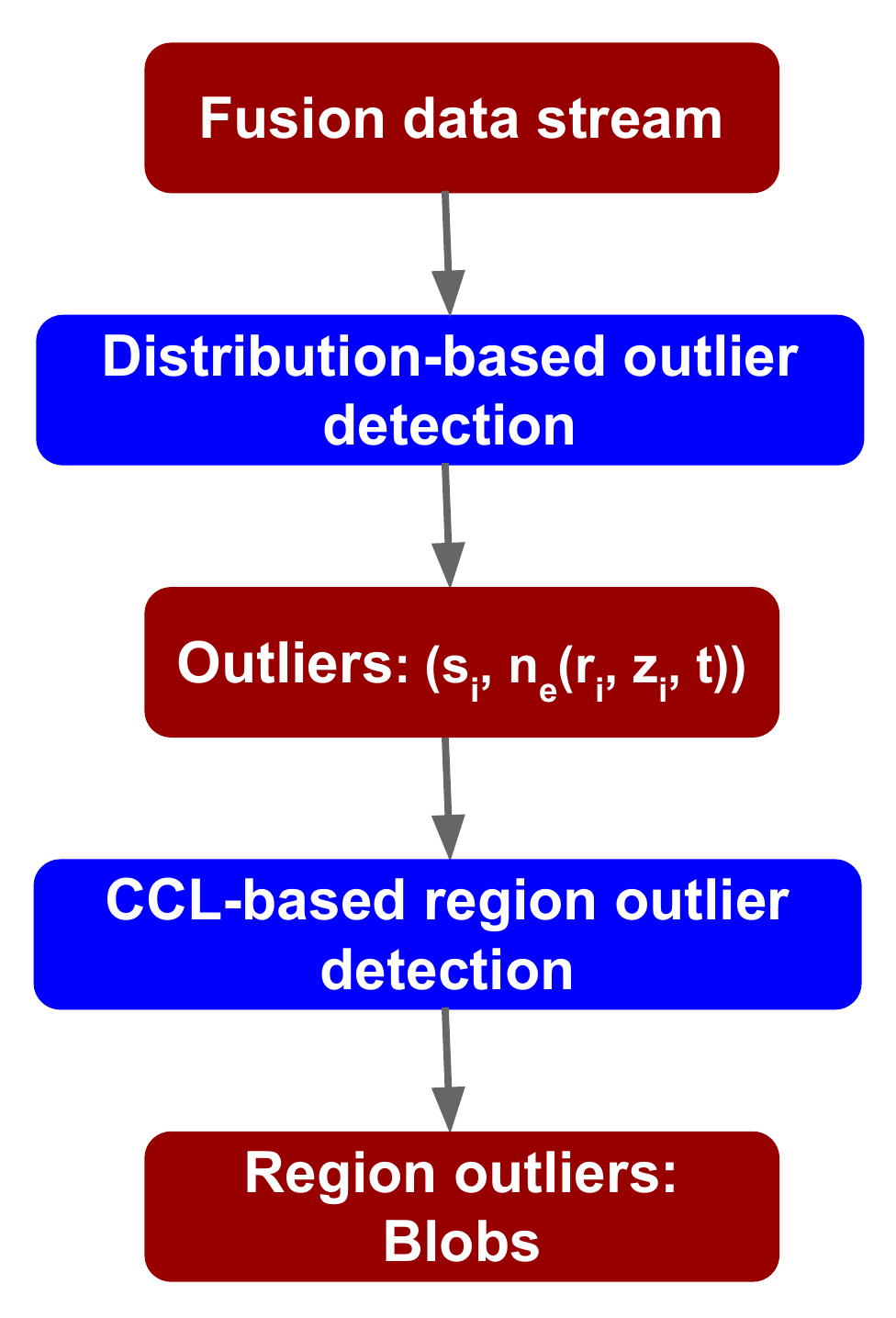}
 \caption{Two-phase region outlier detection for finding blobs}
\label{fig:Two-phase region outlier detection}
\end{figure}

\begin{figure}[!t]
        \centering
        \begin{subfigure}[b]{0.24\textwidth}
                \includegraphics[width=\textwidth]{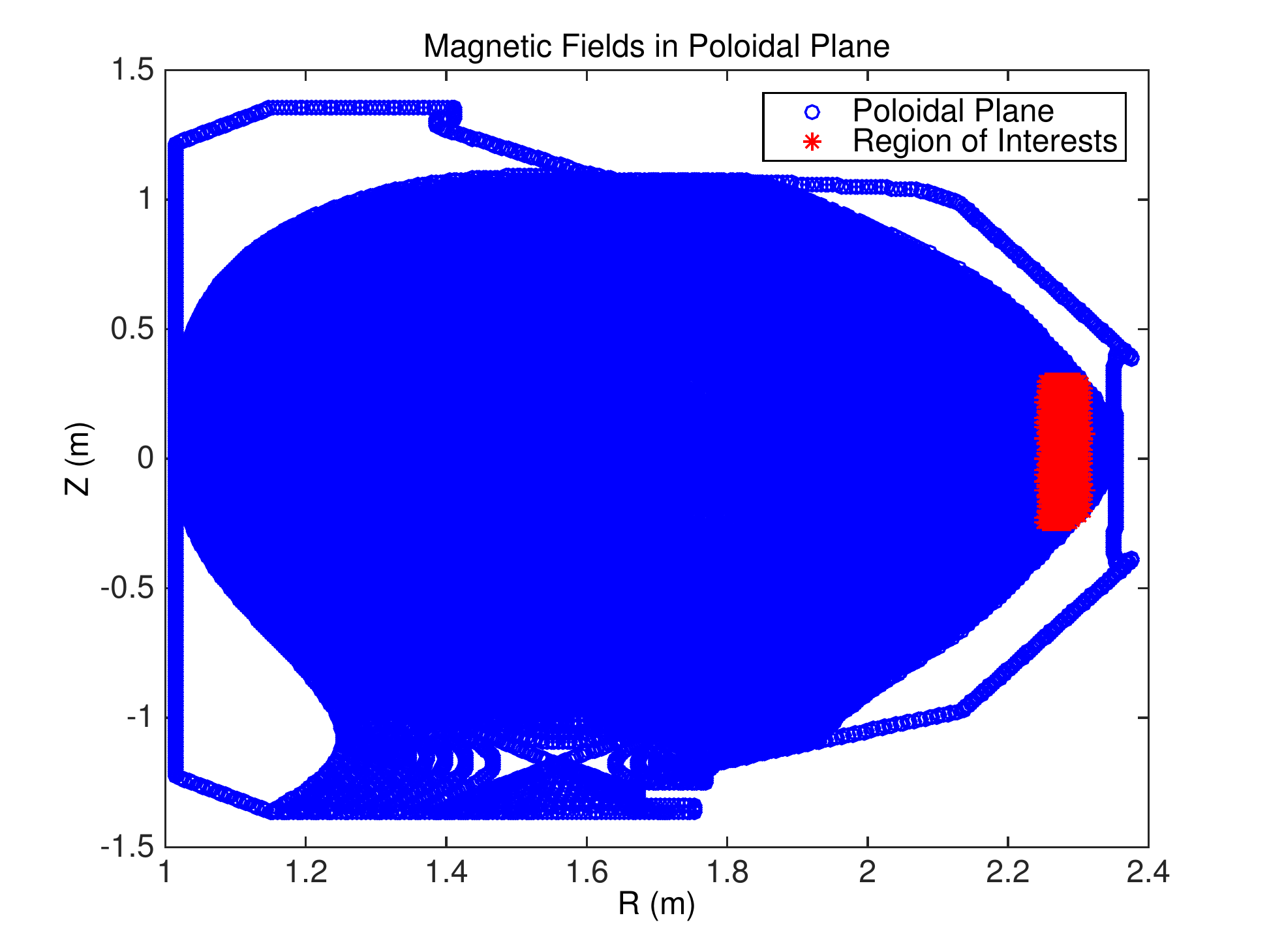}
                \caption{Regions of interest}
                \label{fig:Region of interests}
        \end{subfigure}
        \begin{subfigure}[b]{0.24\textwidth}
                \includegraphics[width=\textwidth]{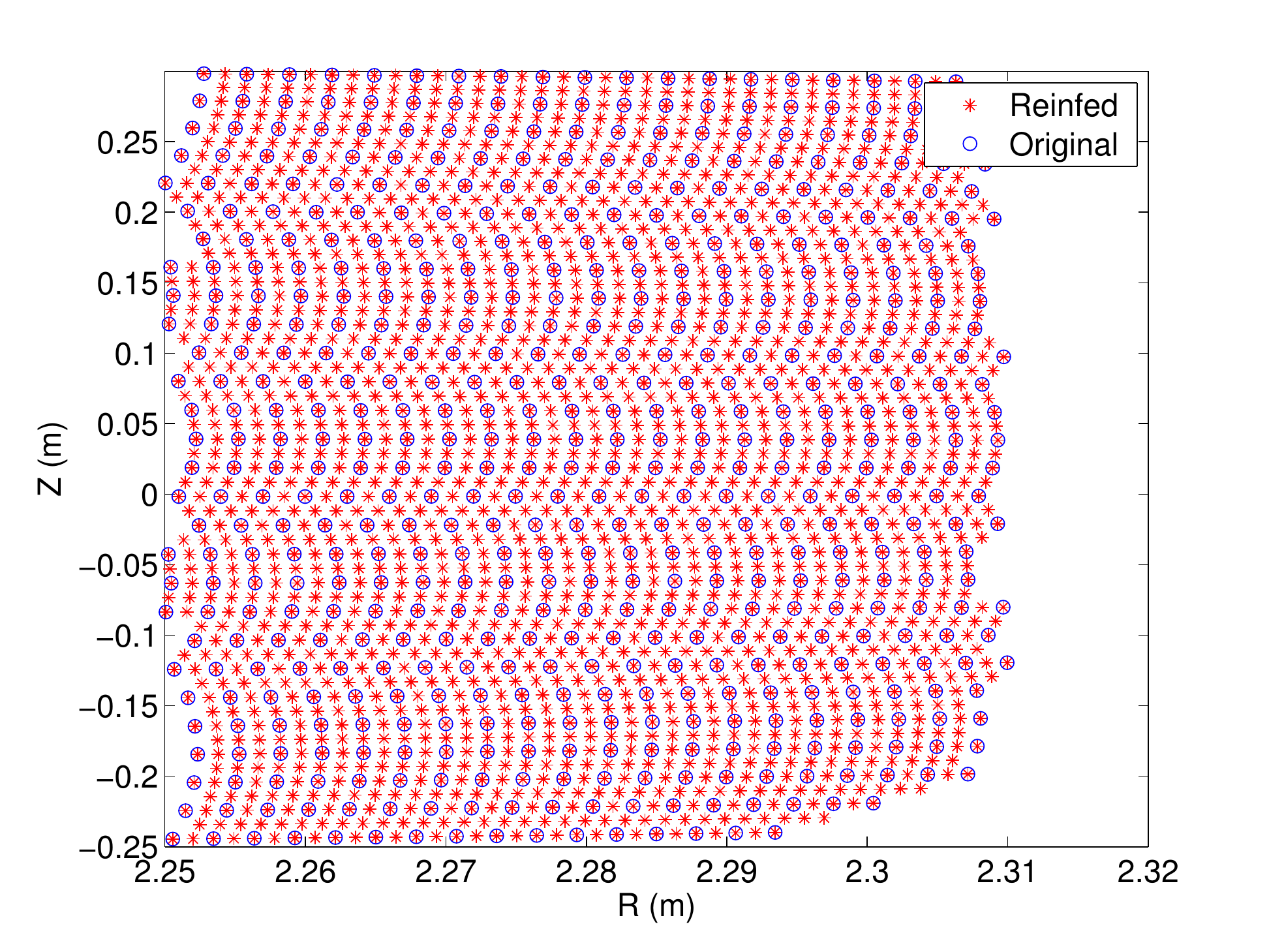}
                \caption{Refined mesh}
                \label{fig:Refined and original vertices}
        \end{subfigure}
        \caption{An example of the regions of interest and the comparison between refined and original triangular mesh vertices in the R (radial) direction and the Z (poloidal) direction.}
        \label{fig:region and refined mesh}
\end{figure}

Our overall goal is to develop an approach to detect and track spatial region outliers (blobs) over time using a stream of fusion data. To achieve this, we break down the process into three steps: (1) find outlier points in the region of interests, (2) group these outlier points as different region outliers, and (3) track these regions outliers by the overlapping in space. We address the first two steps by presenting a two-phase approach, as shown in Figure \ref{fig:Two-phase region outlier detection}. In the first phase, we apply a distribution-based outlier detection algorithm to the fusion data stream in order to detect outlier points which have significantly higher non-spatial attributes than other points. The outputs of this step are tuples $(s_i, n_e(r_i,z_i,t))$, the 2D spatial attributes, and non-spatial attributes such as electron density. These tuples, as well as connectivity information, are used as input for the second phase, where region outlier are detected by applying a fast CCL \cite{wu2009optimizing} to efficiently find different connected components on the triangular mesh. The outputs of the CCL-based region outlier detection algorithm are a set of connected components with outlier points inside, which are associated with blobs if some criteria are satisfied. We address the last step by proposing an efficient blob tracking algorithm by leveraging cues from changes of blobs area and distance of center of blobs.  Note that, by varying the first step, this procedure could be applying to different applications.

%Note that our approach consists of two orthogonal steps, therefore each of the two phases can be replaced by other outlier detection methods. For example, one can leverage density-based outlier detection to find outlier points in the first phase. In addition, edge detection with fuzzy classifier can be used to detect the boundary of region outlier in the second phase \cite{lu2004wavelet}. 

In the following section, we describe the proposed two-phase region outlier detection in detail.

\subsection{Distribution-Based Outlier detection}
\label{subsec:Distribution-Based Outlier detection}
The main task of this phase is to perform efficient outlier detection to determine outlier points which form the region outliers associated with blobs. To facilitate this goal, we propose a novel distribution-based outlier detection algorithm based on the electron density with various criteria for fusion plasma data streams. We separate spatial attributes from non-spatial attributes and consider the statistical distribution of the non-spatial attributes to develop a test based on distribution properties, since it is more suitable for detecting spatial outliers \cite{shekhar2003unified}. As claimed in \cite{subramaniam2006online}, it is very efficient to find outliers by using a data distribution approximation if we estimate the underlying distribution of data accurately. Values for various criteria are determined by domain experts or subjectively by examining the resulting plotting and adjusting them until satisfied. 

In the proposed outlier detection we firstly preprocess the sample frame to compute needed quantities in the region of interests, as shown in Figure \ref{fig:Region of interests}. Then it is analyzed by normalizing the total electron density $n_e(r,z,t)$ (which includes fluctuations) with respect to the initial background electron density, $n_e(r,z,1)$ (if using real diagnostic data from, e.g. GPI, actual emission intensity $I(r,z,t)$ would be used instead of electron density). Note that using the initial time frame as the benchmark is an important factor to achieve real-time blob detection. The normalized electron density in the subsequent time frames can be easily computed, especially compared to the time-average electron density with a long time interval \cite{myra2013edge}.

\begin{algorithm}
\caption{Triangular mesh refinement algorithm}
\label{alg:triangular mesh refinement algorithm}
\begin{algorithmic}[1]
\STATEx \textbf{Input/output:} 
\STATEx	\quad $triGrid$: connectivity array of the triangular mesh 
\STATEx	\quad $(r,z)$: spatial coordinate of each point
\STATEx \quad $n_e$: normalized electron density of each point
\STATE Compute unique edges $E$ and indices vector $I_E$ by sorting and removing duplicates based on $triGrid$ 
\STATE Compute spatial coordinate of each new vertices in the middle of $E$ based on $(r,z)$
\STATE Compute electron density of each new vertices on $E$ by performing linear interpolation based on $n_e$
\STATE Compute indices for each new vertices by adding vector index $I_E$ with the number of original points
\STATE Compute a new triangular mesh by assigning appropriate indices from each new and old vertices
\end{algorithmic}
\end{algorithm}

\begin{figure}[t!]
        \centering
        \begin{subfigure}[b]{0.24\textwidth}
                \includegraphics[width=\textwidth]{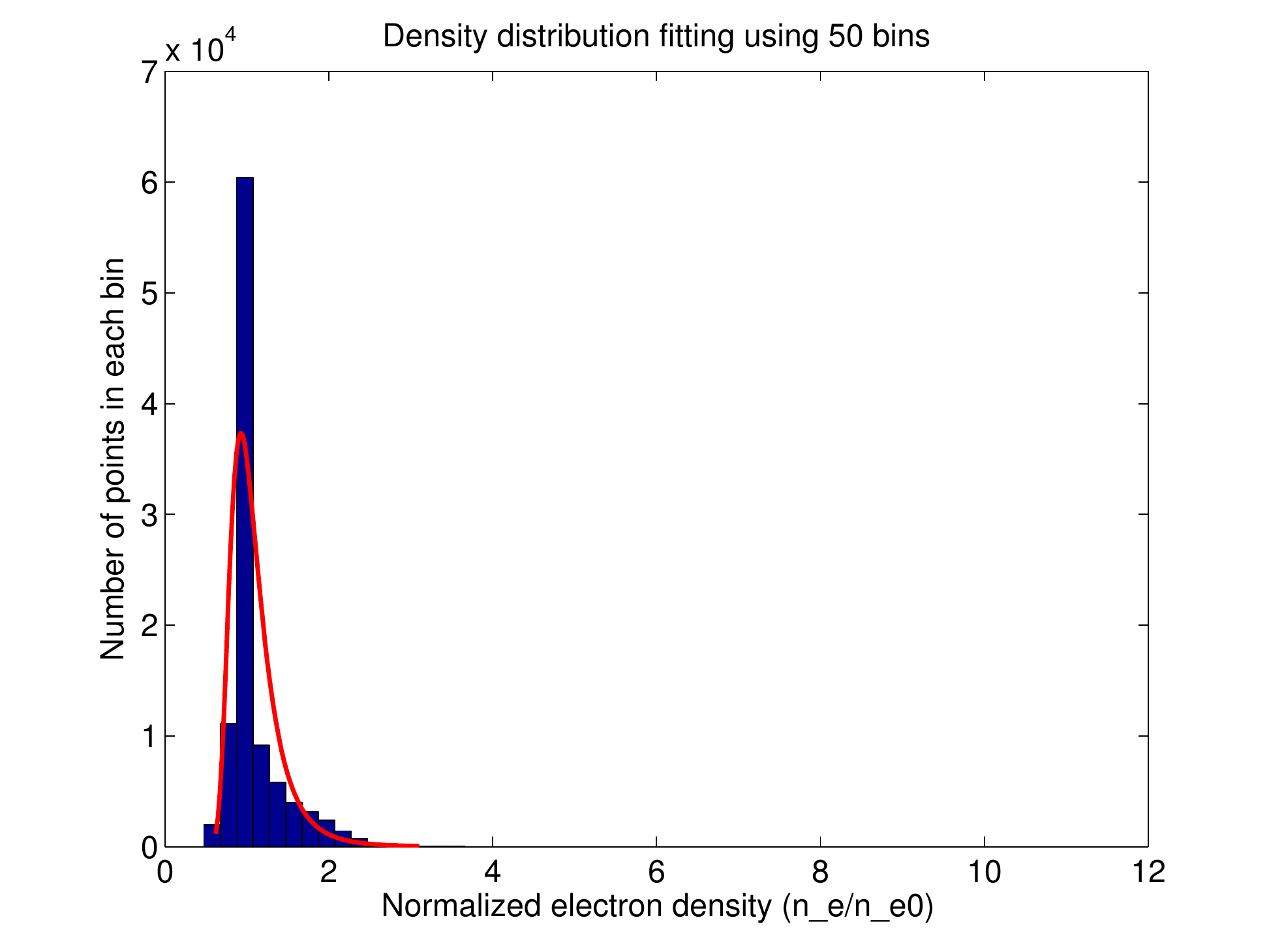}
                \caption{Extreme Value}
                \label{fig: Extreme Value Distribution}
        \end{subfigure}
        \begin{subfigure}[b]{0.24\textwidth}
                \includegraphics[width=\textwidth]{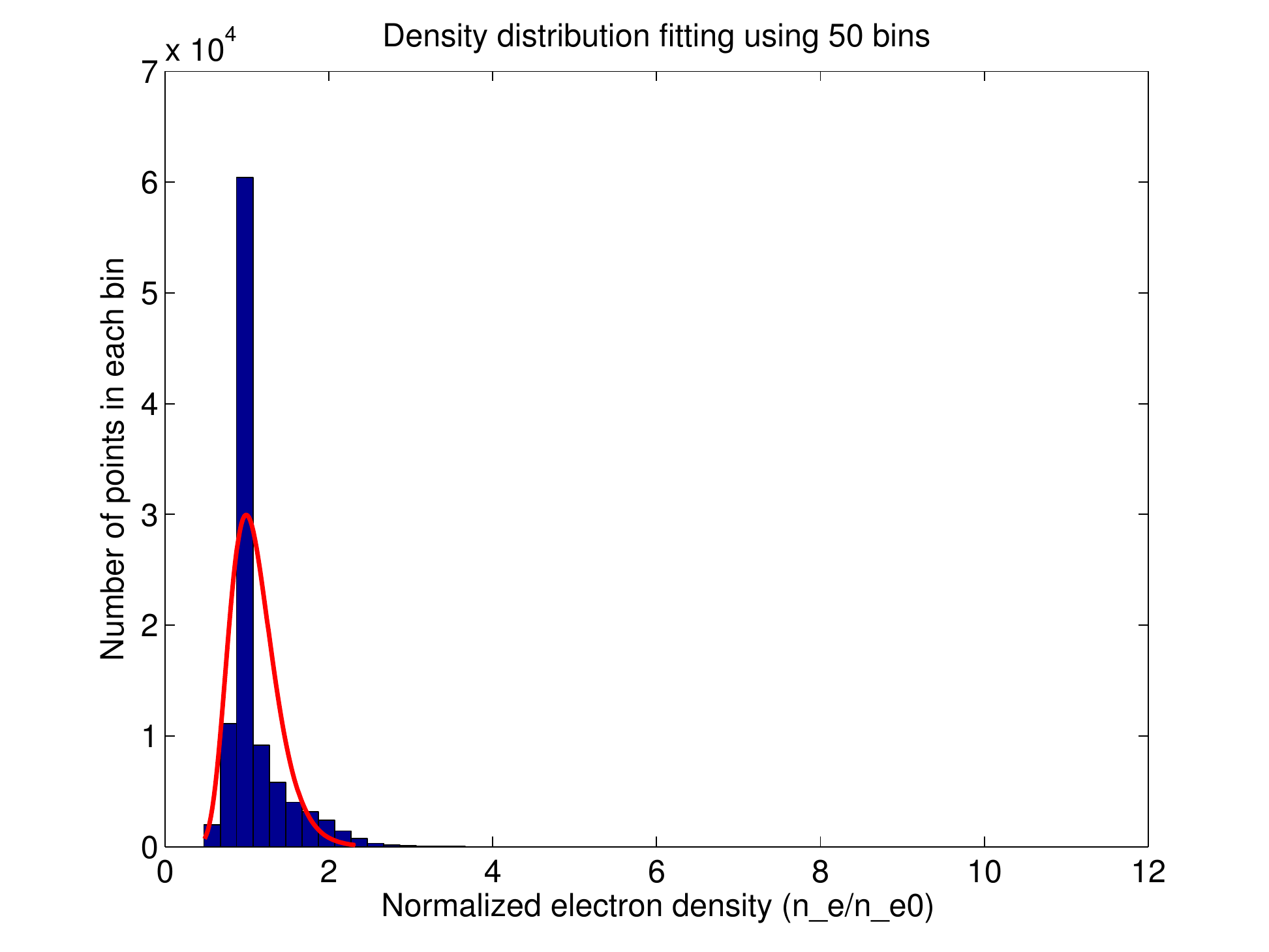}
                \caption{Log Normal}
                \label{fig: Log Normal Distribution}
        \end{subfigure}
        \caption{An example of exploratory data analysis to analyze the underlying distribution of the local normalized density over all poloidal planes and time frames.}
        \label{fig:exploratory data analysis}
\end{figure}

To obtain meaningful region outliers using the CCL method, it is necessary to have fine grained connectivity information. This particular simulation mesh has coarse vertical resolution, so resolution enhancement techniques are applied to generate a higher resolution triangular mesh based on the original triangulated mesh. 
As shown in Algorithm \ref{alg:triangular mesh refinement algorithm}, the resulting triangular mesh is refined to achieve four times better granularity. We split each original triangle into four smaller ones by linking three middle points of the original mesh edges in each triangle. The corresponding density of generated vertices can be obtained using linear interpolation of the original triangular mesh. This step can be applied recursively until the satisfactory resolution of the triangular mesh is computed. Figure \ref{fig:Refined and original vertices} shows the resulting triangular mesh vertices after applying the triangular mesh refinement algorithm once.

In order to apply an appropriate predefined quantile in two-phase distribution-based outlier detection, it is advised to perform exploratory data analysis to exploit main characteristics of the data sets. Figure \ref{fig:exploratory data analysis} reveals that extreme value distribution and log normal distribution are fitted best with one of our sample data sets (after comparing over sixteen different common distributions). After analyzing the underlying distribution, a novel outlier detection is performed to determine outlier points in the regions of interest. The basic idea of the proposed two-step outlier detection is motivated from the observations that there are relatively high density areas (a half banded ellipse area with cyan color) in the edge and several significantly high density small regions (a few small areas with reddish yellow color) in these relatively high density areas, as shown in Figure \ref{fig:density_regions}. The proposed outlier detection method extends the previous approach that applies statistical detection with conditional averaging intensity value \cite{xu2012turbulent, fuchert2013influence}, and applies more intelligent outlier detection with only considering individual time frame data. Compared to traditional single threshold segmentation approach, our approach is more generic, flexible and easier to tune a satisfactory result.

In the first step, the standard deviation $\sigma$ and the expected value $\mu$ are computed over all sixteen poloidal planes in one time frame. Using the best fitted distribution, we apply first step outlier detection to identify the relative high density areas with a specified predefined quantile:
\begin{equation}
N(r_i,z_i,t) - \mu > \alpha \ast \sigma, \forall(r_i,z_i) \in \Gamma
\end{equation}
where $N$ is the normalized electron density, $\alpha$ is the multiple of $\sigma$ associated to the specified predefined quantile and $\Gamma$ is the domain in the region of interests. Once the relative high density regions are determined, we compute another standard deviation $\sigma_2$ and the expected value $\mu_2$ in these areas. Then we employ second step outlier detection to identify the outlier points in the relative high density areas with an appropriately chosen predefined quantile:
\begin{equation}
N(r_i,z_i,t) - \mu_2 > \beta \ast \sigma_2, \forall(r_i,z_i) \in \Gamma_2
\end{equation}
where $\beta$ is the multiple of $\sigma_2$ associated to the judiciously chosen confidence level and $\Gamma_2$ is the domain of blob candidates. In practice, $\alpha$ and $\beta$ can be chosen to be same or different, depending on the characteristics of blob-filaments. In our experience, the $\alpha$ value is generally greater than $\beta$ since the standard deviation $\sigma$ over the region of interests is much smaller than the standard deviation $\sigma_2$ from the relative high density areas. 

However, two-step outlier detection alone cannot be used to distinguish the blob candidates since identified blob candidates may actually have small density, which does not satisfy traditional definition of blobs. Therefore, the density of the mesh points in the outlier points smaller than a certain minimum absolute value criterion need to be filtered out. On the other hand, it is also possible that the middle areas between surrounding plasmas and outlier points have density higher than the given minimum absolute value criterion. Thus, we also apply a minimum relative value criterion to remove these unwanted points. To combine these two rules together, we have a more robust and flexible criterion:
\begin{equation}
N(r_i,z_i,t) > \textbf{max}(d_{ma}, (d_{mr} \ast \mu_2 )), \forall(r_i,z_i) \in \Gamma_3
\end{equation}
where $d_{ma}$ and $d_{mr}$ are minimum absolute value and minimum relative value respectively, and $\Gamma_3$ is the domain of good blob candidates. 

\subsection{CCL-Based Region Outlier Detection}
\label{subsec:CCL-Based Region Outlier Detection}
The main task of the second phase is to apply an efficient connected component labeling algorithm adopted from \cite{wu2009optimizing} on a refined triangular mesh to find different blob candidate components. A connected component labeling algorithm generally considers the problem of labeling binary 2D images with either 4-connectedness or 8-connectedness. It performs an efficient scanning technique, and fills the label array labels so that the neighboring object pixels have the same label. Wu \cite{wu2009optimizing} presents an efficient two-pass labeling algorithm that is much faster than other state-of-the-art methods and theoretically optimal. However, since we process a refined triangular mesh rather than the traditional 2D images, we have modified their CCL algorithm to take the special features of a triangular mesh into account. As shown in Algorithm \ref{alg:connected component labeling algorithm on triangular mesh}, each triangle is scanned first rather than a point. Since we know the three vertices in a triangle are connected, we can reduce unnecessary memory accesses once any vertex in a triangle is found to be connected with another vertex in a different triangle. Then we compute the current minimum parent label in this triangle, and assign each vertex a parent label if its label has already filled or a label if its label has not initialized yet. If all three vertices in a triangle are scanned for the first time, then a new label number is issued and assigned to their labels and the associated parent label. After the label array is filled full, we need flatten the union and find tree. Finally, a second pass is performed to correct labels in the label array, and all blob candidates components are found. Note that to perform efficient union-find operations, the union-find data structure is implemented with a single array as suggested in \cite{wu2009optimizing}.

\begin{algorithm}[h!]
\caption{Connected component labeling algorithm on triangular mesh to find various blob candidates components}
\label{alg:connected component labeling algorithm on triangular mesh}
\begin{algorithmic}[1]
\STATEx \textbf{Input:} 
\STATEx	\quad $triGrid$: connectivity array of the triangular mesh 
\STATEx \textbf{Output:} 
\STATEx	\quad $B_c$: Region structure of each blob candidate
\STATE Initialize $label$, $parentLabel$, and $labnum$
\FOR{Scanning each triangle until the end of $triGrid$} 
	\IF{$label$ of three vertices are all zero}	
	\STATE{Assign a new $labnum$ to all three vertices}
	\STATE{Update $label$ and $parentLabel$ with $labnum$} 
	\ELSE
	\STATE{Find the minimum $parentLabel$ of all three vertices}  
	\STATE{Update their $label$ and $parentLabel$ with this value}	
	\ENDIF
\ENDFOR
\FOR{Scanning until the end of $parentLabel$} 
	\STATE{Update $parentLabel$ by flattening union-find tree}
\ENDFOR
\FOR{Scanning until the end of $Label$} 
	\STATE{Update $Label$ with latest $parentLabel$}
\ENDFOR
\STATE{Find each $B_c$ of points with same $parentLabel$}
\end{algorithmic}
\end{algorithm}

After all blob candidates are determined, a blob is claimed to be found if the median of a blob candidate component satisfies a certain minimum absolute median value criterion. The reason we are setting this constraint to select the blobs is that the minimum value criterion has to be a reasonably small value in order to produce more blob candidate components. It is possible that if the minimum absolute median value criterion is too large, it may also remove the blobs. On the other hand, it is also possible if this value is too small, it does not have effect on filtering out unwanted components. Therefore, with the same philosophy of measurement, a minimum relative median value criterion is also applied to determine the blobs. However, in order to protect the blobs from being removed due to the extremely large mean value $\mu_2$, we also set the maximum absolute median value criterion to limit the power of minimum relative median value criterion. We unify these three rules to be one:
\begin{multline}
N(r_i,z_i,t) > \textbf{max}(\hat{d}_{ma}, min((\hat{d}_{mr} \ast \mu_2 ), \hat{d}_{xa})), \\
\forall(r_i,z_i) \in \Gamma_4
\end{multline}
where $\hat{d}_{ma}$, $\hat{d}_{mr}$ and $\hat{d}_{xa}$ are minimum absolute and relative median values and maximum absolute median value respectively and $\Gamma_4$ is the domain of blobs. 

\subsection{Tracking Region Outliers}
\label{subsec:Tracking Region Outliers}
The objective of the third step is to track the direction and speed of the detected blobs over time. The blob tracking algorithm has to cope with the problem of tracking multiple region outliers simultaneously even when the blobs merge together or split into separated ones. On the other hand, the blob tracking method should be simple and efficient to meet real-time requirements. To achieve this goal, we propose an efficient blob tracking algorithm by leveraging cues from changes of blobs area and distance of center of blobs. We compute the correspondence between previously tracked blobs and currently detected blobs, and then recover the trajectories of the tracked blobs. 

To identify the location center of detected blob, we compute the density-weighted average of the spatial coordinates of each point inside a blob. 
\begin{equation}
(r_c, z_c) = \dfrac{1}{M}\sum_{i=1}^{n} (r,z) n_e 
\end{equation}
where $M$ is summation of $n_e$ of all points in a blob. The density-weighted average is used to better capture the center of density of a blob. We track the movement of these detected blobs by linking the centers in consecutive time frames. In order to obtain the boundary of region outliers (blobs), we compute the convex hull \cite{chan1996optimal} of a set of points in a blob. The area of a blob is computed by counting the number of points in a blob. 

\begin{algorithm}[h!] 
\caption{Efficient blob tracking algorithm}
\label{alg:Efficient blob tracking algorithm}
\begin{algorithmic}[1]
\STATEx \textbf{Input:} 
\STATEx	\quad $B$: Current detected blobs
\STATEx	\quad $T$: Previous blob tracks
\STATEx \textbf{Output:} 
\STATEx	\quad $T$: Updated blob tracks with $B$ appended
\STATE Initialize $hull$, $cen$, and $area$
\STATE $hull$ = getBoundary($B$)
\STATE $cen$ = getCenter($B$)
\STATE $area$ = getArea($B$)
\FOR{Scanning until the end of $B$}
	 \STATE{$cenDis$ = getCenterDis($B$,$T$)}
	 \STATE{$areaDif$ = getAreaDif($B$,$T$)}
	\IF{$cenDis$ $\leq$ maxJump $\wedge$ $areaDif$ $\leq$ maxDif}	
	\STATE{Find a blob track $T$ with smallest $cenDis$} 
	\STATE{Append current blob into this blob track $T$} 
	\ENDIF
\ENDFOR
\STATE{Update $T$ with $hull$, $cen$, $area$, and computed $speed$}
\end{algorithmic}
\end{algorithm}

As shown in Algorithm \ref{alg:Efficient blob tracking algorithm}, the input parameters are current detected blobs and the previous blob tracks. The data structure of a blob track is composed of the track ID, the length of track, the area of previous blob, the time-stamps, the center points, the boundary points, and the velocity. There are two heuristics to verify whether a blob is associated with an existing blob track. The first heuristic is based on the fact that the area of a blob between consecutive time frames cannot decrease or increase significantly. The second heuristic takes into account the distance of the centers of a blob does not change dramatically over very short time period (microseconds). The proper thresholds for these two heuristics are provided by domain experts. Since blobs can appear, disappear, merge together or split, a greedy scheme is applied to find the best matching pair of blob and track based on closest distance of the centers of current detected blob and the latest blob in a blob track. Based on computed correspondence between a blob track and the currently detected blobs, existing blob tracks are automatically processed through corresponding operations such as adding a blob into a track, creating a new track, and a track ending. If the length of a track is smaller than 3 consecutive time frames, the track will be treated an anomaly and deleted due to errors in data or inappropriate blob detection thresholds. The speed and direction of the blobs can thus be computed from two consecutive center points. Finally, we can recover the trajectories of the tracked blobs by monitoring the movement of blob centers. 

\section{A real-time blob detection approach}
\label{sec:A real-time blob detection approach}
Existing blob detection approaches cannot tackle the two challenges of the large amount of data produced in a shot and the real-time requirement. In addition, existing data analysis approaches are often operated in a single thread, only for post-run analysis and often take a few hours to generate the results \cite{muller2006probabilistic}. In order to meet the real-time feedback requirement, we address these challenges by developing a high performance blob detection approach, which can leverage in situ raw data and find blob-filaments efficiently in fusion experiments or numerical simulations.

\subsection{A hybrid MPI/OpenMP parallelization}
\begin{figure}[!t]
\centering
\includegraphics[scale=0.12]{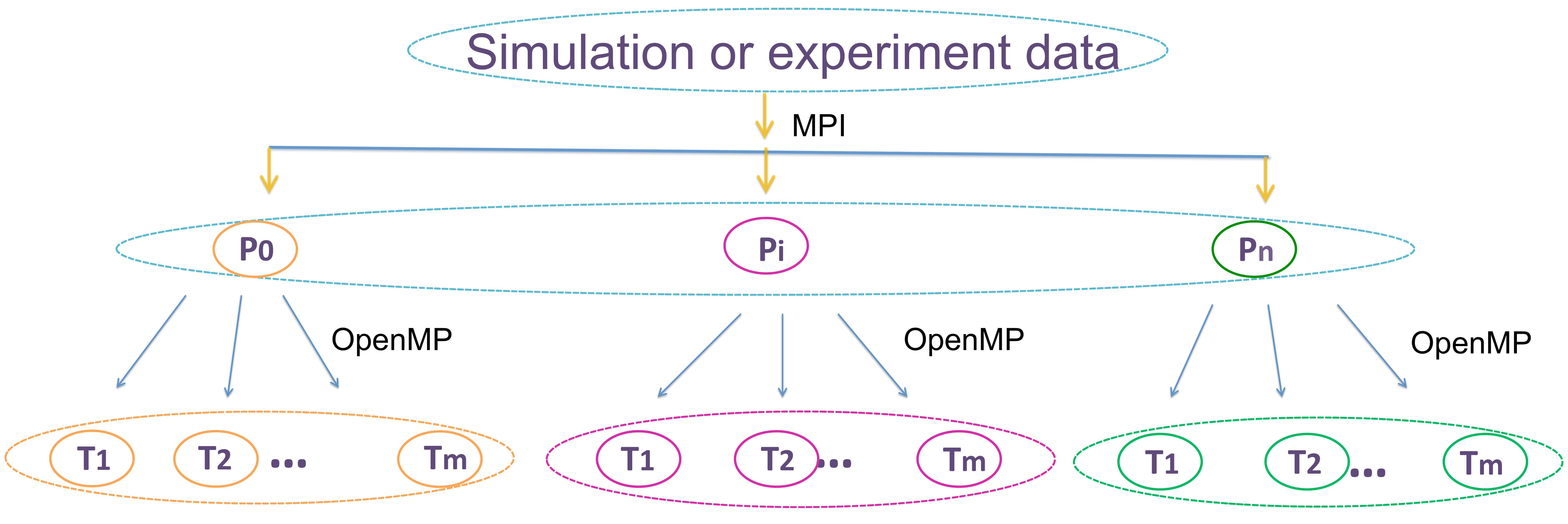}
\caption{Hybrid MPI/OpenMP parallelization}
\label{fig: blob_hybridChat}
\end{figure}

The key idea is to exploit many cores in a large cluster system by running MPI to allocate $n$ processes to process the data in one or several time frames at the high level, and by leveraging OpenMP to accelerate the computations using $m$ threads at the low level. Figure \ref{fig: blob_hybridChat} shows our hybrid MPI/OpenMP parallelization for blob detection. Using this approach, we can complete our blob detection in a few milliseconds with in situ evaluation. 

In order to achieve blob detection in real time, the goal is to minimize data movements in the memory and speed up computation. Ideally, the performance is optimal without any communication if we can perform the job correctly. The proposed blob detection algorithm in the previous section supports embarrassingly parallel since we only need the initial time frame and the target time frame to do the computation. This is an important difference between our blob detection method and recently developed methods \cite{davis2014fast,myra2013edge} in terms of real-time requirement. Furthermore, we explore many-core processor architectures to speed up the computation of each MPI task by taking full advantage of multithreading in the shared memory. Therefore, our real-time blob detection approach based on hybrid MPI/OpenMP parallelization is a natural choice and is expected to provide the optimal performance for fusion plasma data streams.

\begin{figure}[!h]
\centering 
\includegraphics[width=3.0in]{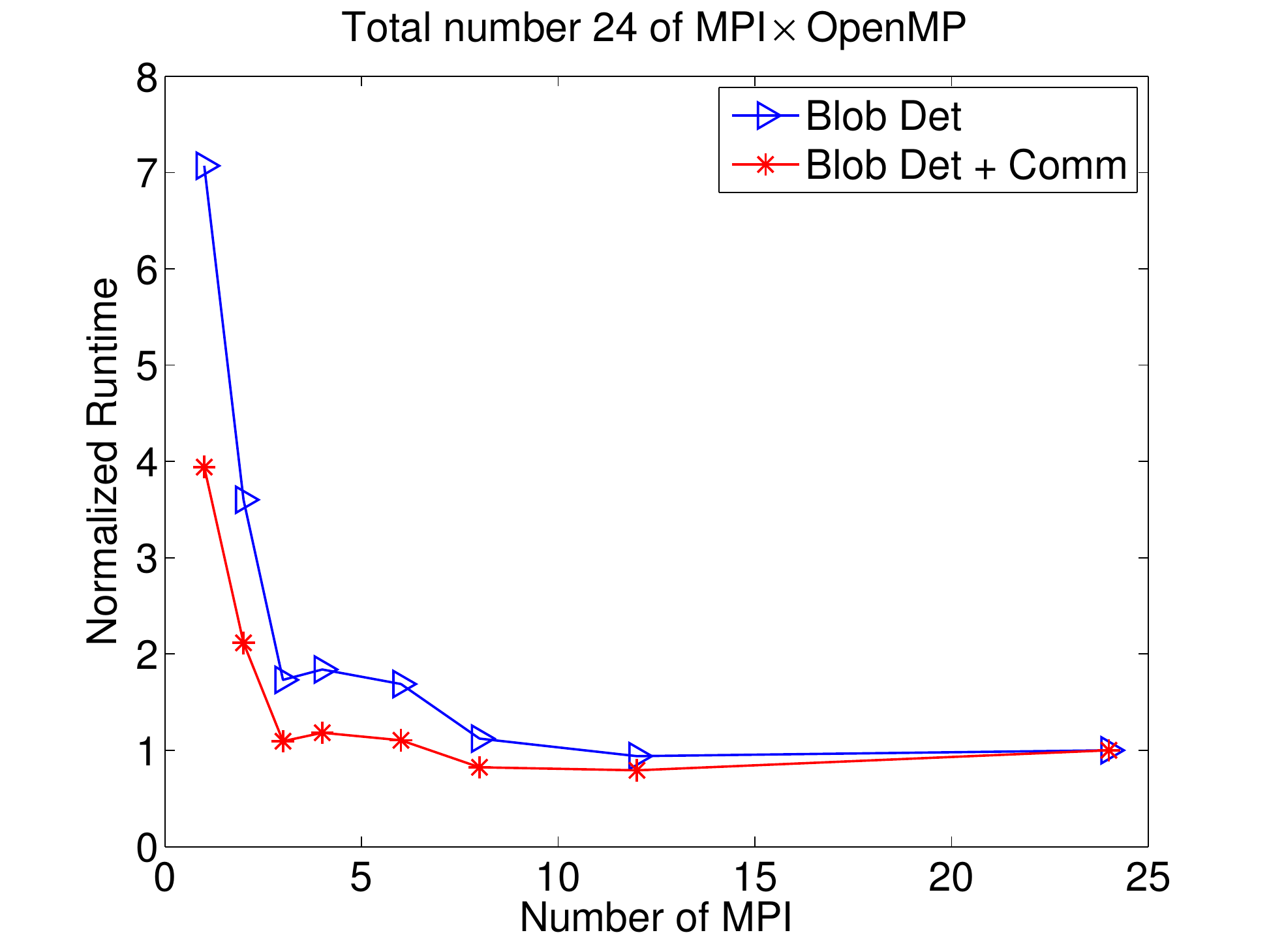}
 \caption{Investigate the performance of hybrid MPI/OpenMP parallelization when varying number of MPI processes and OpenMP threads. The blue triangle denotes only normalized blob detection time. The red star denotes the normalized total time including both blob detection time and initial communication time for broadcasting the first time frame to all analysis nodes for normalization.}
\label{fig:performance of hybrid MPI/OpenMP parallelization}
\end{figure}

A practical interesting issue is how to tune the number of MPI processes and OpenMP threads for the best performance by taking both analysis speed and memory size into account. As shown in Figure \ref{fig:performance of hybrid MPI/OpenMP parallelization}, we vary the number of MPI processes and OpenMP threads but fix the total number to be 24 for investigating the performance when processing the same amount of time frames data. A faster analysis speed is achieved when increasing the number of MPI processes since more data frames can be processed simultaneously. On the other hand, the analysis speed remains constant with a few OpenMP threads and degrades with more OpenMP threads due to lack of enough computation in one time frame. However, more OpenMP threads could significantly reduce the memory demands. Therefore, in this study, we choose the number of OpenMP threads to be four for each MPI task, to achieve a good trade off between analysis speed and memory savings.

\subsection{Outline of the implementation}
We implement our blob detection algorithm in C with a hybrid MPI/OpenMP parallelization. Algorithm \ref{alg:real-time outlier detection algorithm for finding blobs} summarizes the proposed blob detection algorithm without considering OpenMP. Users can specify the regions of interest by (Rmin, Rmax, Zmin, Zmax), the range of target time frames by (t\_start, t\_end), and the location of the data sets. However, with in situ evaluation, there is no need to specify the file location since all data are already in memory. We use static scheduling to evenly divide the number of time frames for each MPI task for efficiency. The $n$ MPI processes are allocated to process one or several time frames and $m$ OpenMP threads are launched to accelerate the computation in one time frame. Note that the MPI process is also the master thread in the runtime environment. At the beginning, the initial time frame data is broadcasted to all MPI processes so that normalization can be performed with new coming time frames. Then each MPI process embarrassingly process the data in each time frame with multithreading in the shared memory. Only detected blobs information are maintained and added into local database. Since these local blobs information are very compact, they can be efficiently transmitted over internet to remote servers for real-time analysis by domain scientists. 

\begin{algorithm}
\caption{A real-time blob detection approach}
\begin{algorithmic}[1]
\STATEx \textbf{Input:} 
\STATEx	\quad Rmin, Rmax, Zmin, Zmax: specify region of interest
\STATEx	\quad t\_start, t\_end : start and end time frames 
\STATEx	\quad FileDir: location where data sets locate
\STATEx \textbf{Output:} 
\STATEx	\quad $B$: Detected region outliers (blobs)
	\STATE Apply static scheduling to assign equal amount of $n$ time frames data to each MPI process
	\STATE Broadcast the initial time frame to all MPI processes
	\FOR{$t = 1:n$}
		\STATE Process $i$ loads raw data in one frame and computes normalized density $n_e(r,z,t)$ in region of interest
		\STATE Refine the triangular mesh. See Algorithm \ref{alg:triangular mesh refinement algorithm}
		\STATE Apply two-phase distribution-based outlier detection to identify outliers with various criteria
		\STATE Apply CCL-based region outlier detection on a triangular mesh to find blob components. See Algorithm \ref{alg:connected component labeling algorithm on triangular mesh}
		\STATE A blob is added into $B$ if certain criteria is satisfied
	\ENDFOR
\end{algorithmic}
\label{alg:real-time outlier detection algorithm for finding blobs}
\end{algorithm}

\section{Experiments and Results}
In this section we present experimental evaluations of our blob detection and tracking algorithms, and report the performance of the real-time blob detection under both strong and weak scaling. Before showing experimental results in the next section, we briefly introduce our experimental environment, data sets, and parameter setting in our blob detection and tracking algorithms. We have tested our implementation on the NERSC's newest supercomputer Edison, where each compute node has two Intel ``Ivy Bridge'' processors (2.4GHz with 12 cores) and 64 GB of memory. Our base data sets are simulation data sets with 1024 time frames based on the XGC1 simulation \cite{chang2009compressed}\cite{ku2009full} from the Princeton Plasma Physics Laboratory, which last around 2.5 milliseconds. One of our main goals is that we can control analysis speed by varying the number of processes to complete the blob detection on the entire data set in a time close to 2.5 milliseconds. It would indicate that our algorithm could monitor fusion experiments in real time (neglecting data transfer latency). If we consider internet transfer latency in real experiments or numerical simulation, the system needs at least 1 to 25 milliseconds to transfer one time frame data depending on size of data, which may give us more time for data analysis.   

Another goal is to validate the effectiveness of the proposed algorithms. In Algorithm \ref{alg:real-time outlier detection algorithm for finding blobs}, we apply various criteria to identify the blobs. The parameters for blob detection and tracking in our experiments are given in Table \ref{ta: blob criteria}. One criterion we have not mentioned in the previous section is parameter ``minArea''. This parameter is used to decide how many points a blob should have, which is used to remove impossibly small blobs. In our experiment, this parameter is set to three since there are at least three vertices connected as a 2D component in a triangular mesh. Another criteria are parameters ``maxFrames'' and ``minFrames'', which are used to control the length of a blob track and remove noisy tracks. It is important to note that these parameters need to be tuned in order to achieve optimal performance in different fusion experiments or numerical simulations. The reasons for this uncertainty in the context of blob detection are from the intrinsic variability and complexity of the blob structures observed in different experiments \cite{d2011convective}.

\begin{table}[htbp]
\centering
\caption{Parameters setting for the proposed blob detection and tracking algorithms on XGC1 simulation data sets}
\label{ta: blob criteria}
\small
\begin{center}  
    \begin{tabular}{ |rr|rr|}
    \hline
    \multicolumn{2}{|c}{{\tt detection criteria}}
	& \multicolumn{2}{c|}{{\tt tracking criteria}} \\  \hline
    minArea					   & 3    &  maxAreaChange  & 25 \\ 
    minRden 	($d_{ma}$)		   & 1.2  &  maxJump   	   & 0.04  \\  
    minAbsden ($d_{mr}$)		   & 2.05 &  maxFrames	   & 100  \\
    maxAbsMden ($\hat d_{xa}$) & 2.75 &  minFrames	   & 3\\ 
    minMden 	($\hat d_{ma}$)	   & 1.3  &  			   &\\ 
    minAbsMden ($\hat d_{mr}$) & 2.15 &   			   &\\ \hline
    \end{tabular}
\end{center}
\end{table}

\subsection{Performance comparison}
\label{subsec:Performance comparison}
We first conduct experiments to compare our method with recently developed state-of-the-art blob detection methods in \cite{davis2014fast,myra2013edge}. Since their methods are based on the contouring methods and thresholding, we call their methods the contouring-based methods. We have to point out that strictly quantitative comparisons are not possible since the blob itself is not well-defined \cite{d2011convective}. Due to this reason, there are rarely direct comparisons between any new proposed method and existing ones in the literature in the domain of fusion plasma \cite{xu2012turbulent,fuchert2013influence,zweben1985search, muller2006probabilistic,love2007image,davis2014fast,kube2013blob, myra2013edge}. However, in order to demonstrate that our methods is more robust than the contouring-based methods, we compare these two methods in two typical cases to shed light on their performance in terms of the detection accuracy. 

\begin{figure*}[t!]
        \centering
        \begin{subfigure}[b]{0.42\textwidth}
                \includegraphics[width=\textwidth]{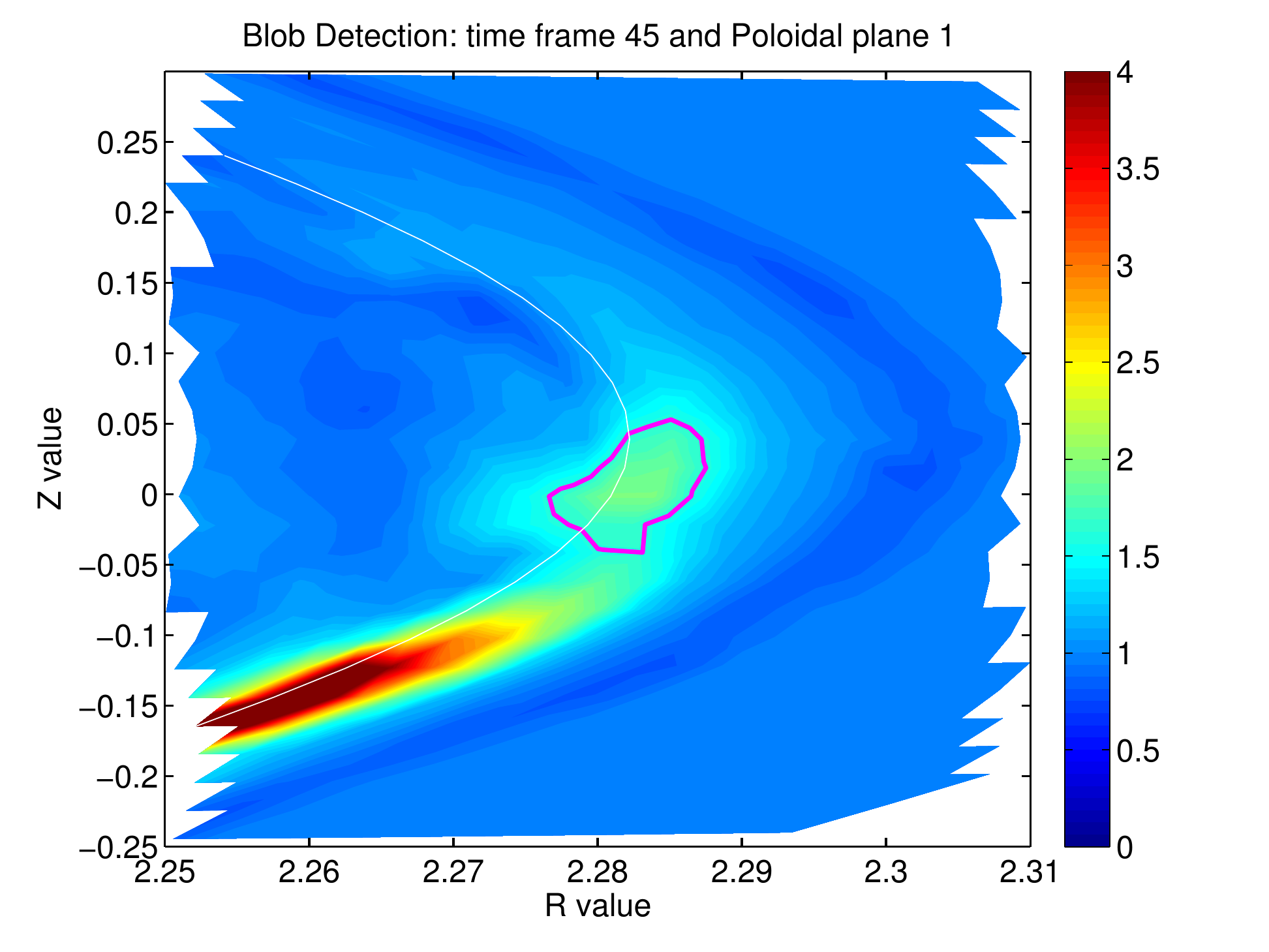}
                \caption{Contouring-based methods}
                \label{fig:contouring method in time frame 45}
        \end{subfigure}
        \begin{subfigure}[b]{0.40\textwidth}
                \includegraphics[width=\textwidth]{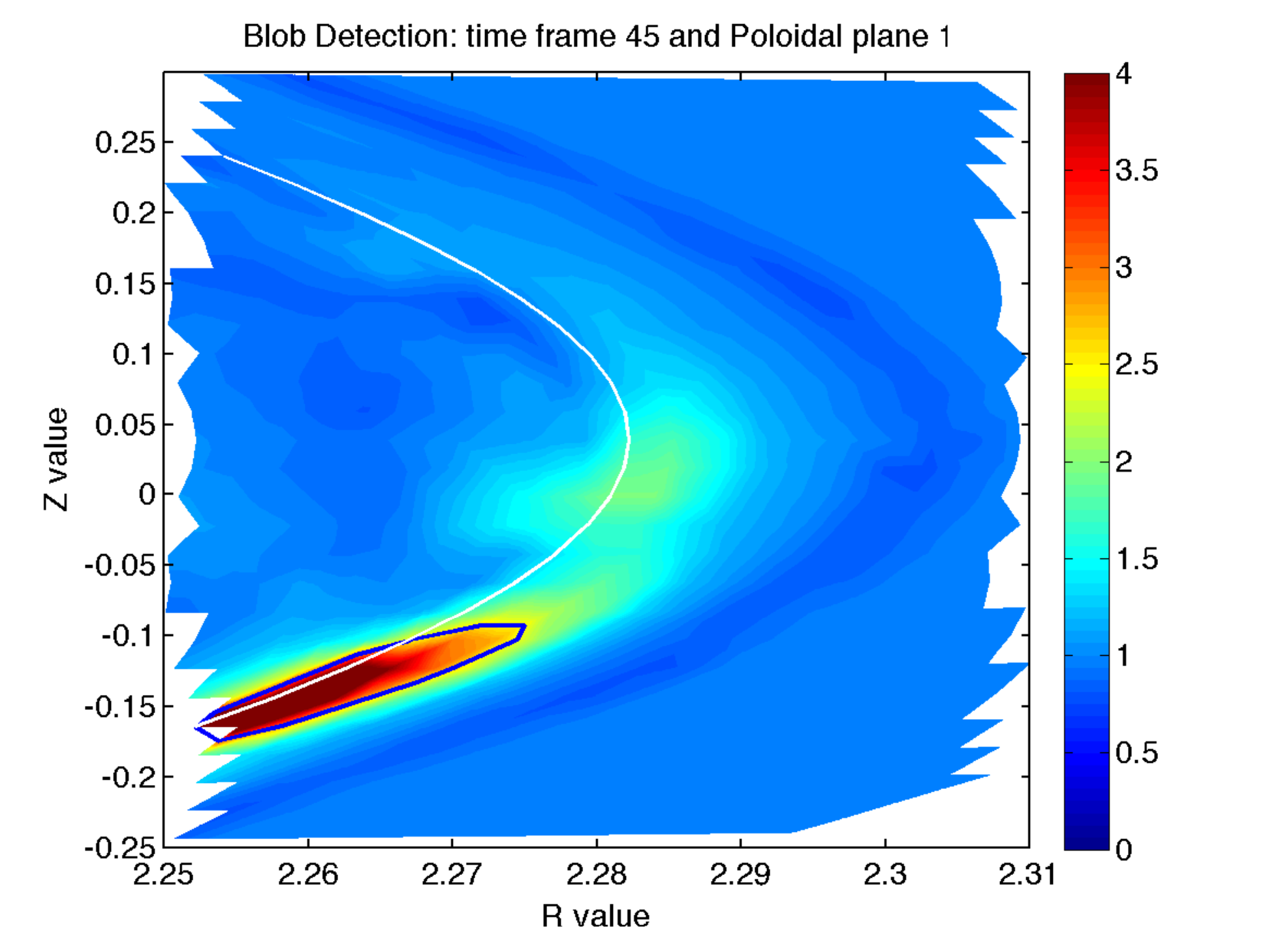}
                \caption{Region outlier detection method}
                \label{fig:region outlier detection method in time frame 45}
        \end{subfigure}
        \begin{subfigure}[b]{0.40\textwidth}
                \includegraphics[width=\textwidth]{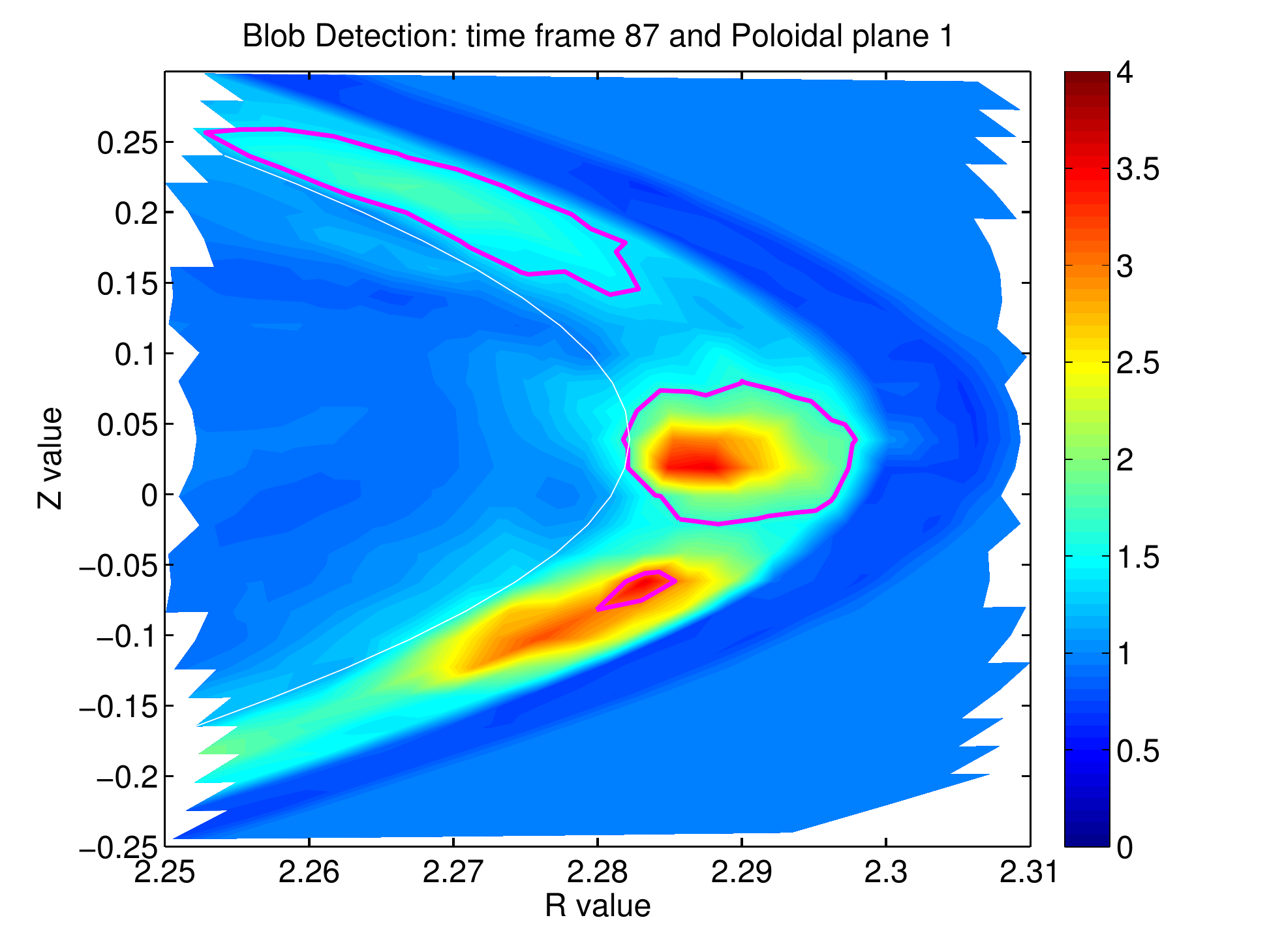}
                \caption{Contouring-based methods}
                \label{fig:contouring method in time frame 87}
        \end{subfigure}        
        \begin{subfigure}[b]{0.40\textwidth}
                \includegraphics[width=\textwidth]{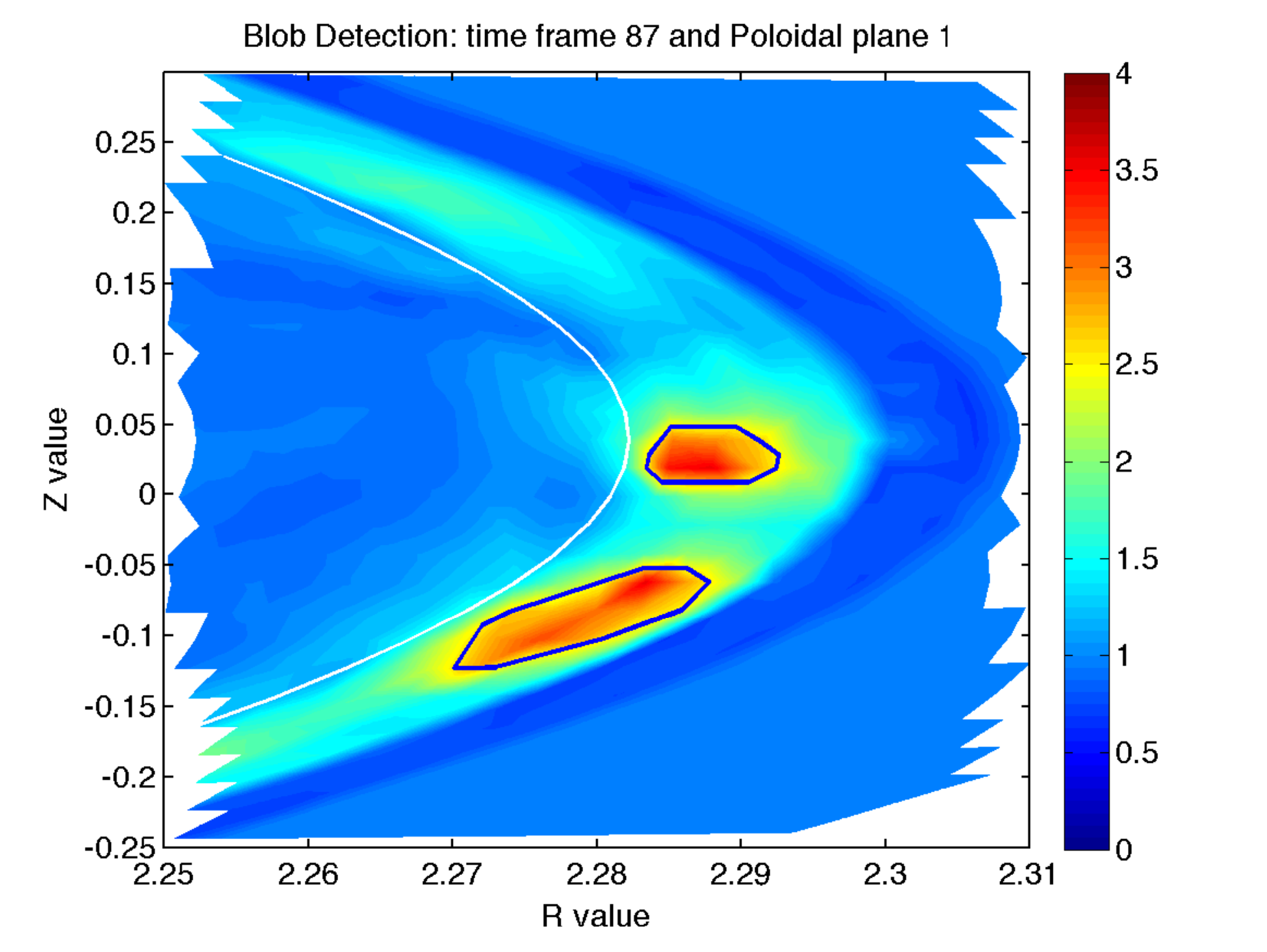}
                \caption{Region outlier detection method}
                \label{fig:region outlier detection method in time frame 87}
        \end{subfigure}        
        \caption{Two examples of comparing our region outlier detection method with the Contouring-based methods in the R (radial) direction and the Z (poloidal) direction. The separatrix position is shown by a white line and the different pink and blue circles denote blobs.}
        \label{fig:comparing our region outlier detection method with the Contouring-based methods}
\end{figure*}

Figure \ref{fig:comparing our region outlier detection method with the Contouring-based methods} shows the comparison of the blob detection results between our region outlier detection method and the contouring-based methods in two different time frames. As shown in Figures \ref{fig:contouring method in time frame 45} and \ref{fig:region outlier detection method in time frame 45}, we can see that our region outlier detection method does not miss detecting the blob at the edge of the regions of interest while the contouring-based methods fail the detection. The reason is that the contouring-based methods require the computed contours are closed, which do not exist at the edge of the regions of interest. In Figures \ref{fig:contouring method in time frame 87} and \ref{fig:region outlier detection method in time frame 87}, we notice that our region outlier detection method can accurately detect all blobs. However, the contouring-based methods either yield the blobs with incorrect areas (much larger or smaller), or misdetect the wrong area as a blob. This is because that it is hard to use one single threshold to identify the blobs for various time frames even in the same experimental data. Our region outlier detection method does not have such problem since we use more flexible distribution-based outlier detection. 

\subsection{More blob detection results}
\label{subsec:More blob detection results}
\begin{figure*}
        \centering
        \begin{subfigure}[b]{0.19\textwidth}
                \includegraphics[width=\textwidth]{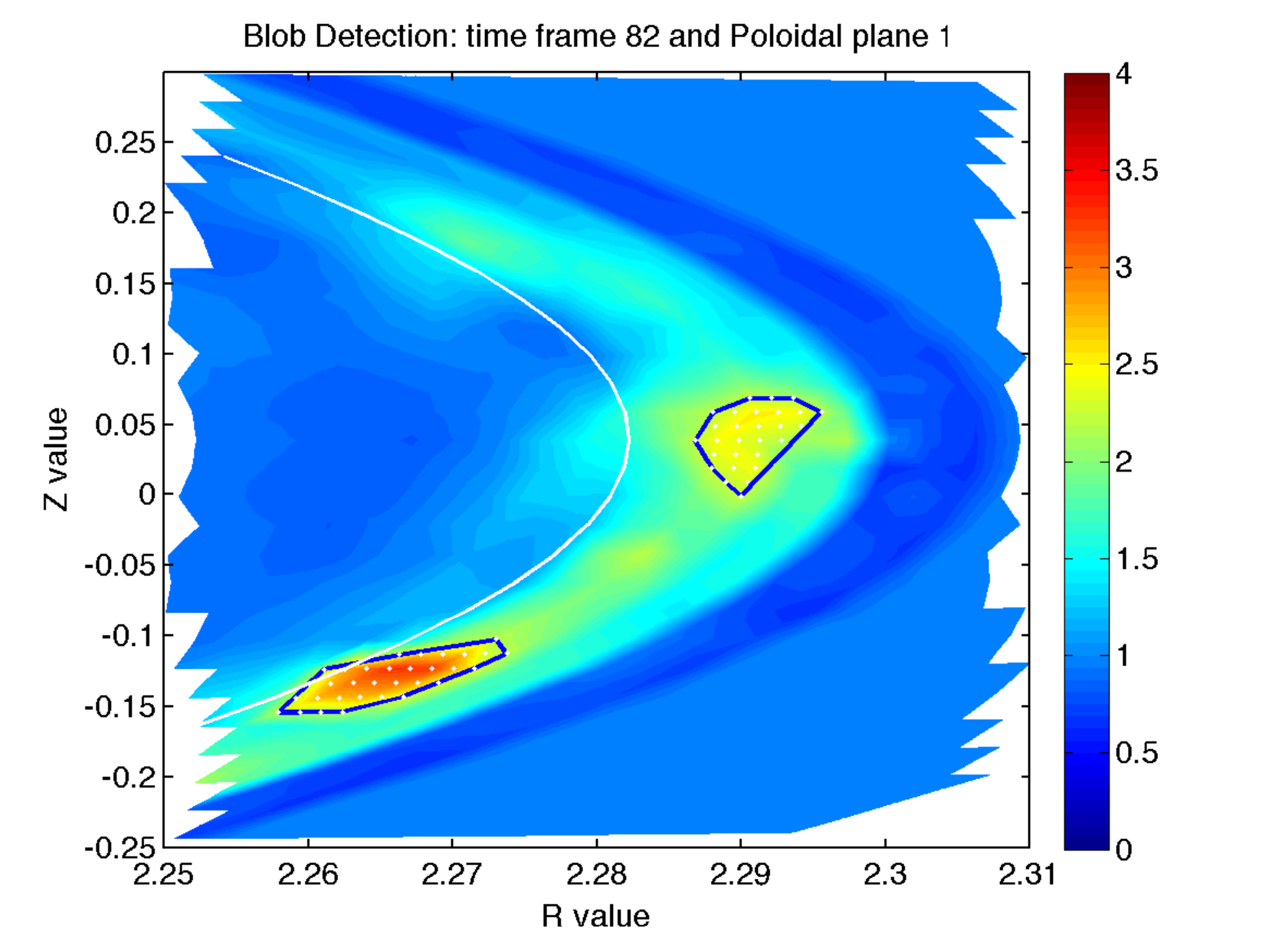}
                \caption{Frame 82 and plane 1}
                \label{fig:Time frame 82 and poloidal plane 1}
        \end{subfigure}
        \begin{subfigure}[b]{0.19\textwidth}
                \includegraphics[width=\textwidth]{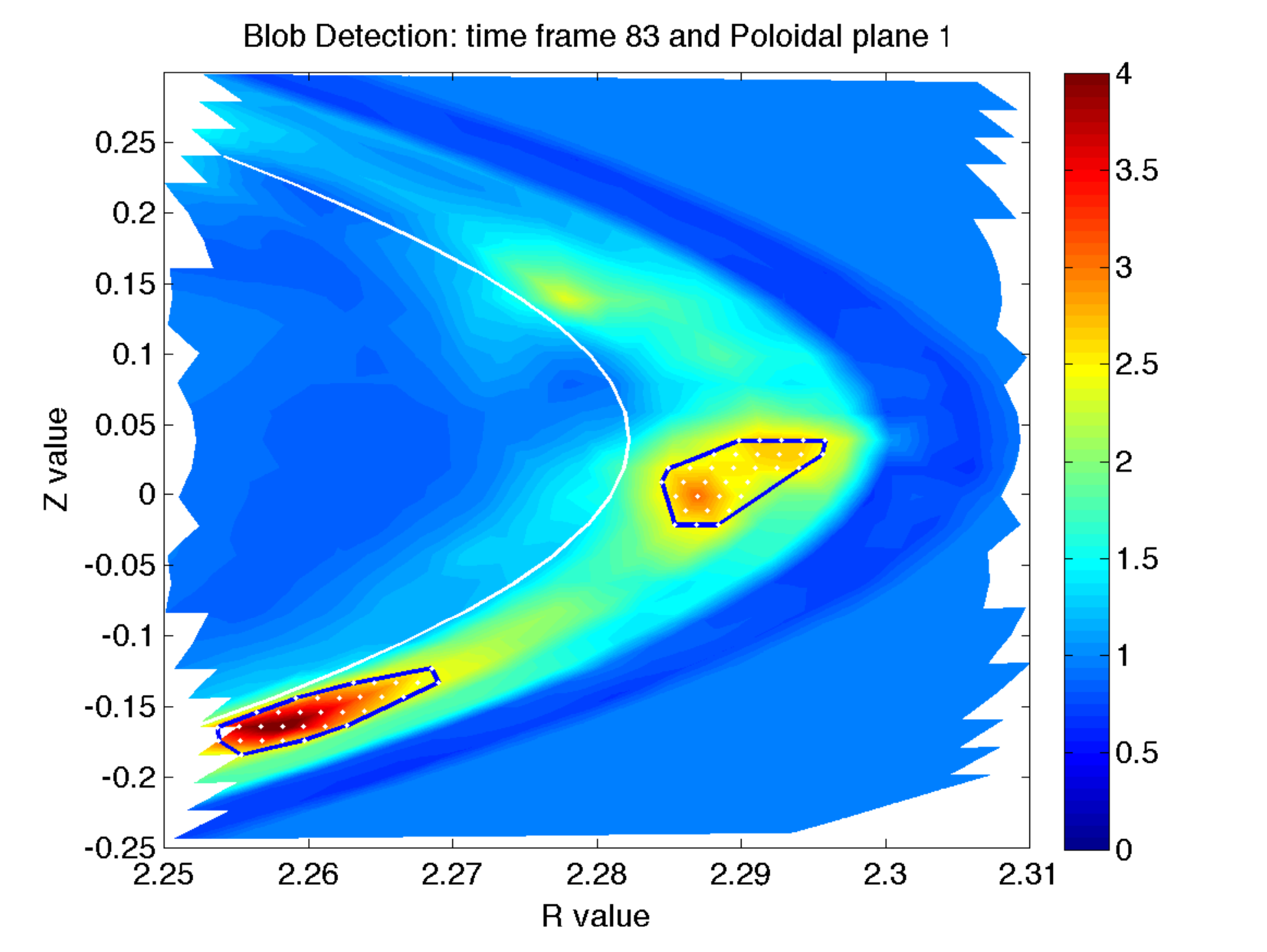}
                \caption{Frame 83 and plane 1}
                \label{fig:Time frame 83 and poloidal plane 1}
        \end{subfigure}
        \begin{subfigure}[b]{0.19\textwidth}
                \includegraphics[width=\textwidth]{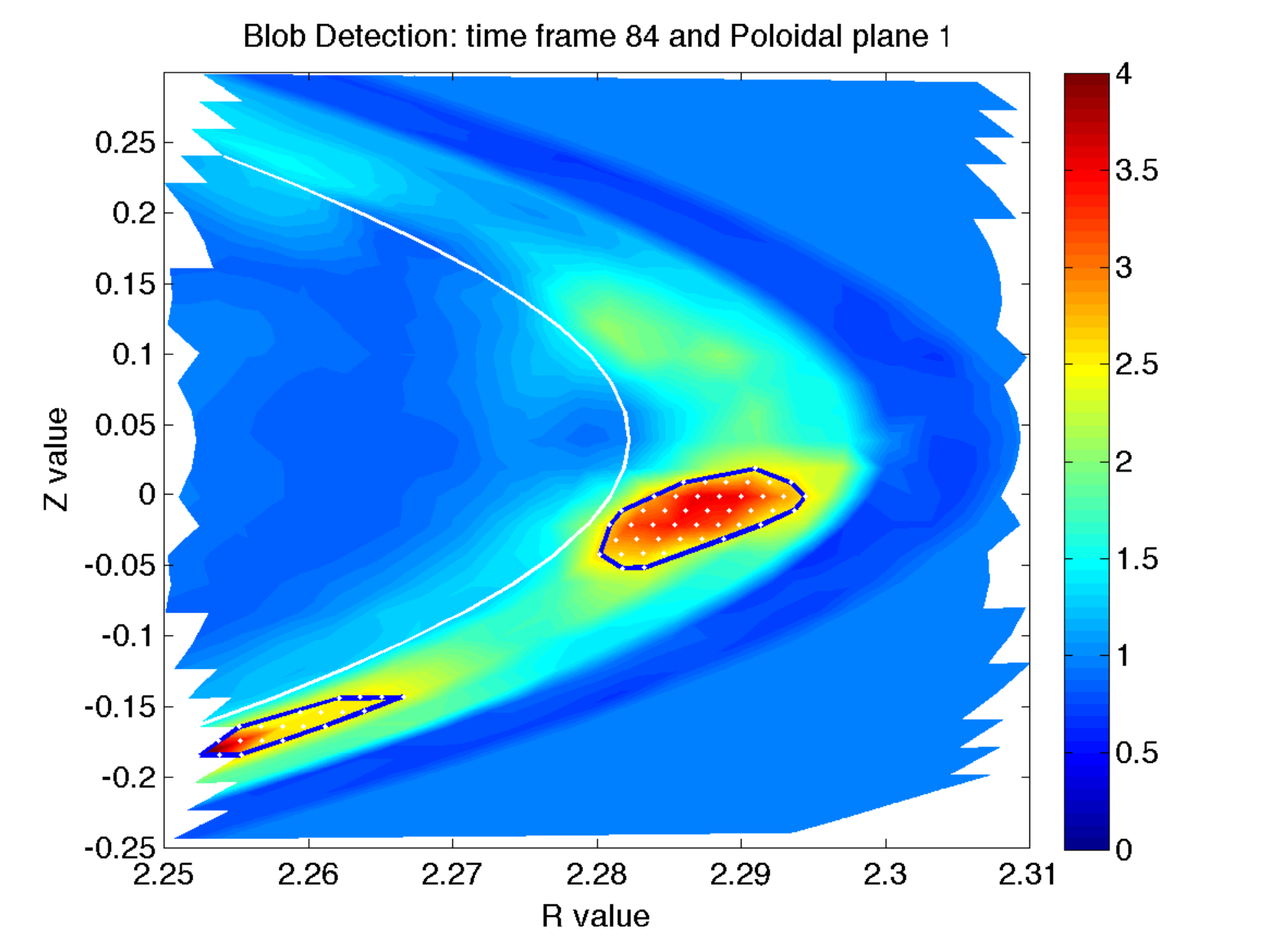}
                \caption{Frame 84 and plane 1}
                \label{fig:Time frame 84 and poloidal plane 1}
        \end{subfigure}
        \begin{subfigure}[b]{0.19\textwidth}
                \includegraphics[width=\textwidth]{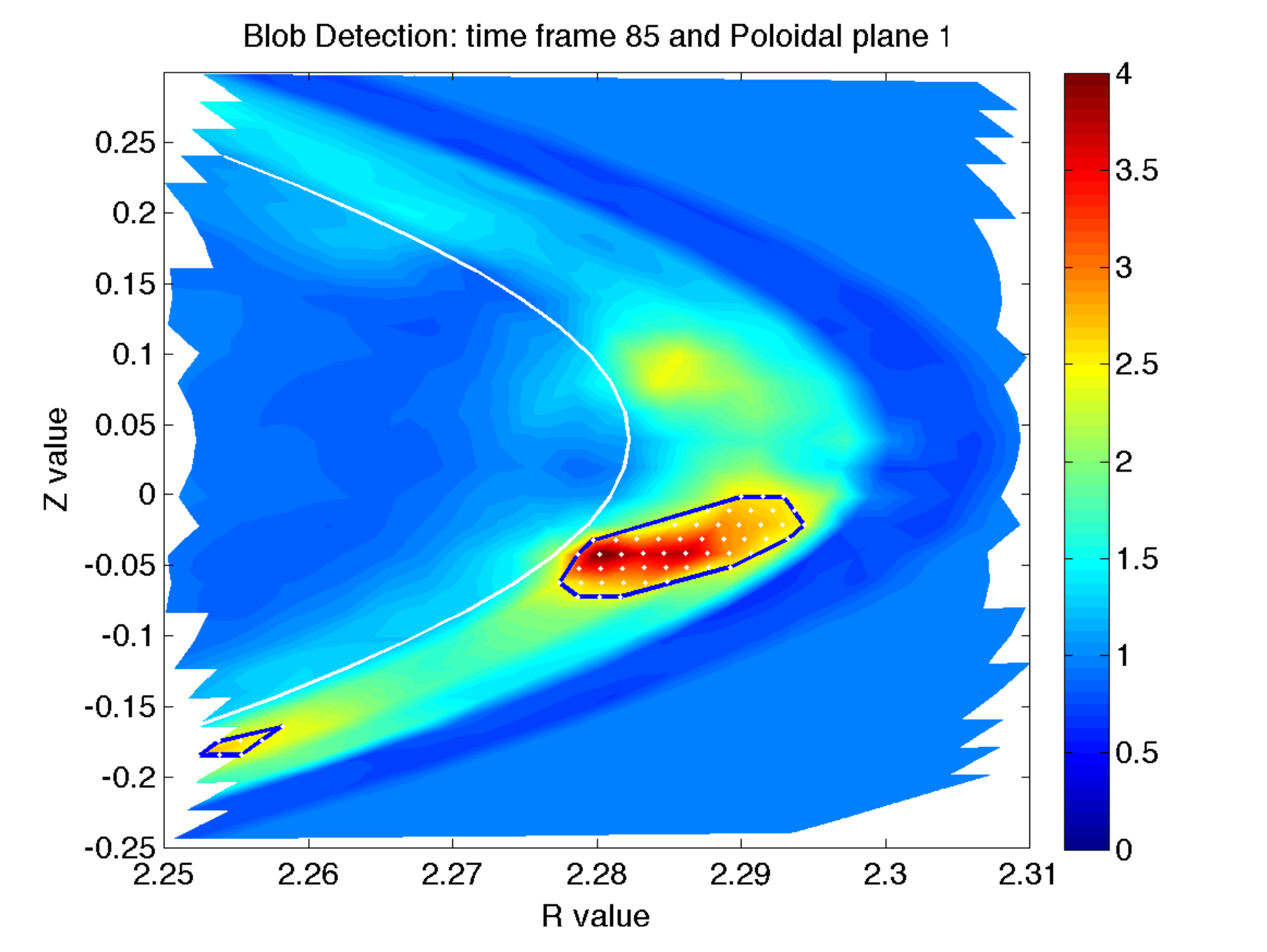}
                \caption{Frame 85 and plane 1}
                \label{fig:Time frame 85 and poloidal plane 1}
        \end{subfigure}
        \begin{subfigure}[b]{0.19\textwidth}
                \includegraphics[width=\textwidth]{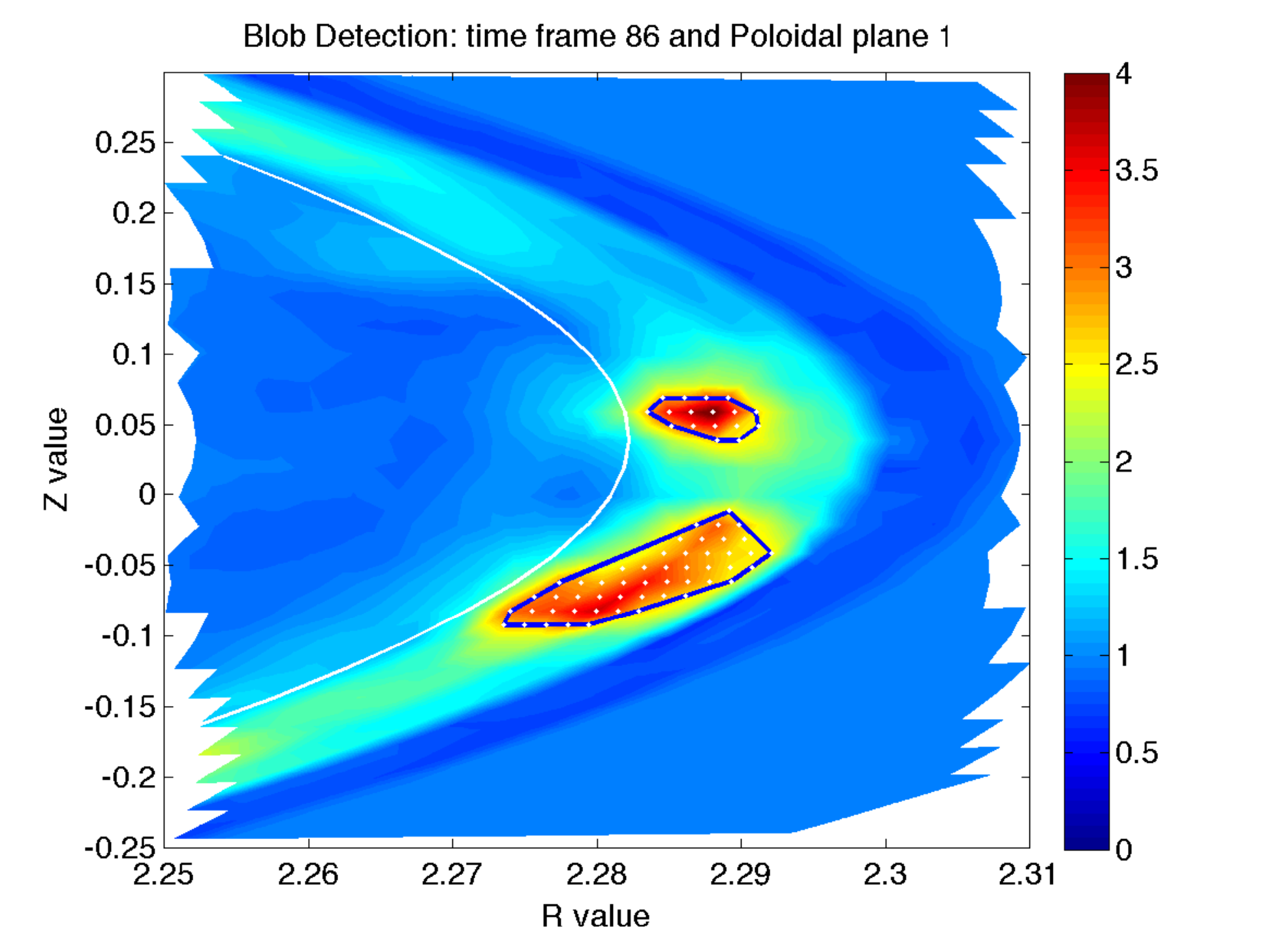}
                \caption{Frame 86 and plane 1}
                \label{fig:Time frame 86 and poloidal plane 1}
        \end{subfigure}
                \centering
        \begin{subfigure}[b]{0.19\textwidth}
                \includegraphics[width=\textwidth]{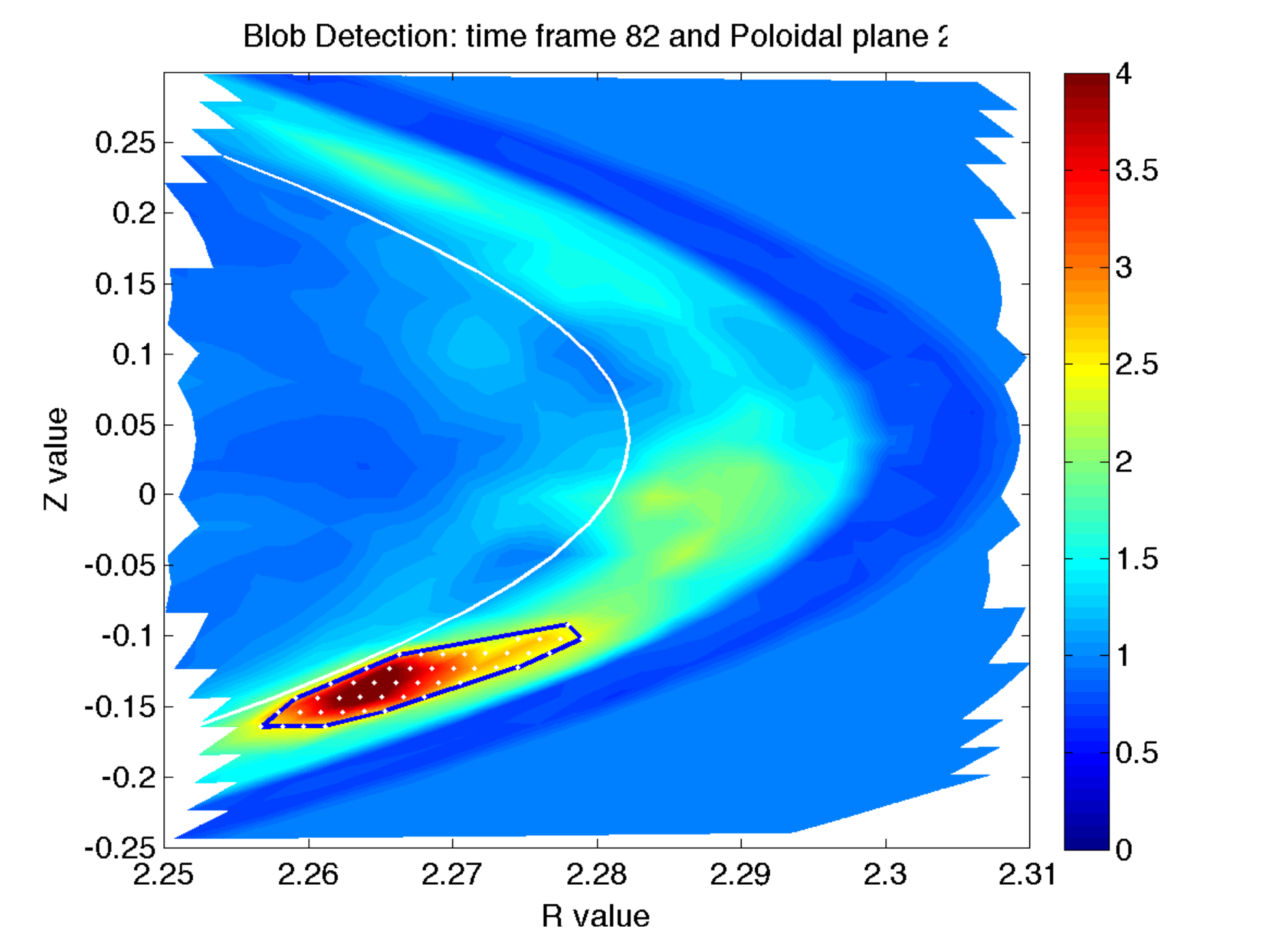}
                \caption{Frame 82 and plane 2}
                \label{fig:Time frame 82 and poloidal plane 2}
        \end{subfigure}
        \begin{subfigure}[b]{0.19\textwidth}
                \includegraphics[width=\textwidth]{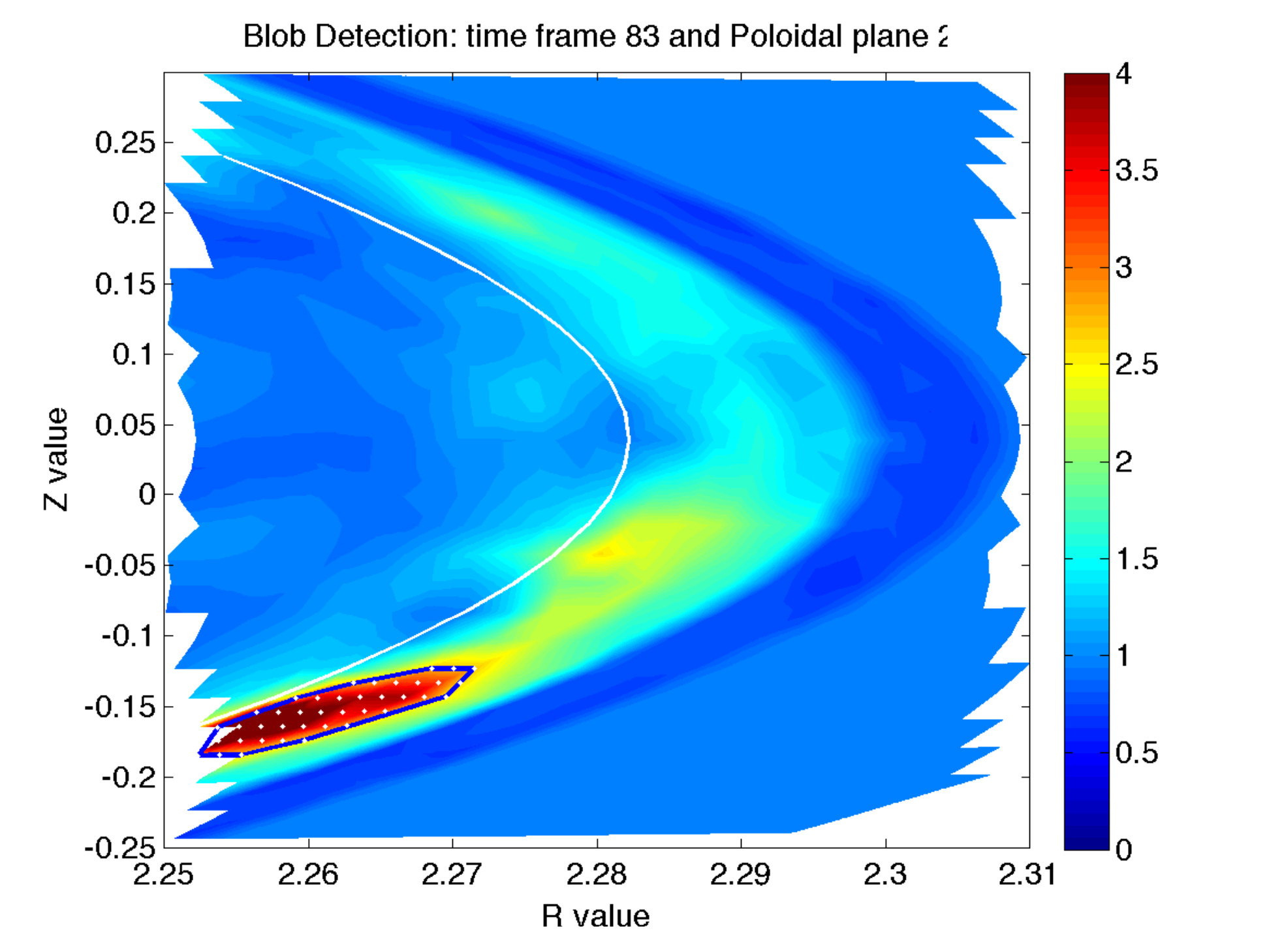}
                \caption{Frame 83 and plane 2}
                \label{fig:Time frame 83 and poloidal plane 2}
        \end{subfigure}
        \begin{subfigure}[b]{0.19\textwidth}
                \includegraphics[width=\textwidth]{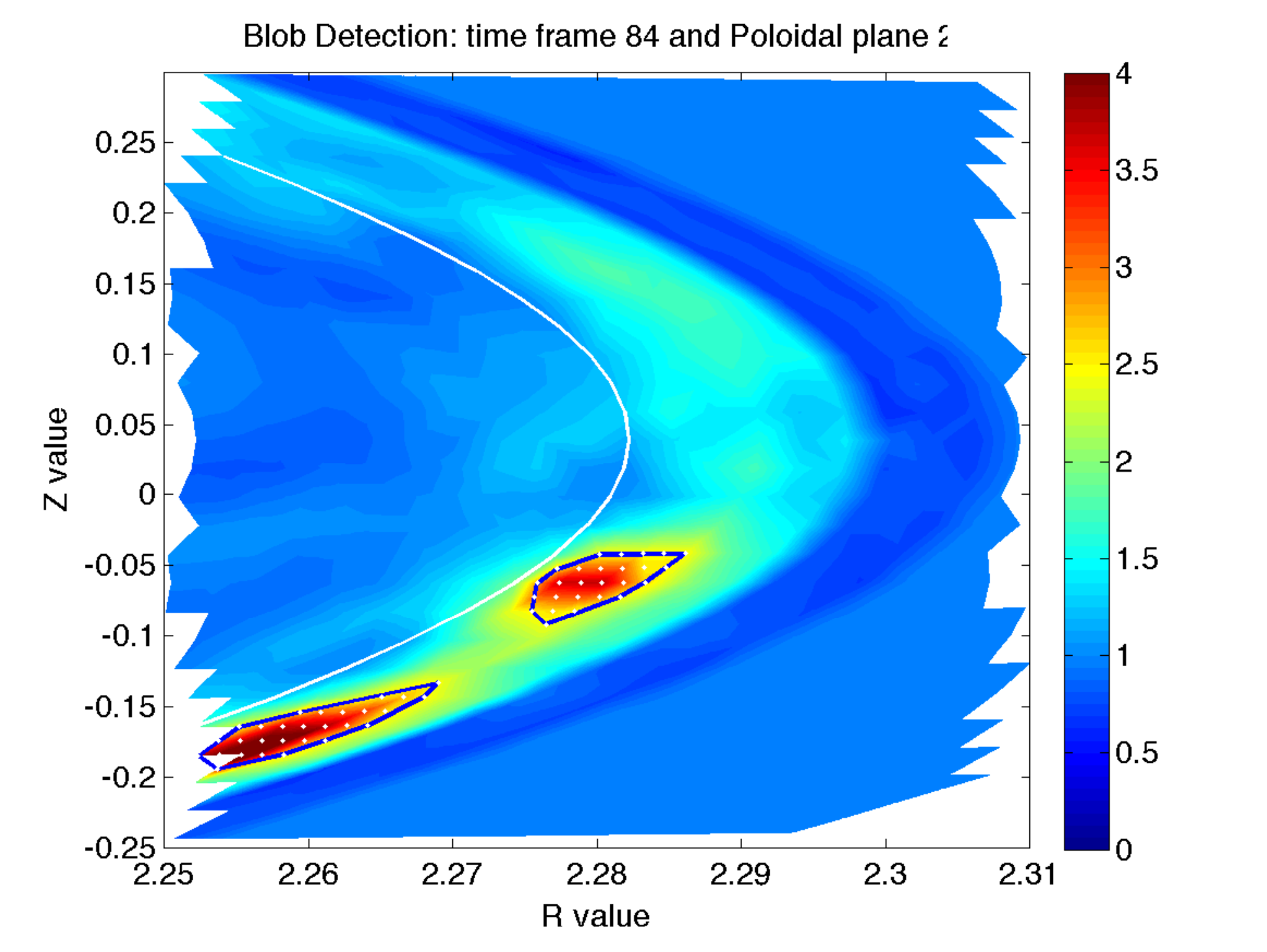}
                \caption{Frame 84 and plane 2}
                \label{fig:Time frame 84 and poloidal plane 2}
        \end{subfigure}
        \begin{subfigure}[b]{0.19\textwidth}
                \includegraphics[width=\textwidth]{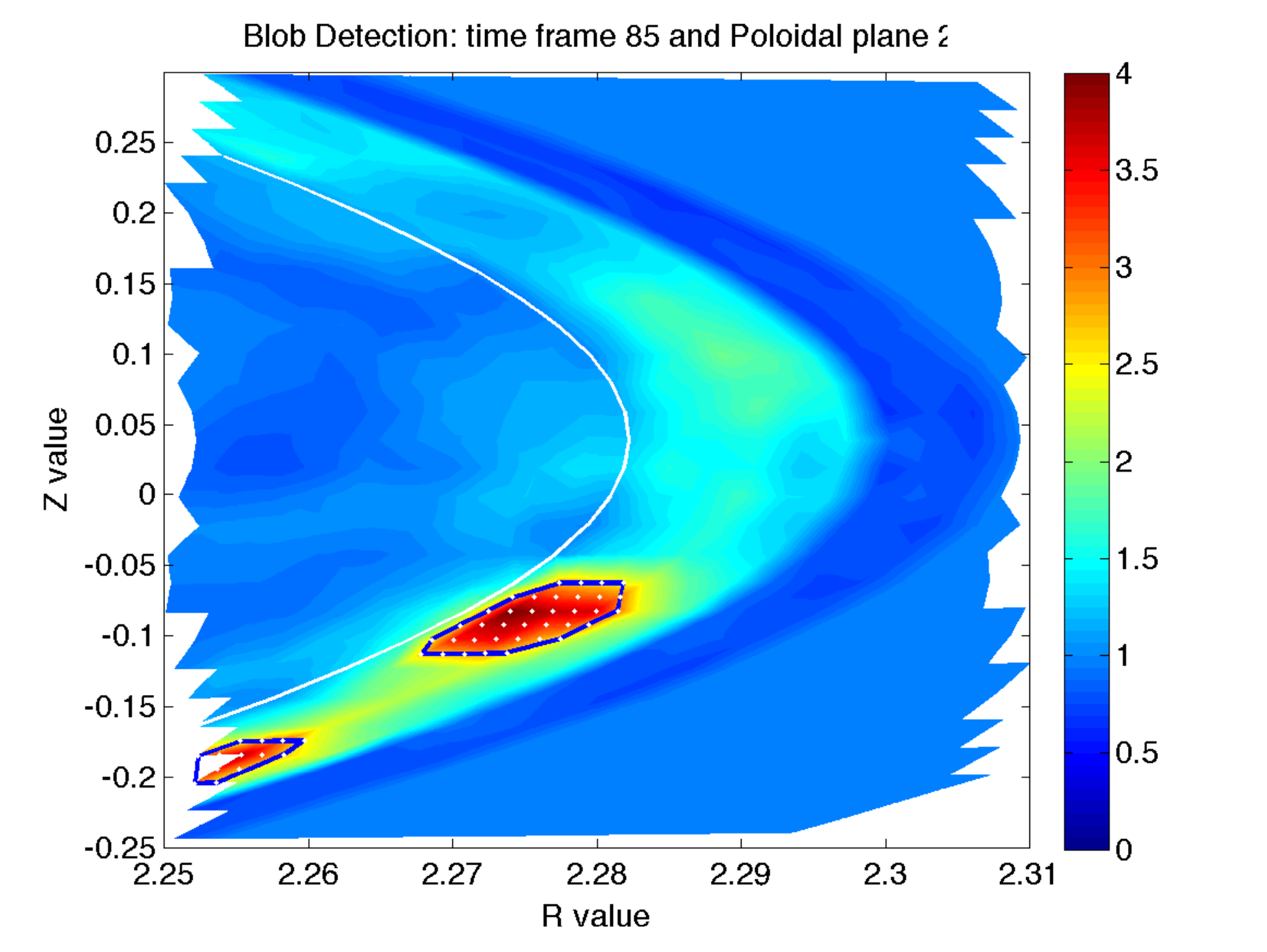}
                \caption{Frame 85 and plane 2}
                \label{fig:Time frame 85 and poloidal plane 2}
        \end{subfigure}
        \begin{subfigure}[b]{0.19\textwidth}
                \includegraphics[width=\textwidth]{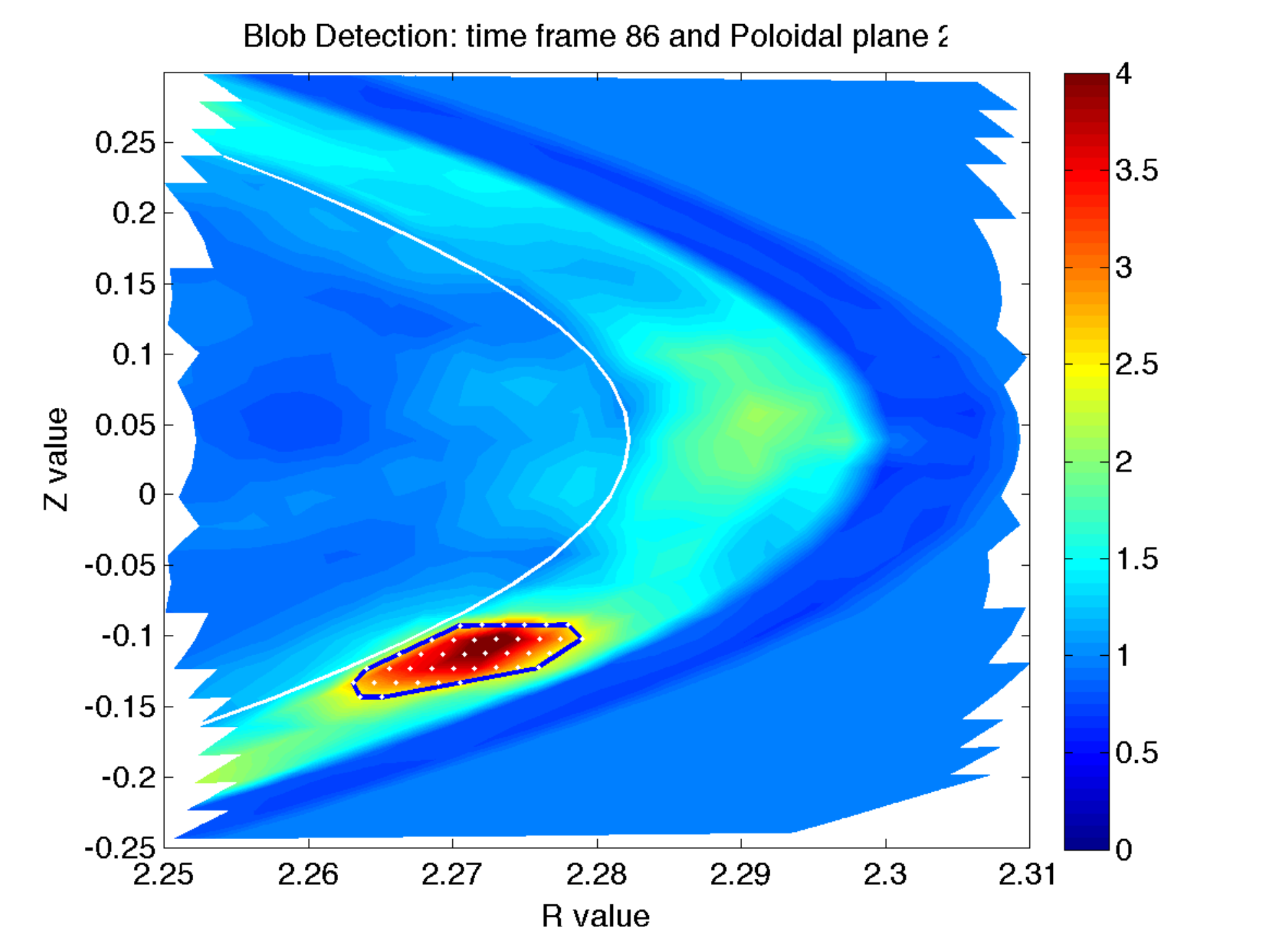}
                \caption{Frame 86 and plane 2}
                \label{fig:Time frame 86 and poloidal plane 2}
        \end{subfigure}
                \begin{subfigure}[b]{0.19\textwidth}
                \includegraphics[width=\textwidth]{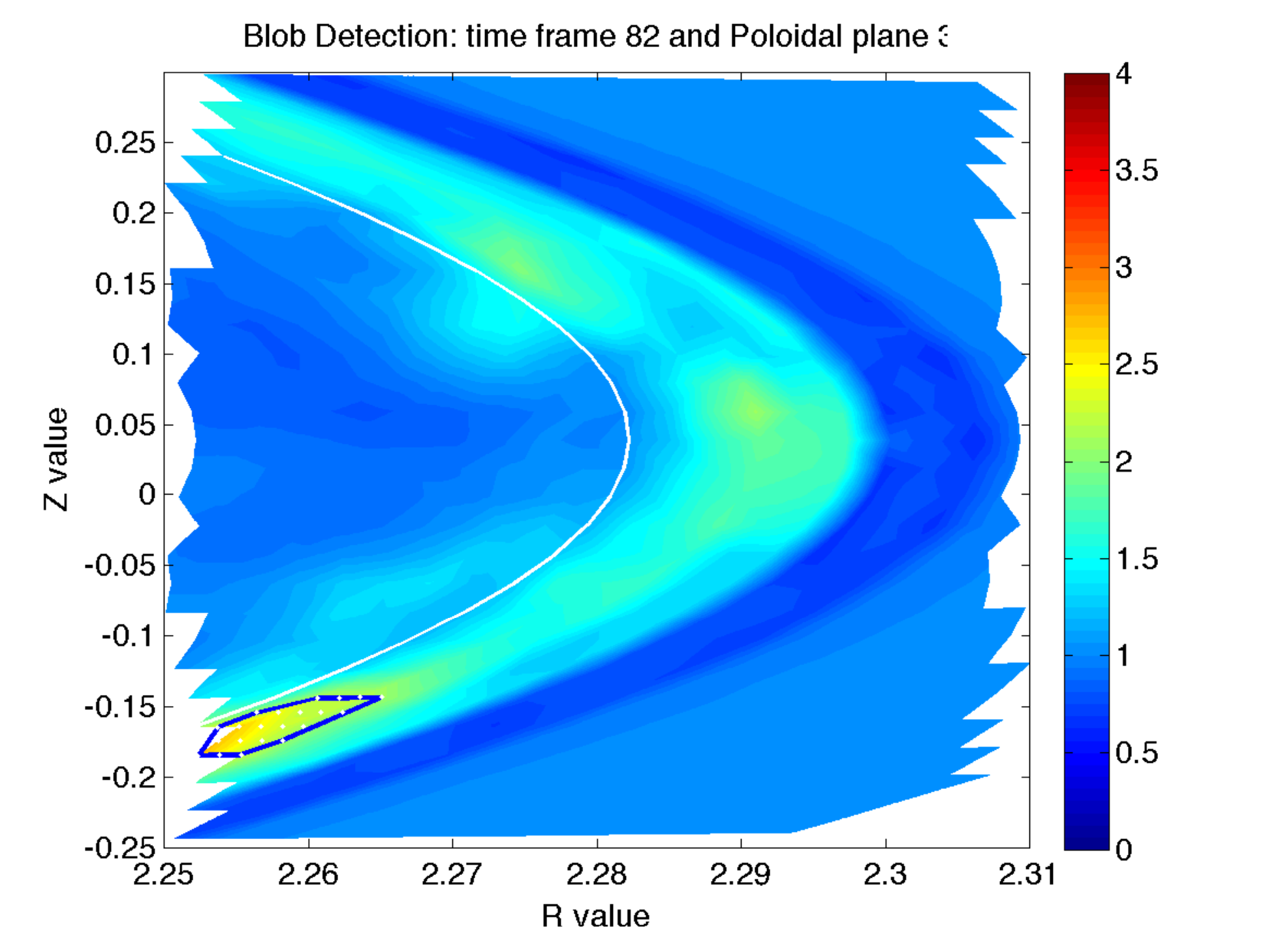}
                \caption{Frame 82 and plane 3}
                \label{fig:Time frame 82 and poloidal plane 3}
        \end{subfigure}
        \begin{subfigure}[b]{0.19\textwidth}
                \includegraphics[width=\textwidth]{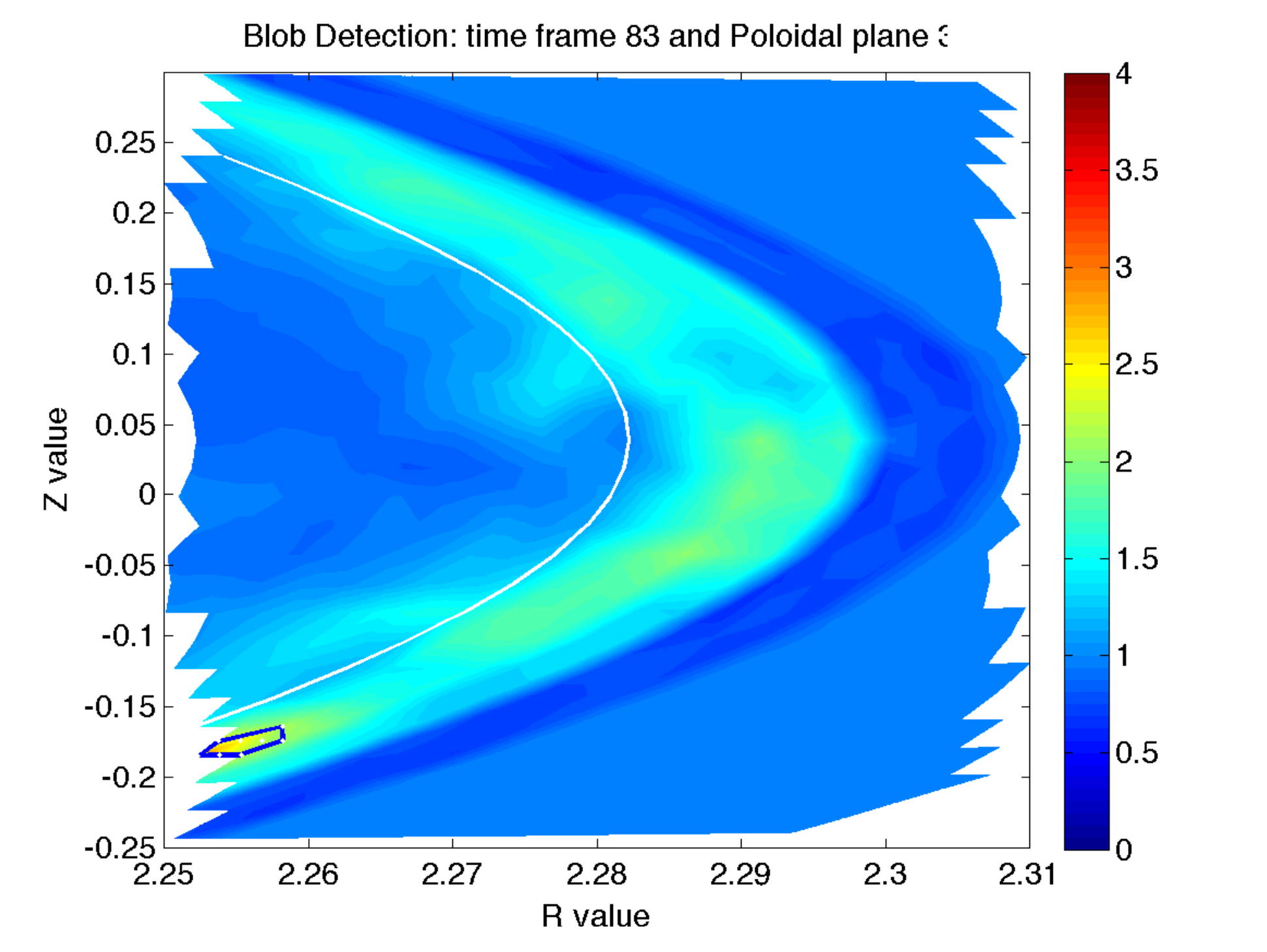}
                \caption{Frame 83 and plane 3}
                \label{fig:Time frame 83 and poloidal plane 3}
        \end{subfigure}
        \begin{subfigure}[b]{0.19\textwidth}
                \includegraphics[width=\textwidth]{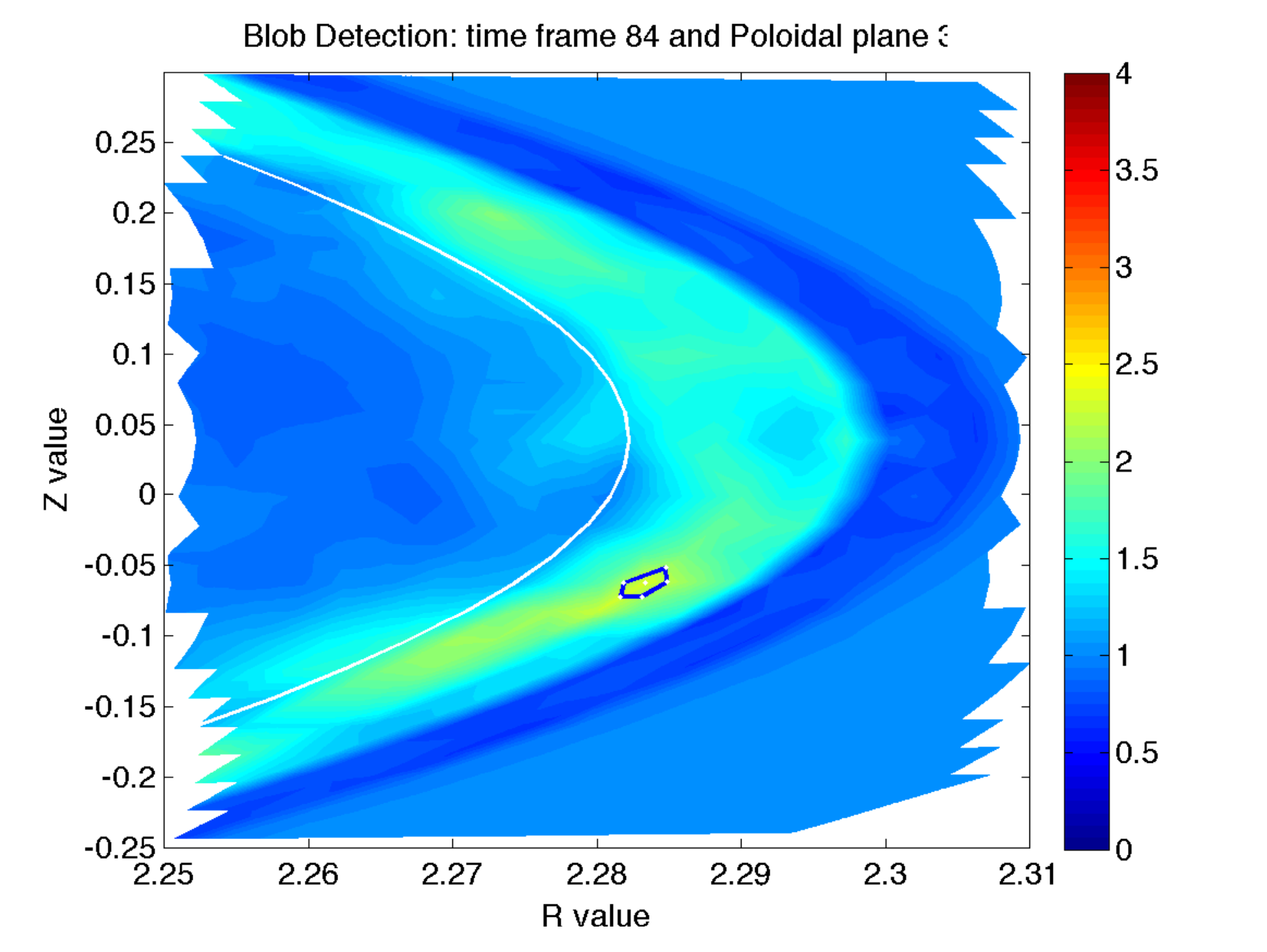}
                \caption{Frame 84 and plane 3}
                \label{fig:Time frame 84 and poloidal plane 3}
        \end{subfigure}
        \begin{subfigure}[b]{0.19\textwidth}
                \includegraphics[width=\textwidth]{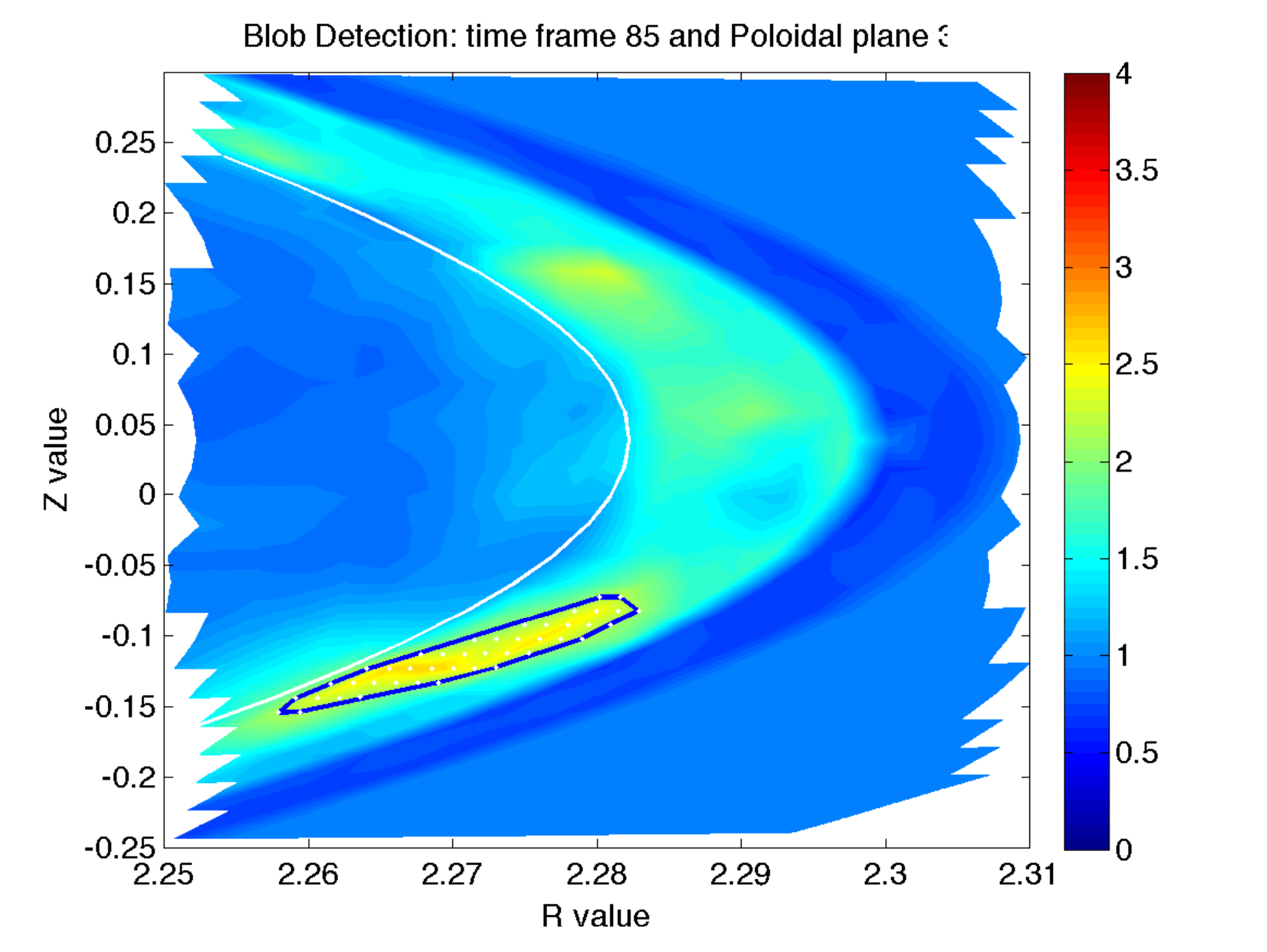}
                \caption{Frame 85 and plane 3}
                \label{fig:Time frame 85 and poloidal plane 3}
        \end{subfigure}
        \begin{subfigure}[b]{0.19\textwidth}
                \includegraphics[width=\textwidth]{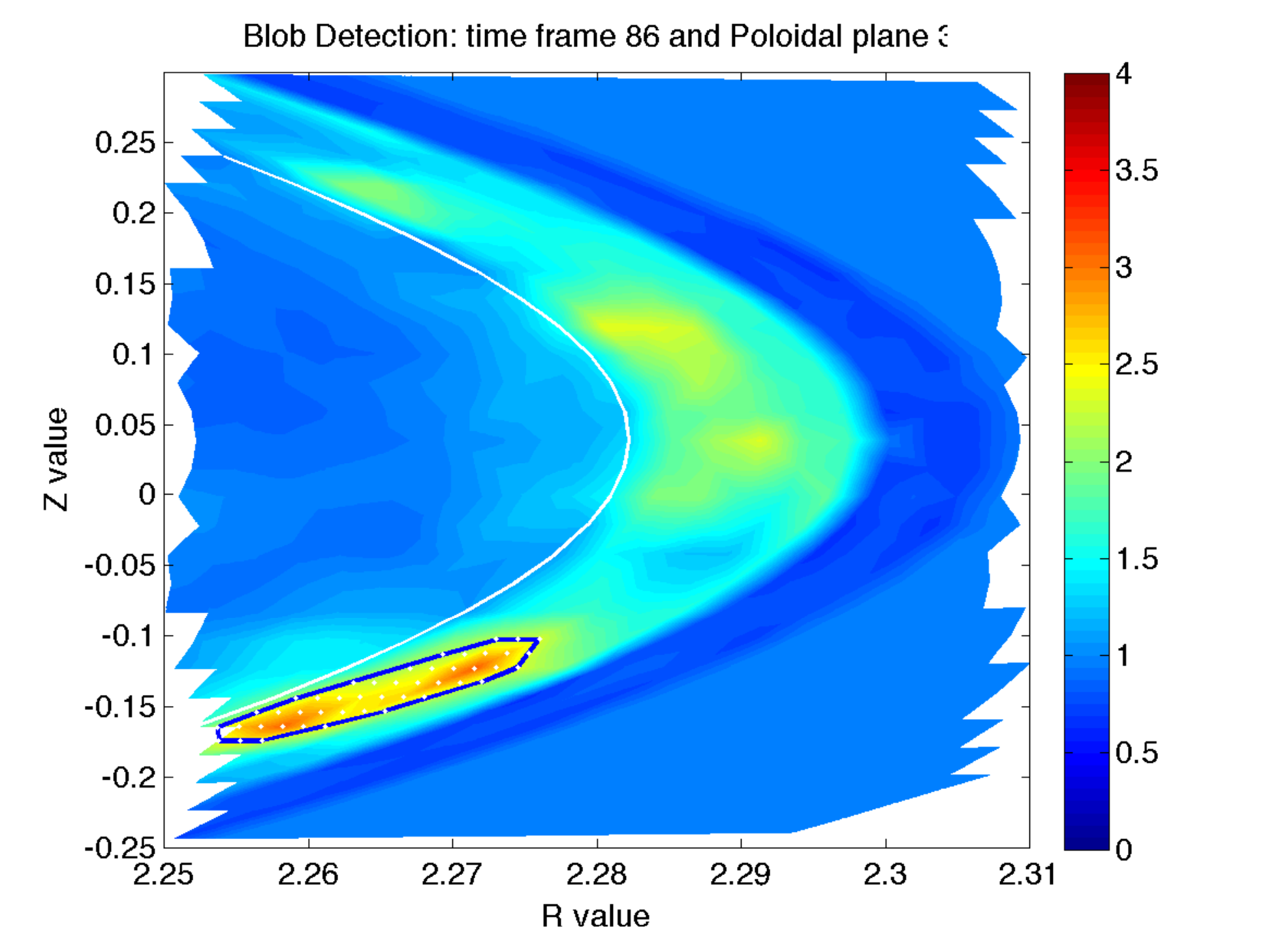}
                \caption{Frame 86 and plane 3}
                \label{fig:Time frame 86 and poloidal plane 3}
        \end{subfigure}
                \begin{subfigure}[b]{0.19\textwidth}
                \includegraphics[width=\textwidth]{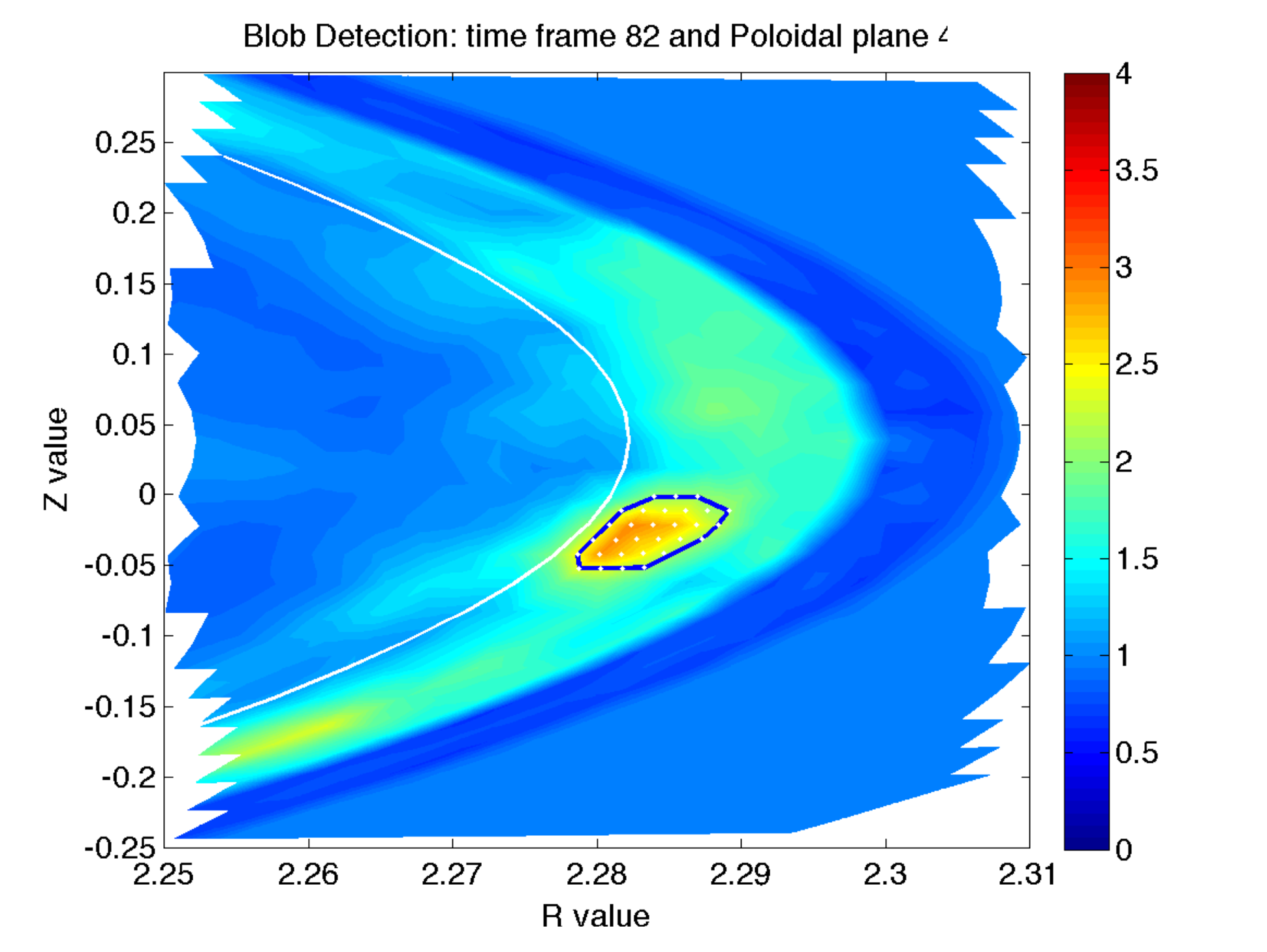}
                \caption{Frame 82 and plane 4}
                \label{fig:Time frame 82 and poloidal plane 4}
        \end{subfigure}
        \begin{subfigure}[b]{0.19\textwidth}
                \includegraphics[width=\textwidth]{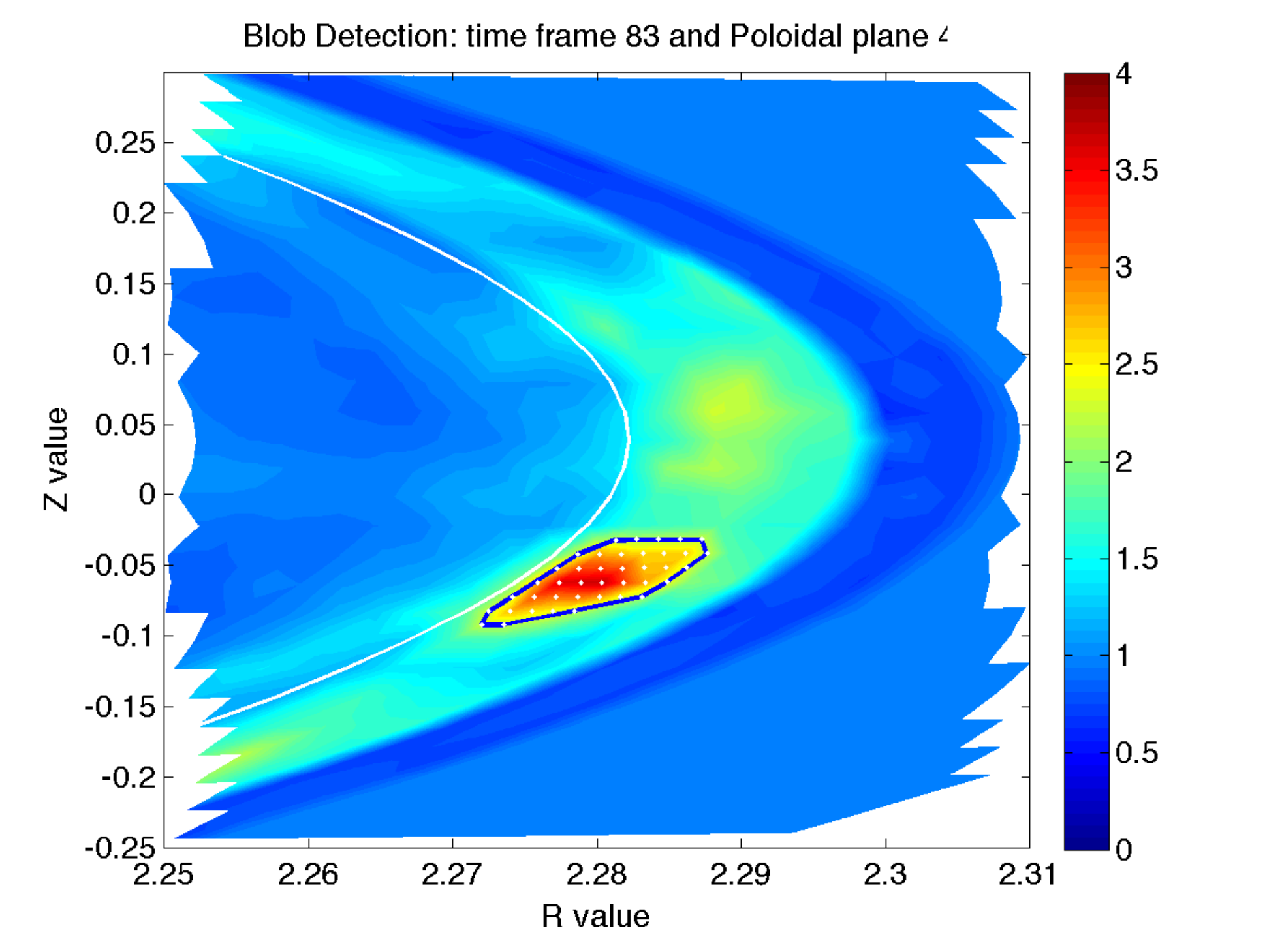}
                \caption{Frame 83 and plane 4}
                \label{fig:Time frame 83 and poloidal plane 4}
        \end{subfigure}
        \begin{subfigure}[b]{0.19\textwidth}
                \includegraphics[width=\textwidth]{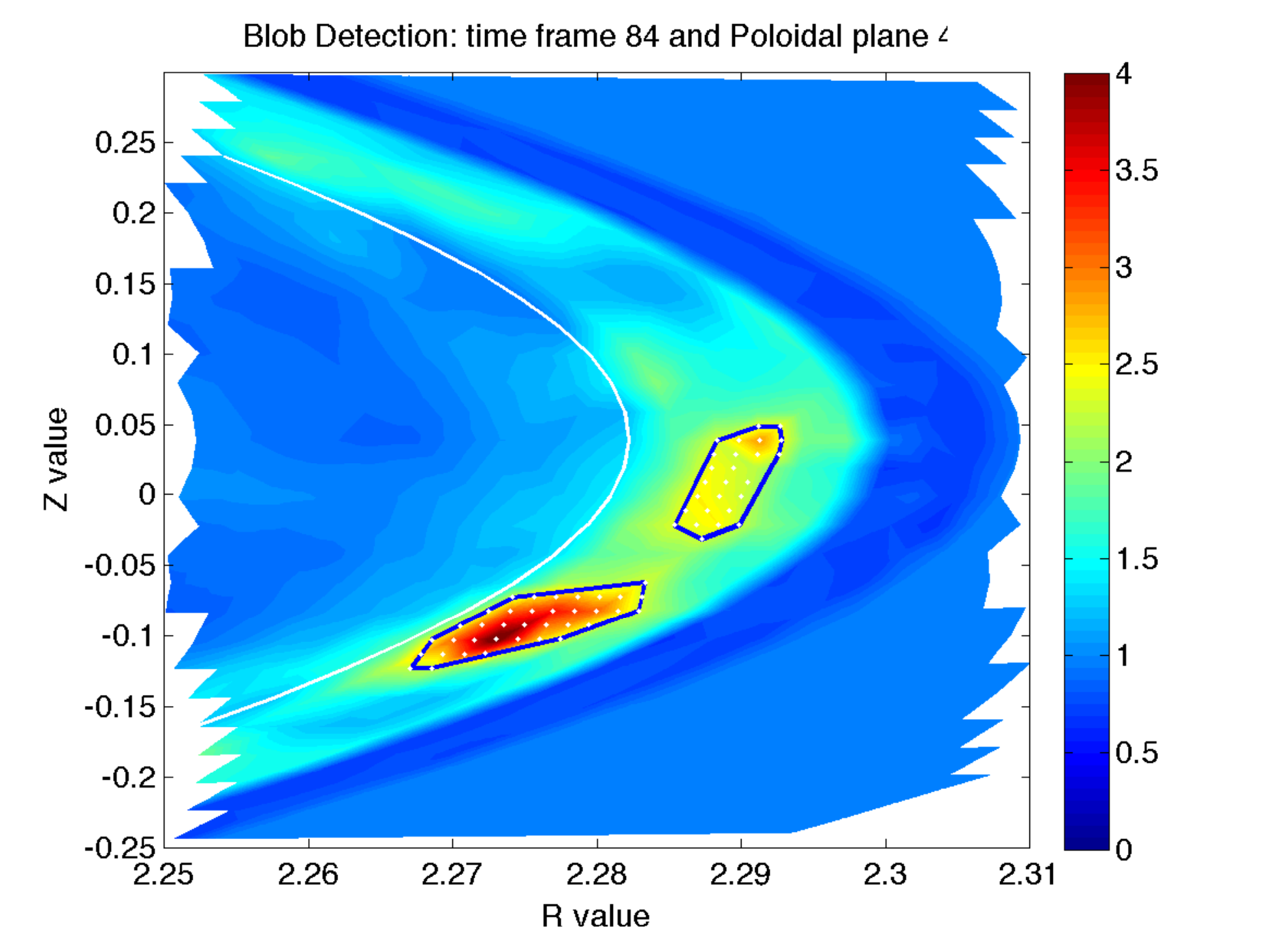}
                \caption{Frame 84 and plane 4}
                \label{fig:Time frame 84 and poloidal plane 4}
        \end{subfigure}
        \begin{subfigure}[b]{0.19\textwidth}
                \includegraphics[width=\textwidth]{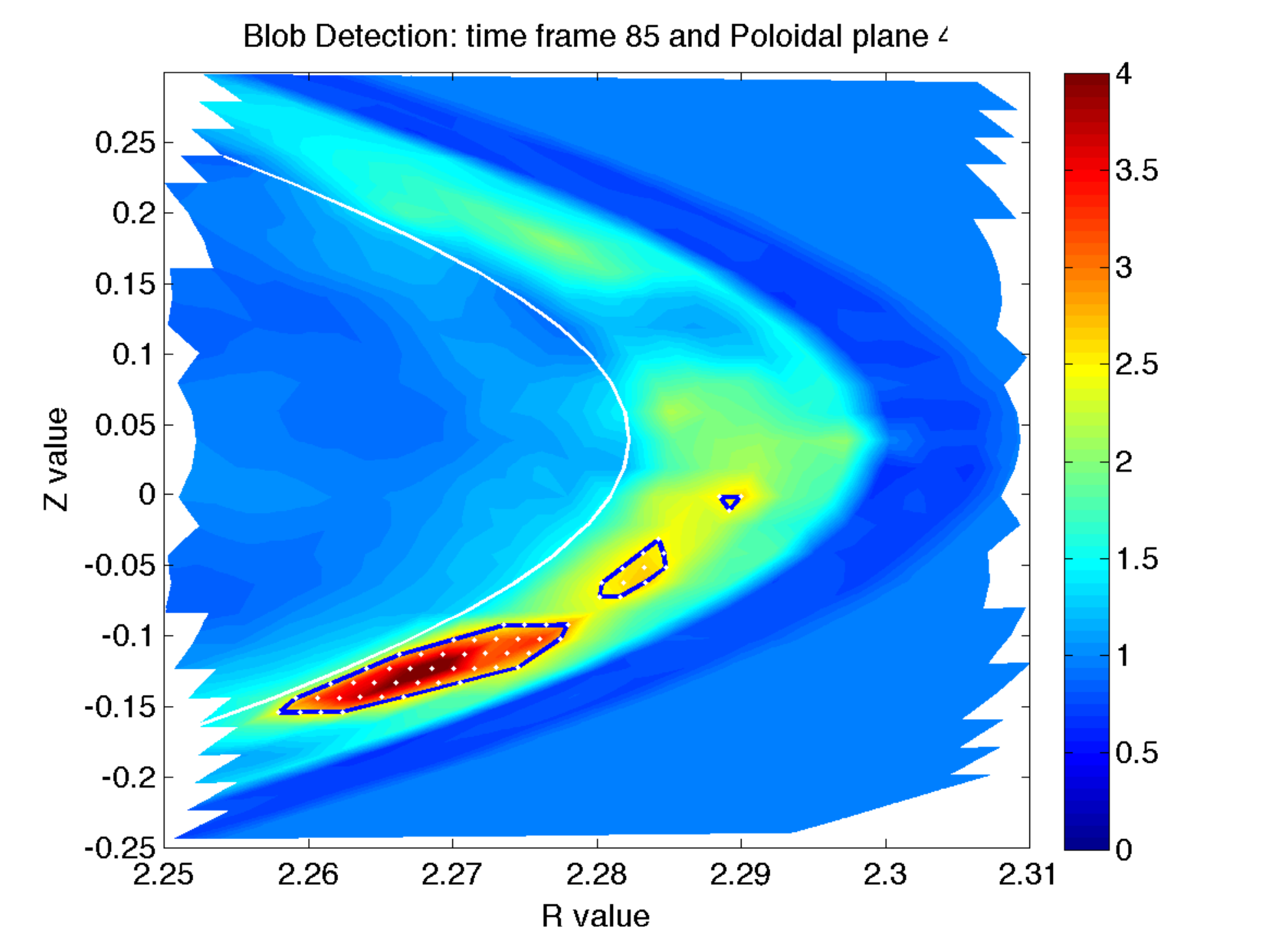}
                \caption{Frame 85 and plane 4}
                \label{fig:Time frame 85 and poloidal plane 4}
        \end{subfigure}
        \begin{subfigure}[b]{0.19\textwidth}
                \includegraphics[width=\textwidth]{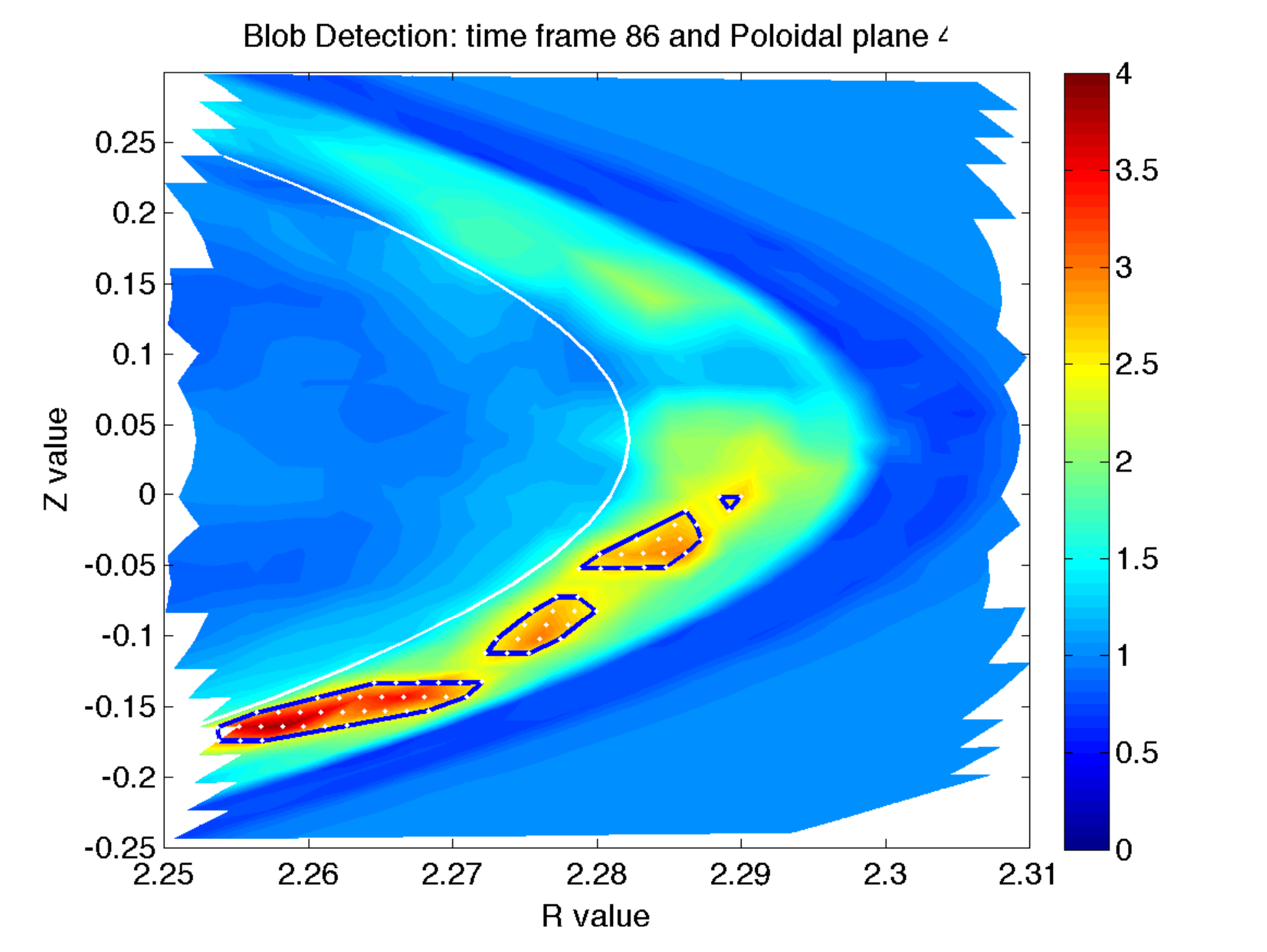}
                \caption{Frame 86 and plane 4}
                \label{fig:Time frame 86 and poloidal plane 4}
        \end{subfigure}
        \caption{An example of the blob detection in five continuous time frames and four different poloidal planes in the R (radial) direction and the Z (poloidal) direction. The separatrix position is shown by a white line and the different blue circles denote blobs.}
        \label{fig:blob detection}
\end{figure*}

We perform more experiments to comprehensively examine the blob detection results in five continuous time frames and four different poloidal planes as shown in Figure \ref{fig:blob detection}. As we can see from the figure, our region outlier detection method can provide consistently good results in different situations. In addition, our method does not miss any blobs at the edge of the regions of interest, as shown in subfigures \ref{fig:Time frame 83 and poloidal plane 1}, \ref{fig:Time frame 83 and poloidal plane 2}, \ref{fig:Time frame 84 and poloidal plane 1} and \ref{fig:Time frame 84 and poloidal plane 2}. It is interesting to see that large-scale blob structures are often generated, which could cause substantial plasma transport \cite{zweben1985search}. As pointed out in \cite{xu2006multiscale}, these large-scale structures are mainly contributed by the low-frequency and long-wavelength fluctuating components, which may be responsible for the observations of long-range correlations. We also noticed that different poloidal planes may display significant diversity in edge turbulence, even in the same time frame. We have shown that we are able to effectively detect the blobs and reveal some interesting results to help physicists improve their understanding of the characteristic of blobs and their correlation with other plasma properties.

\subsection{Blob tracking results}
\label{subsec:Blob tracking results}
\begin{figure*}[t!]
        \centering
        \begin{subfigure}[b]{0.40\textwidth}
                \includegraphics[width=\textwidth]{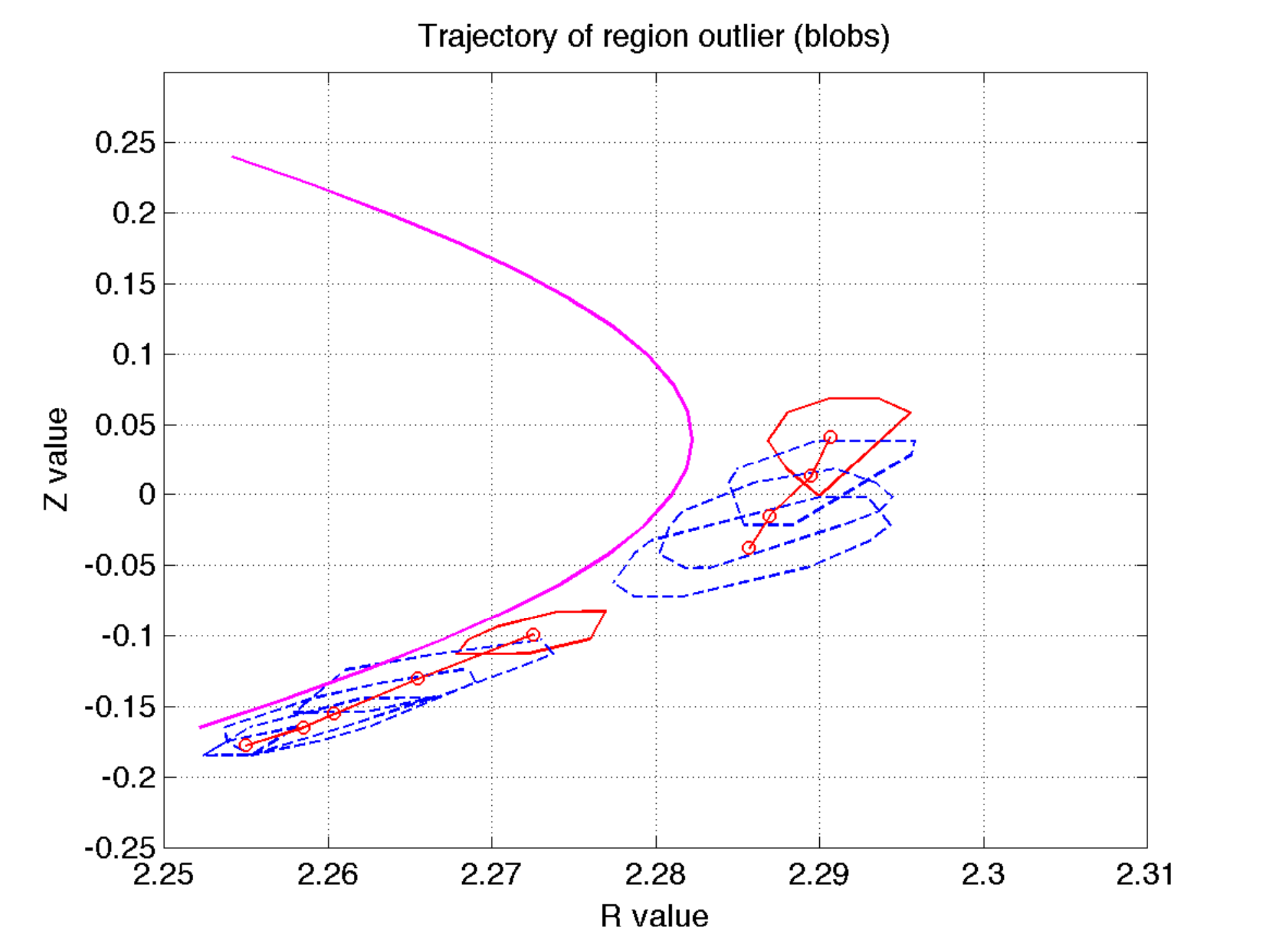}
                \caption{2D trajectory for detected blobs}
                \label{fig:2D trajectory for detected blobs}
        \end{subfigure}
        \begin{subfigure}[b]{0.40\textwidth}
                \includegraphics[width=\textwidth]{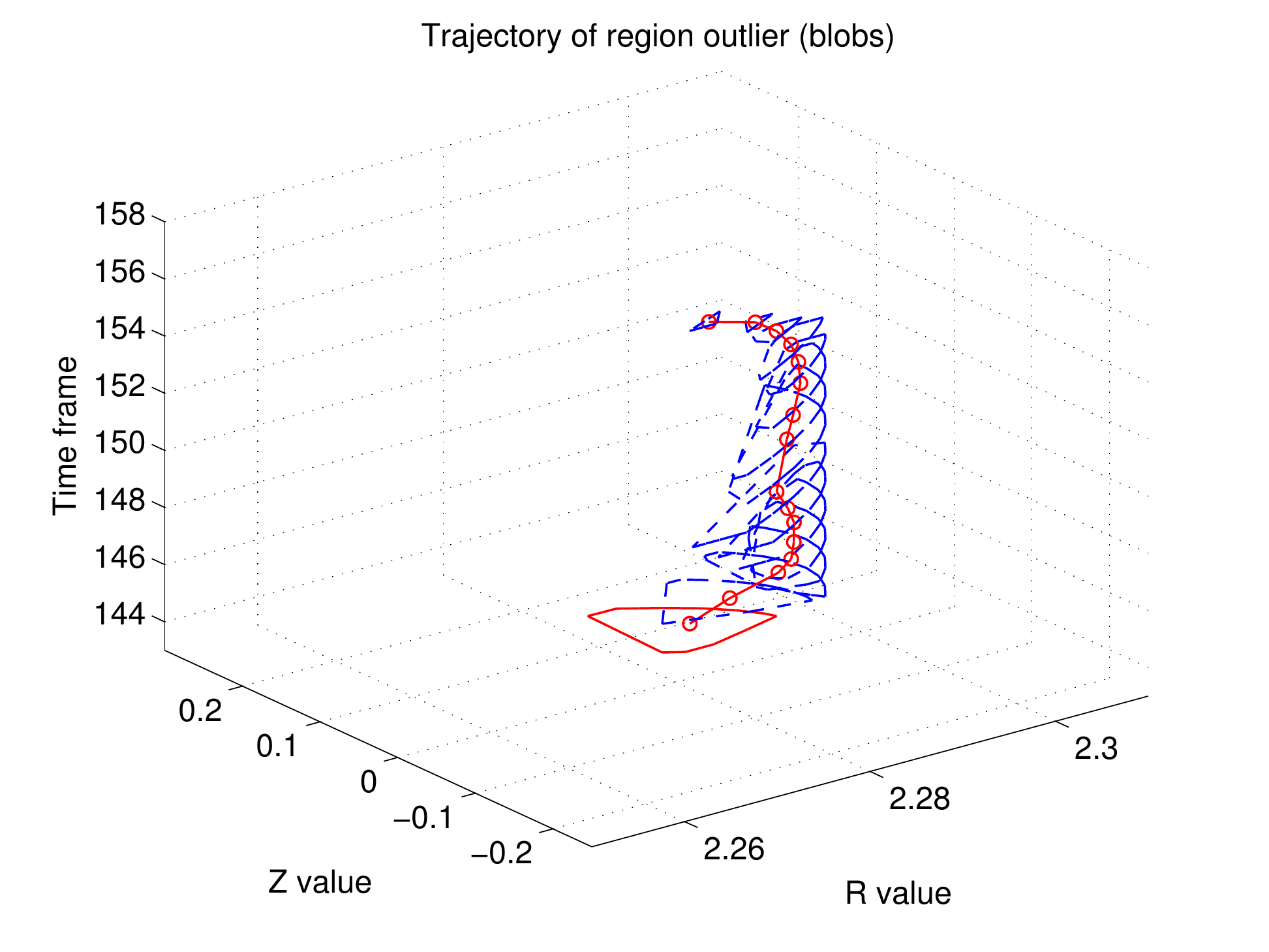}
                \caption{3D trajectory for detected blobs}
                \label{fig:3D trajectory for detected blobs}
        \end{subfigure}
        \caption{2D and 3D center trajectories for detected blobs over consecutive time frames. The red solid polygon indicates the starting times of the blobs tracked while the blue broken polygons indicate subsequent times of the same blobs tracked. The centers of the moving blobs are linked to show their trajectories of the blob motion. The pink line represents the separatrix.}
        \label{fig:2D/3D trajectory for detected blobs}
\end{figure*}

We investigate the blob tracking results in two different situations. Figure \ref{fig:2D trajectory for detected blobs} exhibits a 2D trajectory of a blob. Again, the trajectory is generated by plotting the location of the density peak of the detected blobs over five consecutive time frames. We can see that our blob tracking algorithm can track two separate blobs simultaneously. The blob size can grow when they move towards confined plasma in the right region of separatrix. Figure \ref{fig:3D trajectory for detected blobs} shows a 3D trajectory for a detected blob over fifteen consecutive time frames. In this case, the blob seems to maintain its size for a few time frames, then gradually  decreases, and eventually disappears. Through these interesting results, physicists may be able to understand the characteristics of blobs better.

\subsection{Real-time blob detection under strong scaling}
\label{subsec:Real-time blob detection under strong scaling}
We have illustrated the effectiveness and robustness of the proposed blob detection and tracking methods. Next, we perform a set of experiments to demonstrate the performance of our real-time blob detection approach under strong scaling and weak scaling. We define the speedup of our parallel implementation on heterogeneous multi-core architecture as follows:  
\[ speedup = 
\dfrac{\textit{runtime of Blob detection using single core}}{\textit{runtime using $\mathbb{P}$ cores}}
\]

\begin{figure*}
        \centering
        \begin{subfigure}[b]{0.40\textwidth}
                \includegraphics[width=\textwidth]{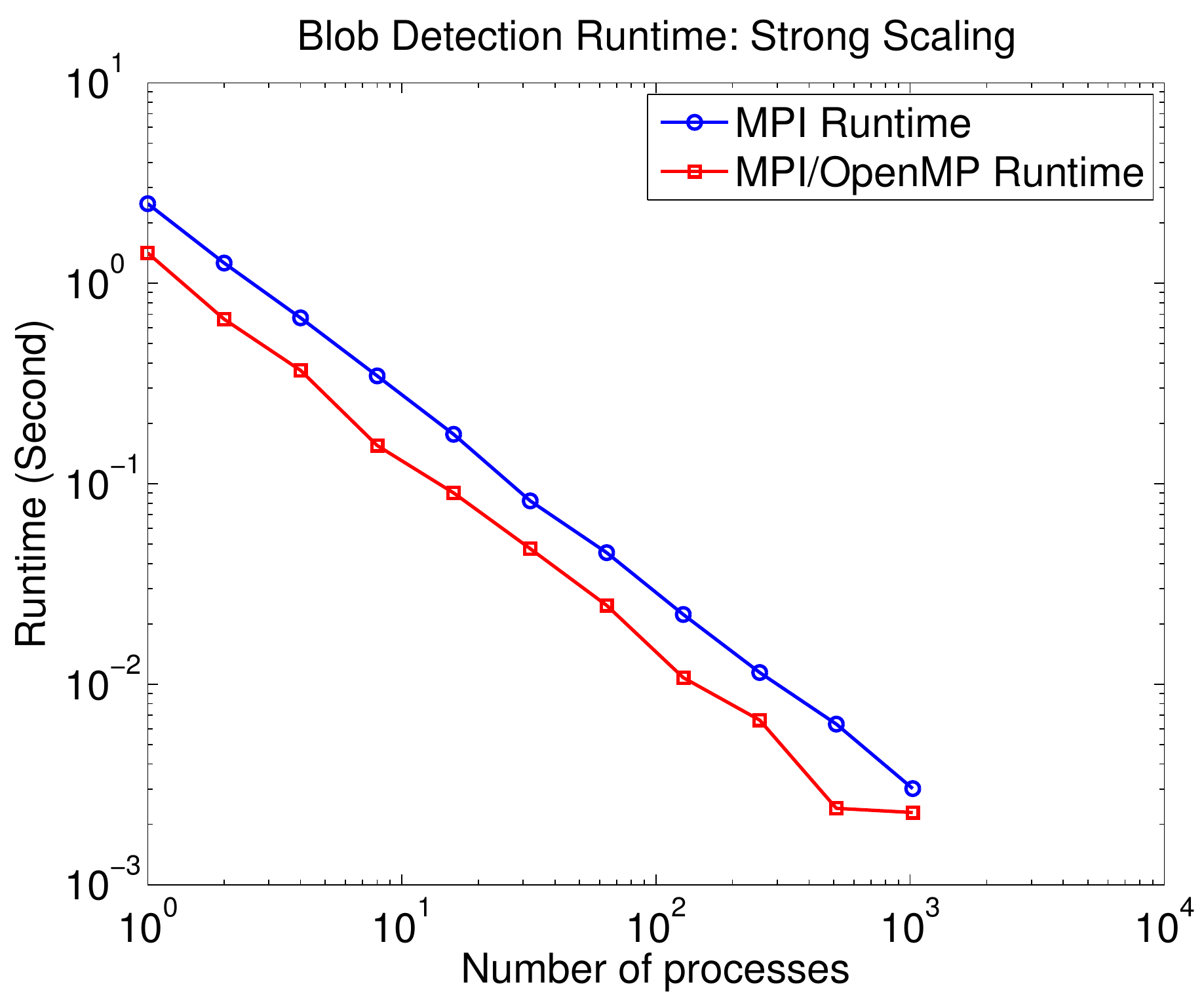}
                \caption{Time}
                \label{fig:blob_all_time_strongscaling}
        \end{subfigure}
        \begin{subfigure}[b]{0.40\textwidth}
                \includegraphics[width=\textwidth]{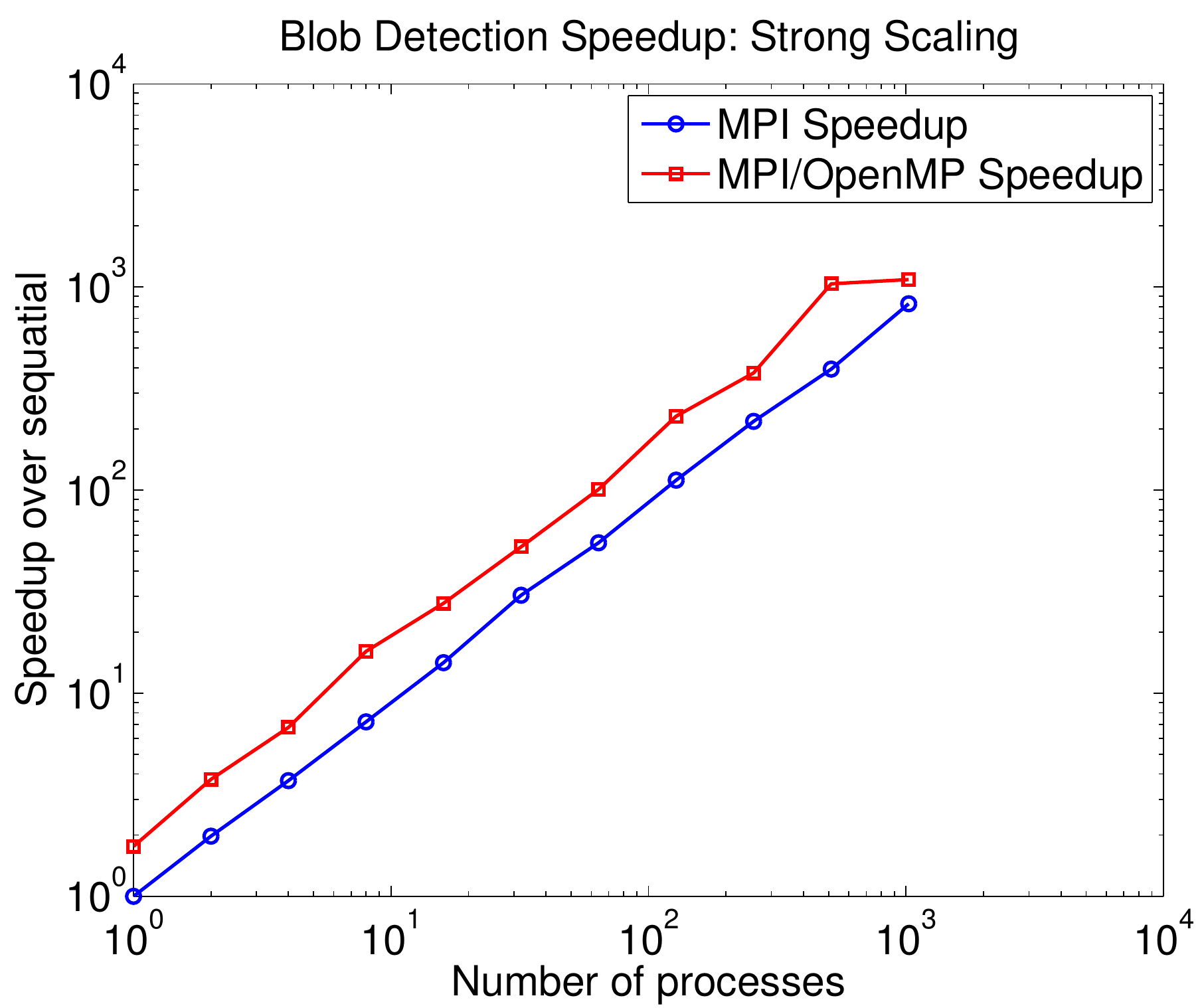}
                \caption{Speedup}
                \label{blob_all_speedup_strongscaling}
        \end{subfigure}
        \caption{Blob detection time and speedup with MPI and MPI/OpenMP varying number of processes under strong scaling}
        \label{fig:Real time blob detection with MPI/OpenMP under strong scaling}
\end{figure*}

Our most encouraging results are that we can complete blob detection on the simulation data set described above in around 2 milliseconds with MPI/OpenMP using 4096 cores and in 3 milliseconds with MPI using 1024 cores. In Figure \ref{fig:Real time blob detection with MPI/OpenMP under strong scaling}, we can achieve linear time speedup in blob detection time under strong scaling. The MPI and the MPI/OpenMP implementations accomplish 800 and 1200 times speedup respectively, when the number of processes is scaled to 1024. Also, we can see that the hybrid MPI/OpenMP implementation is about two times faster than the MPI implementation when varying the number of processes from 1 to 512. With 1024 processes, both of them achieve similar performance, but the MPI/OpenMP one is slightly better. This demonstrates that we are able to control analysis speed by varying the number of processes to achieve real-time analysis. 

\subsection{Real-time blob detection under weak scaling}
\label{subsec:Real-time blob detection under weak scaling}
\begin{figure*}
        \centering
        \begin{subfigure}[b]{0.40\textwidth}
                \includegraphics[width=\textwidth]{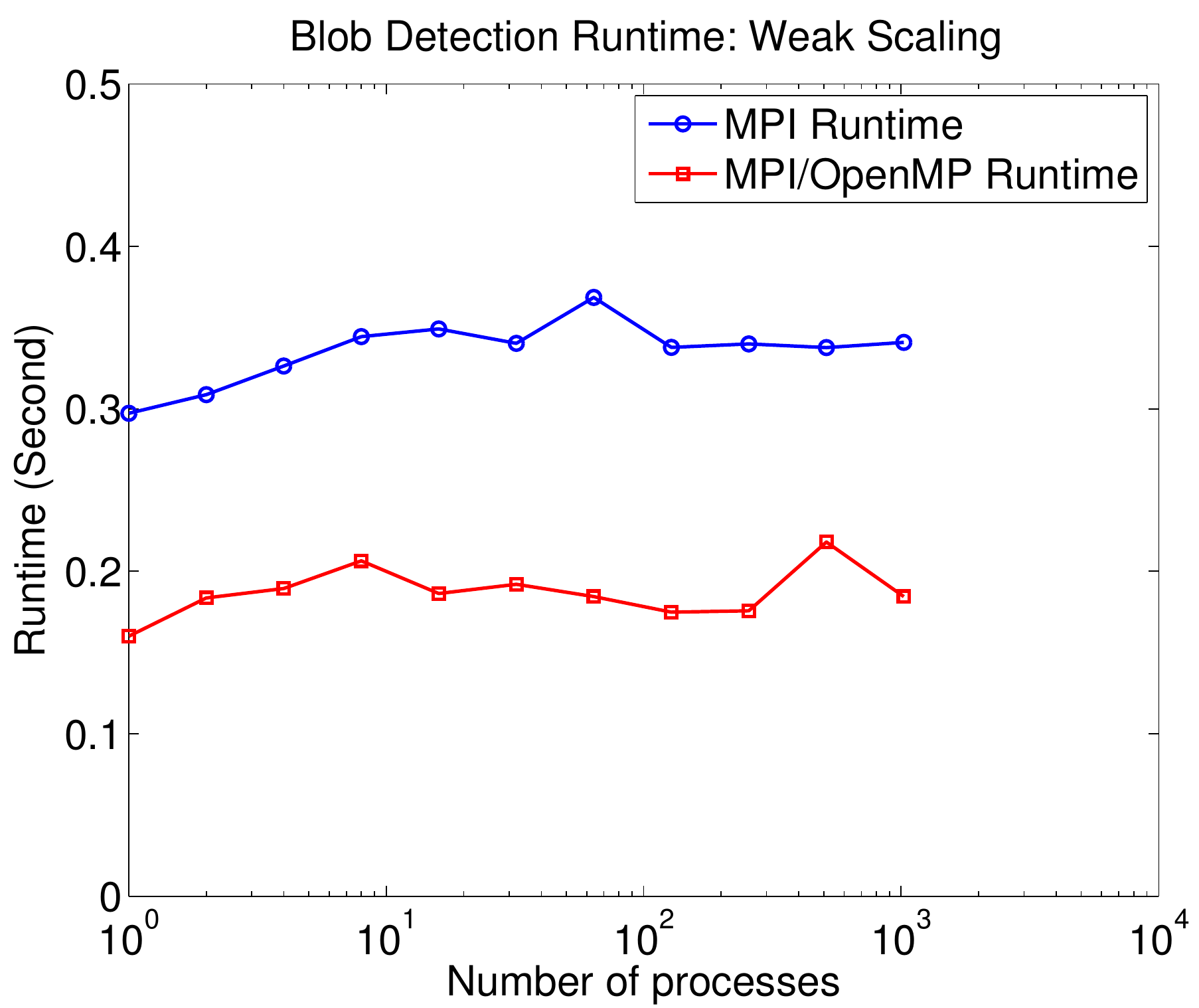}
                \caption{Time}
                \label{fig:blob_all_time_weakscaling}
        \end{subfigure}
        \begin{subfigure}[b]{0.40\textwidth}
                \includegraphics[width=\textwidth]{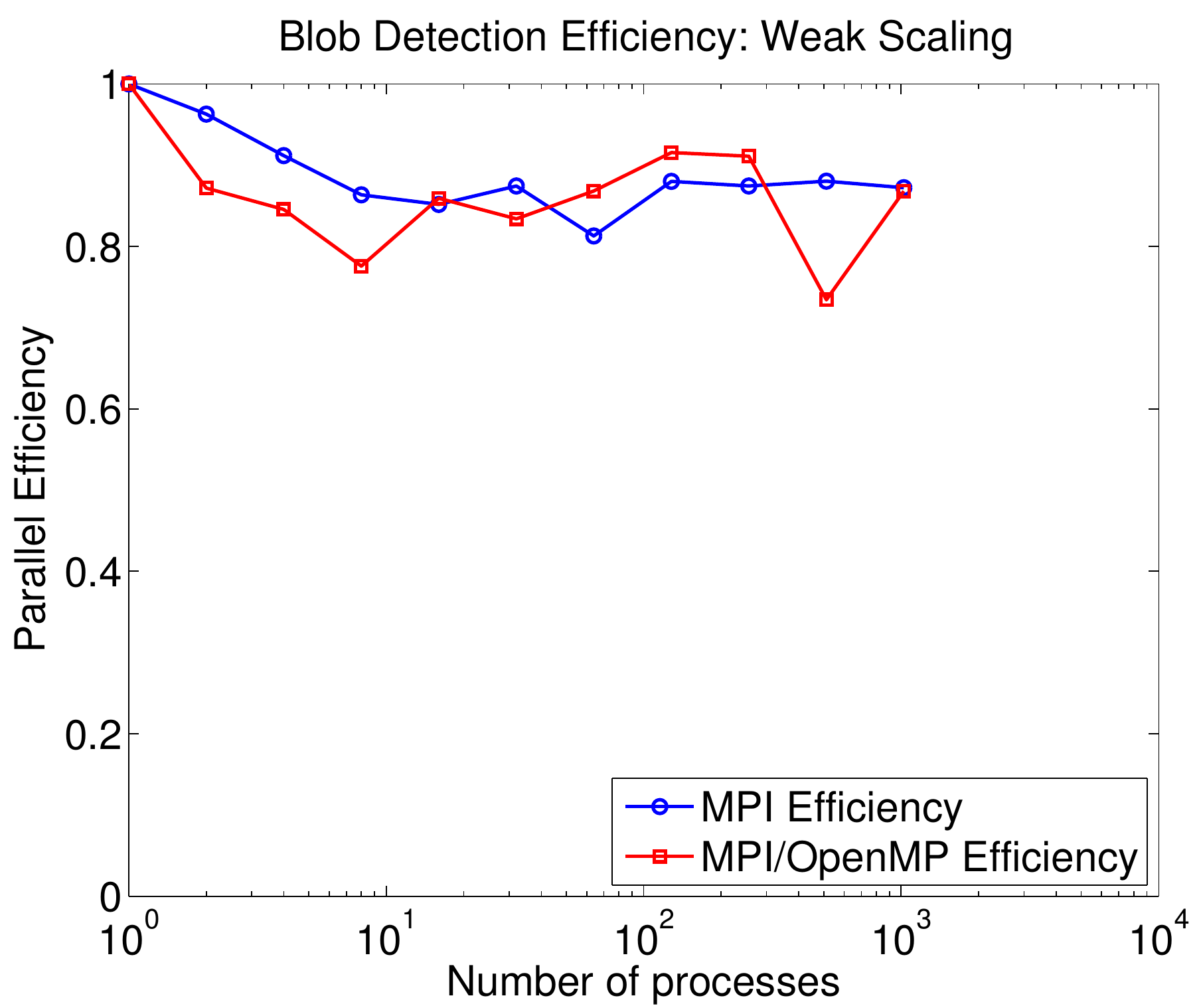}
                \caption{Speedup}
                \label{blob_all_efficiency_weakscaling}
        \end{subfigure}
        \caption{Blob detection time and speedup with MPI and MPI/OpenMP varying number of processes under weak scaling}
        \label{fig:Real time blob detection with MPI/OpenMP under weak scaling}
\end{figure*}

In this experiment, we evaluate the performance of our real-time blob detection under weak scaling. We replicate existing data sets (30GB) in order to obtain adequate experimental data sets (4.3TB). The basic unit data contains 128 time frames and the size of data increases linearly with the number of processes.
In Figure \ref{fig:Real time blob detection with MPI/OpenMP under weak scaling}, the blob detection time remains almost constant under weak scaling, which indicates that our implementations scale very well to solve much larger problems. Also, both MPI and MPI/OpenMP implementations achieve high parallel efficiency as the number of processes increases from 1 to 1024. 

\section{Conclusion and future work} 
Near real-time extraction of spatio-temporal features in very large-scale irregular data presents both opportunities and challenges responding to extreme scale computing and big data in many applications. In this paper, we propose, for the first time, a real-time blob detection and tracking approach for finding blob-filaments in fusion experiments or numerical simulations. The key idea of the proposed approach is to break down the overall process into three steps. The first two steps are based on a distribution-based outlier detection scheme with various criteria and a fast CCL method to find blob components. In the third step, an efficient blob tracking scheme is presented to recover the trajectories of the motions of blobs. Our hybrid MPI/OpenMP implementations demonstrate the effectiveness and efficiency of the proposed approach with a set of fusion plasma simulation data. Our tests show that we can achieve linear time speedup and complete blob detection in two or three milliseconds using a cluster at NERSC.

We are currently working on integrating our blob detection algorithm into the ICEE system for consuming fusion plasma data streams where the blob detection function is used in a central data analysis component and the resulting detection results are monitored and controlled from portable devices, such as an iPad. We plan to test the proposed method in both simulations and real fusion experiments.

% if have a single appendix:
%\appendix[Proof of the Zonklar Equations]
% or
%\appendix  % for no appendix heading
% do not use \section anymore after \appendix, only \section*
% is possibly needed

% use appendices with more than one appendix
% then use \section to start each appendix
% you must declare a \section before using any
% \subsection or using \label (\appendices by itself
% starts a section numbered zero.)
%

%\appendices
%\section{Proof of the First Zonklar Equation}
%Appendix one text goes here.
%
%% you can choose not to have a title for an appendix
%% if you want by leaving the argument blank
%\section{}
%Appendix two text goes here.

% use section* for acknowledgment
\ifCLASSOPTIONcompsoc
  % The Computer Society usually uses the plural form
  \section*{Acknowledgments}
\else
  % regular IEEE prefers the singular form
  \section*{Acknowledgment}
\fi

%The authors would like to thank Scientific Data Management Group at LBNL, and our collaborators in PPPL and ORNL for their contributions to this work. 
%The authors thank Edmund Novak and Daniel Graham for their valuable comments to improve the readability of this paper. 
The authors would like to thank the referees for their valuable comments. This work was supported by the Office of Advanced Scientific Computing Research, Office of Science, of the U.S. Department of Energy under Contract No. DE-AC02-05CH11231 and partially supported by NSF under grants No. CCF 1218349 and ACI SI2-SSE 1440700, and by DOE under a grant No. DE-FC02-12ER41890. The blobby plasma turbulence simulation was performed using resources of the Oak Ridge Leadership Computing Facility, which is a DOE Office of Science User Facility supported under Contract DE-AC05-00OR22.

% Can use something like this to put references on a page
% by themselves when using endfloat and the captionsoff option.
\ifCLASSOPTIONcaptionsoff
  \newpage
\fi

% trigger a \newpage just before the given reference
% number - used to balance the columns on the last page
% adjust value as needed - may need to be readjusted if
% the document is modified later
%\IEEEtriggeratref{8}
% The "triggered" command can be changed if desired:
%\IEEEtriggercmd{\enlargethispage{-5in}}

% references section

% can use a bibliography generated by BibTeX as a .bbl file
% BibTeX documentation can be easily obtained at:
% http://www.ctan.org/tex-archive/biblio/bibtex/contrib/doc/
% The IEEEtran BibTeX style support page is at:
% http://www.michaelshell.org/tex/ieeetran/bibtex/
\bibliographystyle{IEEEtran}
% argument is your BibTeX string definitions and bibliography database(s)
\bibliography{IEEEabrv,blob_TBD}
%
% <OR> manually copy in the resultant .bbl file
% set second argument of \begin to the number of references
% (used to reserve space for the reference number labels box)
%\begin{thebibliography}{1}
%
%\bibitem{IEEEhowto:kopka}
%H.~Kopka and P.~W. Daly, \emph{A Guide to \LaTeX}, 3rd~ed.\hskip 1em plus
%  0.5em minus 0.4em\relax Harlow, England: Addison-Wesley, 1999.
%
%\end{thebibliography}

% biography section
% 
% If you have an EPS/PDF photo (graphicx package needed) extra braces are
% needed around the contents of the optional argument to biography to prevent
% the LaTeX parser from getting confused when it sees the complicated
% \includegraphics command within an optional argument. (You could create
% your own custom macro containing the \includegraphics command to make things
% simpler here.)
\begin{IEEEbiography}[{\includegraphics[width=1in,height=1.25in,clip,keepaspectratio]{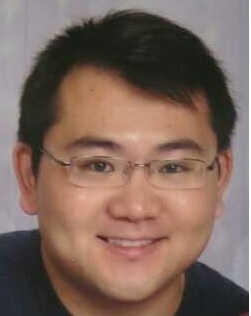}}]{Lingfei Wu}
is a 6th year Ph.D. candidate in the computer science department at College of William and Mary, advised by Dr. Andreas Stathopoulos. His research interests are in the areas of high-performance scientific computing, large-scale machine learning, and big data analytics. In summer 2014 and 2015, Lingfei was a summer research intern at Lawrence Berkeley National Laboratory and IBM T.J.Watson Research Center, respectively. Before joining William and Mary, Lingfei received his M.S. from University of Science and Technology of China (Hefei, 2010), following his B.E. from Auhui University (Hefei, 2007)\end{IEEEbiography}

\begin{IEEEbiography}[{\includegraphics[width=1in,height=1.25in,clip,keepaspectratio]{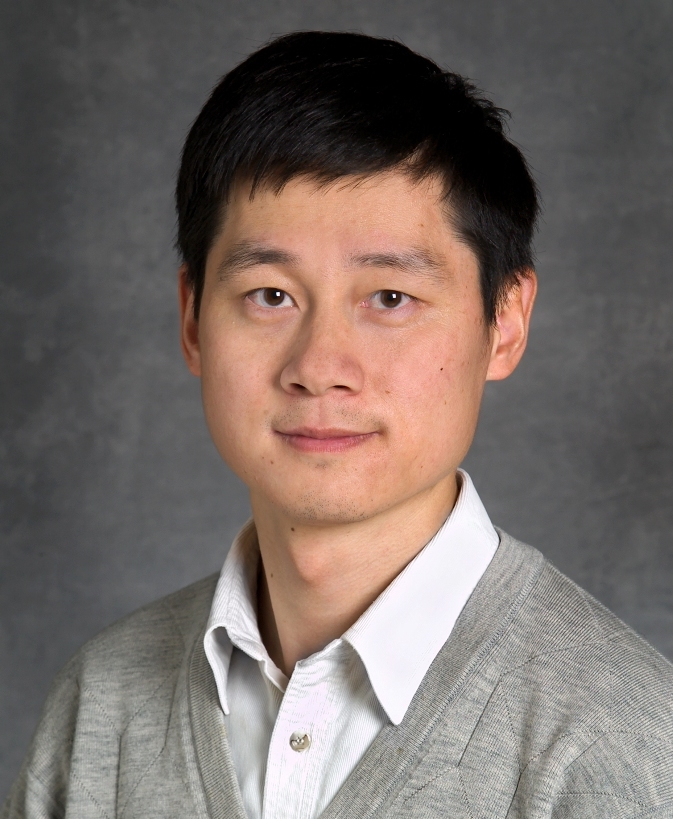}}]{Kesheng (John) Wu}
is a Senior Computer Scientist at Lawrence Berkeley National Laboratory. works actively on a number of topics in data analysis, data
management, and high-performance computing. His recent algorithmic
research work includes bitmap indexing techniques for searching large
datasets, statistical methods for extract features from a variety of
data, and restarting strategies for computing extreme eigenvalues. He
authored and coauthored more than 100 technical publications, nine of
which have more than 100 citations each.
\end{IEEEbiography}

\begin{IEEEbiography}[{\includegraphics[width=1in,height=1.25in,clip,keepaspectratio]{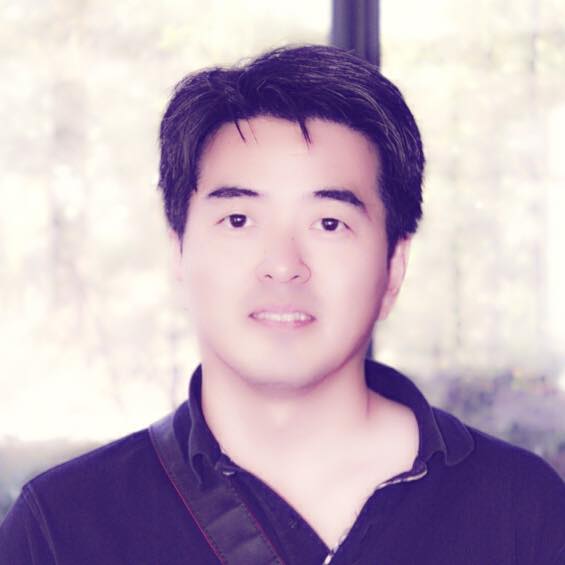}}]{Alex Sim}
is a Senior Computing Engineer at Lawrence Berkeley National Laboratory. His current R\& D activities focus on data mining and modeling, data analysis methods, distributed resource management, and high performance data systems. He authored and coauthored more than 100 technical publications, and released a few software packages under open source license.
\end{IEEEbiography}

\begin{IEEEbiography}[{\includegraphics[width=1in,height=1.25in,clip,keepaspectratio]{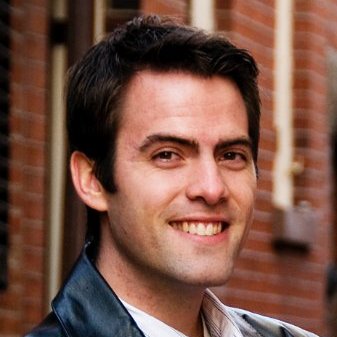}}]{Michael Churchill}
received his PhD in Nuclear Science and Engineering from the Massachusetts Institute of Technology, and his Bachelor's degree in Electrical and Computer Engineering from Brigham Young University. He currently works as a postdoc under Dr. C.S. Chang at the Princeton Plasma Physics Laboratory, focusing on data management and analysis with the large-scale XGC plasma simulation codes.
\end{IEEEbiography}

\begin{IEEEbiography}[{\includegraphics[width=1in,height=1.25in,clip,keepaspectratio]{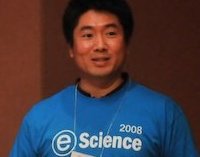}}]{Jong Y. Choi}
is a researcher working in Scientific Data Group, Computer Science and Mathematics Division, Oak Ridge National Laboratory (ORNL), Oak Ridge, Tennessee, USA. He earned his Ph.D. degree in Computer Science at Indiana University Bloomington in 2012 and his MS degree in Computer Science from New York University in 2004. His areas of research interest span data mining and machine learning algorithms, high-performance data-intensive computing, parallel and distributed systems for Cloud and Grid computing. 
\end{IEEEbiography}

\begin{IEEEbiography}[{\includegraphics[width=1in,height=1.25in,clip,keepaspectratio]{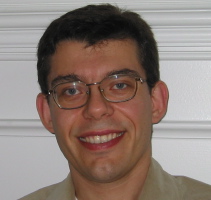}}]{Andreas Stathopoulos}
received a bachelors degree in Mathematics from University of Athens, Greece in 1989, and an M.S. and Ph.D. degrees in Computer Science from Vanderbilt University, USA in 1991 and 1995 respectively. In 1995, he obtained an NSF CISE postdoctoral fellowship to work with Prof. Yousef Saad at the University of Minnesota, USA.  In 1997, he joined the faculty of Computer Science at the College of William and Mary, USA, where he currently holds a full professor position. His research interests span the general area of parallel and high performance scientific computing. He is an expert on solving eigenvalue problems and has developed the state-of-the-art eigenvalue software PRIMME.
% He has also collaborated with scientists in atomic physics, materials science, QCD, and plasma physics.
\end{IEEEbiography}

\begin{IEEEbiography}[{\includegraphics[width=1in,height=1.25in,clip,keepaspectratio]{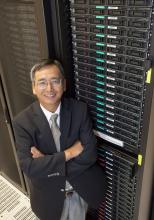}}]{Choong-Seock Chang}
is the head of the multi-institutional multi-disciplinary US SciDAC Center for Edge Physics Simulation (EPSI), headquartered at Princeton Plasma Physics Laboratory, Princeton University, awarded by US Department of Energy, Office of Fusion Energy Science and Office of Advanced Scientific Computing Research, jointly. He has over 150 publications in peer reviewed internationally recognized journals, and has given countless invited talks, keynote speeches and tutorial lectures at major scientific conferences. 
\end{IEEEbiography}

\begin{IEEEbiography}[{\includegraphics[width=1in,height=1.25in,clip,keepaspectratio]{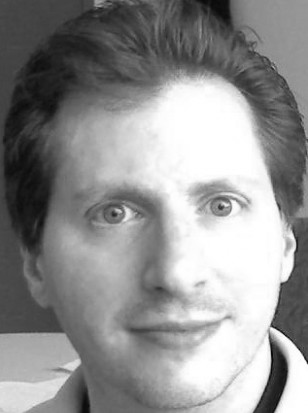}}]{Scott Klasky}
is currently a senior research scientist at Oak Ridge National Laboratory, and head of the end to end task in the scientific computing group at the National Center Computational Sciences. He has 20 years of experience in research and development of middleware for use in high performance computing, and is a author/co-author of over 40 papers in the field of physics and high performance computing. He received his PhD in physics from the University of Texas at Austin, and then went on to Syracuse University, and then the Princeton Plasma Physics Laboratory. 
%His research centers in data intensive computing, high performance computing, workflow automation, I/O, visualization, and fusion.
\end{IEEEbiography}

% if you will not have a photo at all:

%\begin{IEEEbiographynophoto}{Lingfei Wu}
%\end{IEEEbiographynophoto}

%\begin{IEEEbiographynophoto}{Kesheng (John) Wu}
% 
%\end{IEEEbiographynophoto}

%\begin{IEEEbiographynophoto}{Alex Sim}
%
%\end{IEEEbiographynophoto}

%\begin{IEEEbiographynophoto}{Michael Churchill}
%Biography text here.
%\end{IEEEbiographynophoto}

%\begin{IEEEbiographynophoto}{Jong Y. Choi}
%Biography text here.
%\end{IEEEbiographynophoto}
%
%\begin{IEEEbiographynophoto}{Andreas Stathopoulos}
%Biography text here.
%\end{IEEEbiographynophoto}
%
%\begin{IEEEbiographynophoto}{CS Chang, 
%        and~Scott~Klasky}
%Biography text here.
%\end{IEEEbiographynophoto}
%
%\begin{IEEEbiographynophoto}{Scott Klasky}
%Biography text here.
%\end{IEEEbiographynophoto}

% You can push biographies down or up by placing
% a \vfill before or after them. The appropriate
% use of \vfill depends on what kind of text is
% on the last page and whether or not the columns
% are being equalized.

\vfill

% Can be used to pull up biographies so that the bottom of the last one
% is flush with the other column.
%\enlargethispage{-5in}

% that's all folks
\end{document}